\documentclass[11pt, preprint]{aastex}
\usepackage[utf8x]{inputenc}
\usepackage{natbib}
\usepackage{graphicx}
\usepackage{amsmath} 

\title{Broad [CII] line wings as tracer of molecular and multi-phase outflows in infrared bright galaxies}
\author{A.W. Janssen\altaffilmark{1}, 
N. Christopher\altaffilmark{2, 3},
E. Sturm\altaffilmark{1},
S. Veilleux\altaffilmark{4,5},
A. Contursi\altaffilmark{1},
E. Gonz{\'a}lez-Alfonso\altaffilmark{6},
J. Fischer\altaffilmark{7},
R. Davies\altaffilmark{1},
A. Verma\altaffilmark{2},
J. Graci{\'a}-Carpio\altaffilmark{1},
R. Genzel\altaffilmark{1},
D. Lutz\altaffilmark{1},
A. Sternberg\altaffilmark{8},
L. Tacconi\altaffilmark{1},
L. Burtscher\altaffilmark{1}, 
A. Poglitsch\altaffilmark{1}
}
\email{janssen@mpe.mpg.de}

\altaffiltext{1}{Max-Planck-Institute for Extraterrestrial Physics (MPE), Giessenbachstra{\ss}e 1, 85748 Garching, Germany}
\altaffiltext{2}{Oxford University, Dept. of Astrophysics, Oxford OX1 3RH, UK}
\altaffiltext{3}{School of Sciences, European University Cyprus, Diogenes Street, Engomi, 1516 Nicosia, Cyprus}
\altaffiltext{4}{Department of Astronomy, University of Maryland, College Park, MD 20742, USA}
\altaffiltext{5}{Joint Space-Science Institute, University of Maryland, College Park, MD 20742, USA}
\altaffiltext{6}{Universidad de Alcal{\'a} de Henares, 28871 Alcal{\'a} de Henares, Madrid, Spain}
\altaffiltext{7}{Naval Research Laboratory, Remote Sensing Division, 4555 Overlook Ave SW, Washington, DC 20375, USA}
\altaffiltext{8}{Tel Aviv University, Sackler School of Physics \&\ Astronomy, Ramat Aviv 69978, Israel}

\begin{document}

\begin{abstract}
{We report a  tentative correlation between the outflow characteristics derived from OH absorption at $119\,\mu\text{m}$ and [CII] emission at $158\,\mu\text{m}$ in a sample of 22 local and bright ultraluminous infrared galaxies (ULIRGs). For this sample we investigate whether [CII] broad wings are a good tracer of molecular outflows, and how the two tracers are connected. Fourteen objects in our sample have a broad wing component as traced by [CII], and all of these also show OH119 absorption indicative of an outflow (in 1 case an inflow). The other eight cases, where no broad [CII] component was found, are predominantly objects with no OH outflow or a low-velocity ($\leq 100\,\text{km s}^{-1}$) OH outflow. The full width at half maximum (FWHM) of the broad [CII] component shows a trend with the OH119 blue-shifted velocity, although with significant scatter. Moreover, and despite large uncertainties, the outflow masses derived from OH and broad [CII] show a 1:1 relation. The main conclusion is 
therefore that broad [CII] 
wings can be used to trace molecular outflows. This may be particularly relevant at high redshift, where the usual tracers of molecular gas (like low-J CO lines) become hard to observe. 
Additionally, observations of blue-shifted Na I D $\lambda\lambda 5890,5896$ absorption are available for ten of our sources. Outflow velocities of Na I D show a trend with OH velocity and broad [CII] FWHM. These  observations suggest that the atomic and molecular gas phases of the outflow are connected. }
\end{abstract}

  \keywords{ISM: jets and outflows ---
                galaxies: active ---
                galaxies: evolution ---
		galaxies: starburst
               }

\maketitle

\section{INTRODUCTION}

Feedback via powerful winds is thought to play a critical role in galaxy evolution. Without any form of feedback, galaxy evolution models overpredict star formation in the early universe, and spheroid dominated remnants would predominate in the current universe, which is not observed \citep{Cole-1991}. In order to match observations, current models implement, among other things, gas inflows from outside the halo and feedback from stars and active galactic nuclei (AGN). The output is then compared to observables like the galactic stellar mass function, the Tully-Fisher relation, the mass-metallicity relation, and the apparent co-evolution of the central Black Hole (BH) mass with the stellar mass (\cite{Murray-2005, Sales-2010, Fabian-2012, Vogelsberger-2013, Heckman-2014}, also see \cite{Silk-2011} for a short review on this topic).  Negative feedback, in the form of winds or outflows, is usually required to match the observations.  

In the local universe, outflows have been observed in all types of galaxies and in different gas phases. Ionized gas outflows have been observed in dwarf galaxies \citep{Marlowe-1995, Martin-1998} and Seyfert 2 AGNs \citep{Cecil-1990, Harrison-2014}. In ultraluminous infrared galaxies (ULIRGs), outflows are observed in several phases: ionized \citep{Spoon-2009, Westmoquette-2012, Arribas-2014}, molecular \citep{Sakamoto-2009, Fischer-2010, Feruglio-2010, Sturm-2011b, Veilleux-2013, Spoon-2013, Cicone-2014} and atomic \citep{Rupke-2013a}. \citet{Morganti-2005} discuss neutral gas outflows in radio galaxies. Possible drivers of the outflows are stars and/or AGN. Theoretically, star formation has the power to remove significant amounts of gas from a Milky-Way sized galaxy \citep{Efstathiou-2000, Hopkins-2012}. Indeed, star formation has been observed to drive outflows in the form of H-alpha bubbles above disk galaxies \citep{Rand-1996}. Blue-shifted absorption of Na I D and Mg II in star-forming galaxies may 
also be driven by star formation \citep{Heckman-2011, Diamond-2012}, although the star formation rate must reach a certain threshold before any correlations with the outflow properties become apparent \citep{Heckman-2015}. 
In massive galaxies however, star formation alone may not be sufficient to power the outflows which can reach velocities $>> 500\,\text{km}\,\text{s}^{-1}$. In ULIRGs outflow velocities can reach $1500\,\text{km}\,\text{s}^{-1}$ and AGN are the most probable driver (e.g. \cite {Sturm-2011b, Cicone-2014, Spoon-2013, Veilleux-2013}). 

Detections of such outflows at high redshift would confirm that feedback also played an important role in the early universe. These outflows have been detected at $z > 1$, in type 2 AGN \citep{Alexander-2010, Nesvadba-2011, Harrison-2012, Cano-Diaz-2012}, in massive star-forming galaxies \citep{Weiner-2009, Banerji-2011, Genzel-2014, Natasha-2014} and in radio galaxies \citep{Nesvadba-2008}, however mostly in ionized or atomic gas. Since most mass of a gas-rich galaxy is in molecular form, observations of molecular outflows at high redshift would be most helpful in constraining the role of feedback in the early universe. Candidates for these observations are CO rotational lines and OH rotational lines in absorption, preferably observed with sensitive telescopes like ALMA or NOEMA. However, molecular outflows have not yet been detected in high-J CO lines, and low-J CO lines quickly shift out of the ALMA and NOEMA bands.  

The ground state OH $^{2}\Pi_{3/2}$  $ 5/2 \rightarrow 3/2$ transition at 119$\,\mu\text{m}$, or OH119 for short, probes molecular and to a lesser extent atomic gas \citep{Meijerink-2005}. OH119 becomes observable with ALMA at z=2 (in the $450\,\mu\text{m}$ atmospheric window in Band 9),  but a high S/N and bright continuum is needed to  detect the outflow in absorption. ULIRGs are intrinsically bright in the far infrared (FIR), providing the continuum for the OH119 absorption, but not all galaxies have such a bright FIR continuum. Detections of blue-shifted OH119 emission are possible but rare at high redshift \citep{Riechers-2014}. Direct observations of molecular outflows at high redshift are therefore challenging. 

As an alternative, the collisionally-excited fine structure transition of C$^{+}$  $^{2}\text{P}_{3/2} \rightarrow ^{2}\text{P}_{1/2}$ at rest frame wavelength of 157.741 $\mu\text{m}$ (hereafter [CII]) may be used. It is one of the brightest lines in spectra of star-forming galaxies, as it is the dominant coolant of the ISM over a wide range of physical conditions. Most [CII] emission is thought to arise in Photodissociation Regions (PDRs) around starforming regions, where the intensity of $> 13.6\,\text{eV}$ photons has dropped sufficiently for hydrogen to stay atomic, while carbon remains ionized (carbon has a slightly lower ionization potential of 11.26 eV). \citet{Pineda-2013} find that in our galaxy about 50\% of [CII] emission arises from these PDRs, about 30\% arises from CO-dark H$_{2}$ gas, about 20\% from atomic gas, and only very little from hot ionized gas. 
[CII] shifts into the $450\, \mu\text{m}$ windows of ALMA at redshifts $z = 1.85$, and  in the $0.8\, mm$ band of NOEMA at $z = 4$.  High velocity outflows appear as broad wings in  the [CII] line profile, and such broad wings have been detected at high redshift, for example in the QSO SDSS J1148+5251 $z = 6.4$ \citep{Maiolino-2012, Cicone-2015}. 

In this paper, we investigate how well [CII] traces the molecular outflows in ULIRGs, by comparing the outflow velocity and mass retrieved from {\it Herschel}\footnote{Herschel is an ESA space observatory with science instruments provided by European-led Principal Investigator consortia and with important participation from NASA.}/PACS observations of [CII] and OH119. The ULIRGs are all gas-rich and most contain an AGN.  A short description of the sample is given in Section \ref{sample}. We then cover the PACS data reduction in Section \ref{data}. The [CII] line profiles have been fitted with Gaussian line profiles, and the FHWM of the broad component is compared to the OH blue-shifted velocity in Section \ref{analysis}. In the same Section we compare the FWHM of the  broad [CII] component with the AGN luminosity, with the Na I D blue-shifted absorption velocity and with CO(1-0) line profiles.  A discussion of the outflow masses and possible connections between the different phases follows in 
Section \ref{discussion}. We adopt $H_{0} = 70\,\text{km}\,\text{s}^{-1}\,\text{Mpc}^{-1}$ throughout this paper.

\section{SAMPLE DESCRIPTION}
\label{sample}
The sample consists of all 22 local ($z < 0.1$) ULIRGs from the IRAS Revised Bright Galaxy Sample \citep{Sanders-2003}; these are objects with $L_{8-1000\,\mu\text{m}} \geq 10^{12}\,L_{\odot}$. IRAS F09111-1007 lies above this luminosity cutoff, but is faint in the PACS range and therefore excluded. We include, however, NGC 6240 and UGC 5101 which are formally slightly below this cutoff, but in any relevant aspect for this study have similar charateristics as the rest of the sample. See Table \ref{tab:sample} for the redshifts, coordinates and AGN fractions of the sample. The AGN fractions are estimated using the ratio of the 30 micron to 15 micron flux densities, $f_{30}/f_{15}$ as measured with the Spitzer Infrared Spectrograph, following \citet{Veilleux-2009}. 

\begin{deluxetable}{llrrrrr}
 \tablecaption{Sample}
 \tablehead{ \colhead{Name} & \colhead{Alternative name} & \colhead{z} & \colhead{RA} & \colhead{DEC} & \colhead{ $\alpha_{\text{AGN}}$ (\%)} & \colhead{log($L_{AGN}$)} 
 } 
 
\startdata

IRAS F05189-2524 & 		& 0.0426 & 05:21:01.4 & $-$25:21:46 & 72 & 12.07 \\
IRAS 07251-0248 & 		& 0.0876 & 07:27:37.5 & $-$02:54:55 & 30 & 11.92 \\ 
IRAS F08572+3915 & 		& 0.0584 & 09:00:25.0 & 39:03:56 & 70 & 12.05  \\ 
IRAS 09022-3615 & 		& 0.0596 & 09:04:12.8 & $-$36:27:02 & 55 & 12.09  \\ 
UGC 5101 & IRAS F09320+6134 	& 0.0394 & 09:35:48.8 & 61:21:22 & 56	& 11.8  \\ 
IRAS F10565+2448 & 		& 0.0431 & 10:59:17.4 & 24:32:38 & 47  & 11.77 \\ 
IRAS F12112+0305 & 		& 0.0733 & 12:13:47.4 & 02:48:34 & 18 & 11.63  \\ 
Mrk 231 & IRAS F12540+5708 	& 0.0422 & 12:56:15.0 & 56:52:17 & 81  & 12.51 \\ 
IRAS 13120-5453 & 		& 0.0308 & 13:15:06.2 & $-$55:09:24 & 33 & 11.83 \\ 
Mrk 273 & IRAS F13428+5608 	& 0.0378 & 13:44:41.8 & 55:53:14 & 34  & 11.74 \\ 
IRAS F14348-1447 & 		& 0.0830 & 14:37:37.3 & $-$15:00:20 & 17 & 11.64 \\ 
IRAS F14378-3651 & 		& 0.0676 & 14:40:57.8 & $-$37:04:25 & 21 & 11.5 \\ 
IRAS F15250+3608 & 		& 0.0552 & 15:26:59.9 & 35:58:34 & 42  & 11.73 \\ 
Arp 220 &  IRAS F15327+2340 	& 0.0181 & 15:34:57.1 & 23:30:10 & 5.8 & 10.98 \\ 
NGC 6240 & IRAS F16504+0228  	& 0.0245 & 16:52:58.6 & 02:24:03 & 56  & 11.65 \\ 
IRAS F17207-0014 & IRAS 17208-0014& 0.0428 & 17:23:21.4 & $-$00:17:00 & $\le 5$ & $\le11.15$ \\
IRAS F19297-0406 & 		& 0.0857 & 19:32:22.1 & $-$04:00:02 & 23 & 11.81\\ 
IRAS 19542+1110 & 		& 0.0624 & 19:56:35.4 & 11:19:03 & 26 & 11.52 \\
IRAS F20551-4250 & 		& 0.0430 & 20:58:27.4 & $-$42:38:57 & 57 & 11.87\\
IRAS F22491-1808 & 		& 0.0778 & 22:51:49.0 & $-$17:52:27 & 14 & 12.05\\
IRAS F23128-5919 & ESO 148-IG 002& 0.0446 & 23:15:46.6 & $-$59:03:14 & 63 & 11.89\\
IRAS F23365+3604 & 		& 0.0645 & 23:39:01.7 & 36:21:14 & 45 & 11.87 \\

\enddata
\tablecomments{Redshifts, coordinates and AGN fractions and luminosities of the observed sample. AGN fractions are calculated from $f_{30}/f_{15}$, as described in \citet{Veilleux-2009}.}
\label{tab:sample}
\end{deluxetable}

Because ULIRGs represent the IR-bright phase of mergers of gas-rich galaxies \citep{Sanders-1996}, several objects in our sample have resolved double nuclei related to interaction and merging.  All known double nuclei are covered completely by our PACS observations. 

\section{OBSERVATIONS AND DATA REDUCTION}
\label{data}
Observations were made with the Photodetector Array Camera and Spectrometer (PACS) \citep{PACS-2010}, on board the Herschel Space Observatory \citep{Pilbratt-2010}. Both the OH  $119\, \mu\text{m}$ and the [CII] $158\,\mu\text{m}$ spectra have been obtained as part of the \textit{Herschel} guaranteed time key program SHINING (PI: E. Sturm). We reduced and analyzed the OH data in \citet{Veilleux-2013}, (V13). Because the OH119 line appears either in pure absorption, pure emission, or as a  P-Cygni profile, the spectra were fitted with an absorption component, an emission component, or both. (21/22) spectra have absorption features, which are mostly blue-shifted. We will use the velocities of the absorption features of OH in our analysis using the object's  $v_{50}$(abs), $v_{84}$(abs) and $v_{max}$(abs) measured by V13 (see Table \ref{tab:oh_observations}). $v_{50}$(abs) is the median velocity of the fitted absorption profile, i.e. 50\% of the flux arises above (from velocities more positive than) $v_{50}$(
abs), $v_{84}$(abs) is the velocity above which 84\% of the absorption takes place, and $v_{max}$(abs) is the velocity, above which all absorption takes place. $v_{max}$(abs) however depends strongly on the fit to the continuum and the signal to noise ratio, and this introduces typical uncertainties of $\pm 200\,\text{km}\,\text{s}^{-1}$.  
We characterise the molecular outflows by their blue-shifted absorption of a FIR continuum. Therefore, only gas along our line of sight to the ULIRG center is traced. The P-Cygni profiles in OH119 and the large number of objects with blue-shifted OH absorption led \citet{Sturm-2011b} and \citet{Veilleux-2013} to the conclusion that the outflows are isotropic, or have at least a large opening angle. Because [CII] appears in emission, such an isotropic outflow would become visible as blue-shifted and red-shifted wings in the line profile.

\begin{deluxetable}{lrrrcr}
 
\tablecaption{OH and Na I D observations}
\tablehead{ \colhead{Name} & \colhead{OH $v_{50}$} & \colhead{OH $v_{84}$} & \colhead{OH $v_{max}$} & \colhead{Na I D range} & \colhead{OH mass} \\
\colhead{} & \colhead{$\text{km}\,\text{s}^{-1}$} & \colhead{$\text{km}\,\text{s}^{-1}$} & \colhead{$\text{km}\,\text{s}^{-1}$} & \colhead{$\text{km}\,\text{s}^{-1}$} & \colhead{$10^{8} M_{\odot}$}
} 

\startdata
IRAS F05189-2524 &  -327 & -574 & -850 & -402 to -92 (2) & 1.27\\ 
IRAS 07251-0248 & -63 & -255 & -550 & ... & ...\\ 
IRAS F08572+3915 & -489 & -832 & -1100 & −700 to -1000 (3) & 0.49\\ 
IRAS 09022-3615 &  -153 & -297 & -650 & ... & ...\\ 
UGC 5101	& -9 & -225 & -1200 & -16 to 11 (1) & ...\\ 
IRAS F10565+2448 &  -267 & -489 & -950 & -309  to -141 (1) & 2.99\\ 
IRAS F12112+0305 &  -117 & -237 & -400 & ... & ...\\ 
Mrk 231		&  -237 & -610 & -1500 & -386 (2) & 6.66\\ 
IRAS 13120-5453 & -195 & -520 & -1200 & ... & 1.91\\ 
Mrk 273		&  -201 & -495 &  -750 & 29 (2) & 1.27\\ 
IRAS F14348-1447 &  -291:& -508 & -900 & ... & 15.2\\ 
IRAS F14378-3651 &  -219 & -556 & -1200 & ... & 0.91\\ 
IRAS F15250+3608 &  189 & -21 & ... & ... & ...\\ 
Arp 220 	&  21   & -153 & -700 & ... & ...\\
NGC 6240	&  -207 & -544 & -1200 & -99 to -74 (1) & ... \\ 
IRAS F17207-0014 &  51 & -165 & ... & -298 to 27 (1) & ...\\ 
IRAS F19297-0406 &  -231 & -532 & -1000 & -413 to -114 (4) & ...\\ 
IRAS 19542+1110 & -93: & -489: & -700 & ... & ...\\ 
IRAS F20551-4250 &  -381 & -748 & -1200  & ... & 0.91\\ 
IRAS F22491-1808 & 99 & 3:: & ... & ... & ...\\ 
IRAS F23128-5919 &  ... & ... & ... & ... & ...\\ 
IRAS F23365+3604 &  -243 & -604 & -1300 & -384 $\pm$ 96 (4) & 5.34\\ 

\enddata

\tablecomments{OH data adopted from V13. Negative velocities indicate blue-shifted absorption, positive values indicate red-shifted absorption. The uncertainties on the OH velocities are typically $50\,\text{km s}^{-1}$, unless the value is followed by a colon (when the uncertainty is $50-150\,\text{km s}^{-1}$), or a double colon (when the uncertainty is larger than $150\,\text{km s}^{-1}$). $v_{max}$(abs) values have typical uncertainties of $200\,\text{km s}^{-1}$. References for Na I D data (1): \citet{Rupke-2005a} (2): \citet{Rupke-2005b} (3): \citet{Rupke-2013b} (4): \citet{Martin-2005} OH masses are preliminary, final results will be published in Gonz\'{a}lez-Alfonso (in preparation).}
\label{tab:oh_observations}
\end{deluxetable}

The [CII] observations were made in PACS range scan mode with a $2600\,\text{km}\,\text{s}^{-1}$ velocity range. Because of the instanteneous coverage of $\sim 1650\,\text{km}\,\text{s}^{-1}$, the total spectral range is from $-2950$ to $+2950\,\text{km}\,\text{s}^{-1}$, with decreasing S/N towards the edges. The data reduction was done using the standard PACS reduction and calibration pipeline included in HIPE 13, with upsample and oversample set to 2 (resulting in $60\,\text{km}\,\text{s}^{-1}$ sampling in the spectral direction). 

We extracted both the point source corrected spectrum from the central spatial pixel (spaxel;  one spaxel covers $(9.4'')^{2}$ on-sky) and the point source corrected sum of the central 9 spaxels. Although the central spaxel has the best signal to noise ratio, the [CII] emission might be slightly extended, or the telescope badly pointed, in which case the central 9 spaxels were used (Table \ref{tab:cii_obs}). 

The pointing was verified with a footprint plot, where a raster of the 
spectroscopic observations is overlaid on a PACS photometry map with higher spatial resolution. The source was not well centered on the central spaxel in four cases (F10565+2448, F12112+0305, 19542+1110, F23128-5919). In 5 other cases the final spectrum was extracted from the central 9 spaxels because the broad wings in the [CII] line profile differ from the central spaxel, hinting at slightly extended [CII] emission. Among these are the sources with double nuclei 08572+3915, separated by $6''$, and IRAS 23128-5919, separated by $4.5''$  \citep{Surace-1998,Zenner-1993}. This choice does not affect the line fluxes much, because the central spaxel's flux is scaled to the (point source corrected) 9 spaxel flux, to account for the telescope Point Spread Function (PSF; $13''$ at 158 $\mu\text{m}$) and minor mispointings. The uncertainty in the line fluxes is 10\% for flux densities $> 1$ Jy, but could be $\sim 30\%$ for any faint wings. 
 
\begin{deluxetable}{lllrrrr}
 \tablecaption{[CII] observations}
\tablehead{ \colhead{Name} & \colhead{OBSID} & \colhead{Extraction} & \colhead{FWHM} & \colhead{$1\sigma$ error} & \colhead{$L_{\text{outflow}}$} & \colhead{$M_{\text{outflow}}$} \\
\colhead{} & \colhead{} & \colhead{Spaxels} & \colhead{$\text{km}\,\text{s}^{-1}$} & \colhead{$\text{km}\,\text{s}^{-1}$} & \colhead{$10^{8} L_{\odot}$} & \colhead{$10^{8} M_{\odot}$}
} 
\startdata

IRAS F05189-2524 & 1342219442 & central & 1213 & 412 & 0.68 & 1.45 \\
IRAS 07251-0248 &  1342207825 & central & - & 502 & - & - \\
IRAS F08572+3915 & 1342208952 & $3\times 3$  & 1567 & 71 & 0.63 & 1.34 \\ 
IRAS 09022-3615 & 1342209403  &  central & 693 & 1 & 9.28 & 19.86 \\
UGC 5101 	& 1342208949 &  $3\times 3$ & 1158 & 394 & 0.6 & 1.28 \\
IRAS F10565+2448 &  1342207788& $3\times 3$ & 882 & 1 & 2.01 & 4.29 \\
IRAS F12112+0305 & 1342210832 & $3\times 3$ & - & 161 & - & - \\
Mrk 231 	&  1342186811 & central & 940 & 8 & 1.48 & 3.18 \\
IRAS 13120-5453 &  1342214629 & central & 1787 & 601 & 0.62 & 1.33 \\
Mrk 273 	&  1342207802 & central & 852 & 295 & 1.85 & 3.95 \\
IRAS F14348-1447 & 1342224242 &  $3\times 3$ & 887 & 1 & 5.06 & 10.82 \\
IRAS F14378-3651 & 1342204338 &  central & - & 222 & - & - \\
IRAS F15250+3608 & 1342213752 &  central & - & 120 & - & - \\
Arp 220		 & 1342238930 & central  & - & 173 & - & - \\
NGC 6240 	&  1342216623 & $3\times 3$ & 934 & 29 & 8.23 & 17.62 \\
IRAS F17207-0014 & 1342229693 &  central & - & 177 & - & - \\
IRAS F19297-0406 &  1342208891& central & 1025 & 120 & 4.13 & 8.83 \\
IRAS 19542+1110 & 1342208916 & $3\times 3$ & - & 122 & - & - \\
IRAS F20551-4250 & 1342208934 &  $3\times 3$ & 768 & 22 & 1.41 & 3.01 \\
IRAS F22491-1808 & 1342211825 & central & 409 & 158 & 1.98 & 4.23 \\
IRAS F23128-5919 & 1342210395 & $3\times 3$ & 1027 & 18 & 2.27 & 4.87 \\
IRAS F23365+3604 & 1342212515 & central & - & 106 & - & - \\
\enddata
\tablecomments{[CII] observations: PACS obsid, the spaxels from which the spectrum was extracted, the average [CII] broad component FHWM after subtracting $240\,\text{km s}^{-1}$ in quadrature (see Appendix A), the error as explained in Section \ref{fits}, the luminosity of the [CII] broad component and the total neutral gas mass as described in Section \ref{mass}. If no broad component is found, the error is the FHWM of the narrow component divided by 3. Outflow masses above $10^{9} M_{\odot}$ should be regarded as upper limits because the outflow emission is contaminated by emission from the host galaxy.}
\label{tab:cii_obs}
\end{deluxetable}

\section{ANALYSIS AND RESULTS}
\label{analysis}

More than half of all objects in our sample have broad wings in [CII] (14/22), as shown in the figures in Appendix A. Usually the wings appear both red-shifted and blue-shifted from the host galaxy, resulting in a symmetric line profile. This is expected if [CII] traces an outflow with a large opening angle, which looks rather like an expanding shell than a single blob. We will now assume that all OH outflows have a large opening angle, so that we can compare the blue-shifted velocity of the OH absorption with the full width at half maximum (FWHM) of the [CII] outflow.

\subsection{Gaussian fits to the [CII] spectra}
\label{fits}

We have attempted to fit two Gaussian components (a narrow and a broad one) to the [CII] spectra. This should separate host galaxy emission from outflow emission. In most cases it is clear whether a broad component improves the fit or not. In case of doubt, we identify the best fit based on the reduced $\chi^{2}$ value, i.e. we deem an outflow is present when the reduced $\chi^{2}$ value of the two-component fit is lower than that of the one-component fit. We also visually inspect the fits to make sure that the second Gaussian fits the broad wings, not the central component. Figure \ref{fig:appA} in Appendix A shows the fits overplotted on the [CII] line profiles, while Table \ref{tab:appA} lists the Gaussian fits and their parameters.

The spectral resolution of PACS is $\sim 240\,\text{km}\,\text{s}^{-1}$ at $158\,\mu\text{m}$. Any Gaussian fit with a FWHM below $240\,\text{km}\,\text{s}^{-1}$ is therefore rejected. The PSF in PACS is not perfectly Gaussian, but it has no wings that could be confused with an outflow. In some objects, the fit to the continuum significantly affects the FWHM of the second component. Ideally, one would like to take the continuum far away from the systemic velocity of the host galaxy, in order to prevent the removal of broad wings. But the signal to noise ratio drops at (absolute) velocities above $1300\,\text{km}\,\text{s}^{-1}$.  Faint wings extending to $>1300\,\text{km}\,\text{s}^{-1}$ can therefore not be reliably observed in the spectra.  The fits have been done twice in order to estimate the error introduced in our method: once fitting the continuum close to the line, with the potential danger of removing the broad wings, and once fitting the continuum far away from the line,  leaving plenty of room for 
the broad wings.  We have adopted the average FWHM of these two fits to be the width of the [CII] broad component, and its difference to be the $3\sigma$ error. 
The difference can be very large in some cases: F05189-2524 and Mrk 273 are best fitted either with a second narrow Gaussian or with a broad component, while 07251-0248, 13120-5453, F14378-3651 and F22491-1808 can be fitted either with or without a broad component. In the first case, we take the FWHM of the broad component, to be consistent with the rest of our analysis. In the second case, we take the fit with the lowest reduced $\chi^{2}$ value.  The FWHM and error of the broad [CII] components are listed in Table \ref{tab:cii_obs}.  

There are some caveats in the analysis. Contamination by H$_{2}$O (158.3$\mu$m) may be present at $1080\,\text{km}\,\text{s}^{-1}$. It is not clear whether the bump around $1000\,\text{km}\,\text{s}^{-1}$ in F19297-0406 is caused by red-shifted [CII] emission, or by H$_{2}$O emission. Two scenarios are fitted: one where the emission around $1000\,\text{km}\,\text{s}^{-1}$ is a broad component, and another where this emission is included in the continuum fit. The average of the two FHWMs is used. 

Another complication is that NGC 6240, Mrk 273 and UGC 5101 are not well described by a combination of 2 Gaussians. NGC 6240 is an early stage merger in which the two nuclei are  well separated by $1.8''$ in the I and r-band \citep{Fried-1983}. Both nuclei are visible in the 3 mm continuum, but the CO(1-0) and CO(2-1) line intensities peak between the two nuclei \citep{Tacconi-1999, Feruglio-2013b}. The high spatial resolution maps in CO show that the line emission consists of several components. Also the [CII] line profile requires three components for a good fit. The [CII] line profile of Mrk 273 is irregular with a non-Gaussian shape. \citet{Cicone-2014} observe a similar line profile in CO(1-0), and fit the central line with two Gaussians. We also fit two Gaussians to the central line in the [CII] profile, and add a third Gaussian for the broad component. The [CII] line profile in UGC 5101 suggests self-absorption or a 'double horn' profile as is typical for a disk seen edge-on. It could well arise from 
the molecular disk that is observed in CO(1-0) at velocities similar to the ones in the [CII] line profile \citep{Genzel-1998}. The central line profile is well fitted by two Gaussians, and adding a third component slightly improves the overall fit.

\subsection{Comparison with OH }
\label{corr_oh}

All fourteen objects with broad [CII] wings either have an OH outflow (11), or inflow (1), with two exceptions: IRAS 23128-5919 clearly has wings in [CII], but the OH119 spectrum is too noisy to be fitted reliably. A faint P-Cygni profile may however be present with an estimated OH $v_{50}$ $\sim -500\,\text{km}\,\text{s}^{-1} $. We present the [CII] spectra and fits of this object in Appendix A but exclude it from our analysis.
UGC 5101 has a marginally detected broad [CII] component, while OH $v_{50} = -9\,\text{km s}^{-1}$. Faint high-velocity wings are however observed in the OH119 spectrum, which are better represented in this case by $v_{84}$ and $v_{max}$.
The object with an OH inflow is IRAS F22491-1808. Hubble Space Telescope observations of IRAS F22491-1808 reveal bright tails, which extend to 16 kpc from the center, and two nuclei separated by 2.5 kpc \citep{Surace-2000}. We cannot exclude the possibility that the line profiles reflect tidal effects of a merger rather than an inflow. Inflows are excluded from the sample during the analysis.

\begin{figure}
\centering
\includegraphics[scale=0.45]{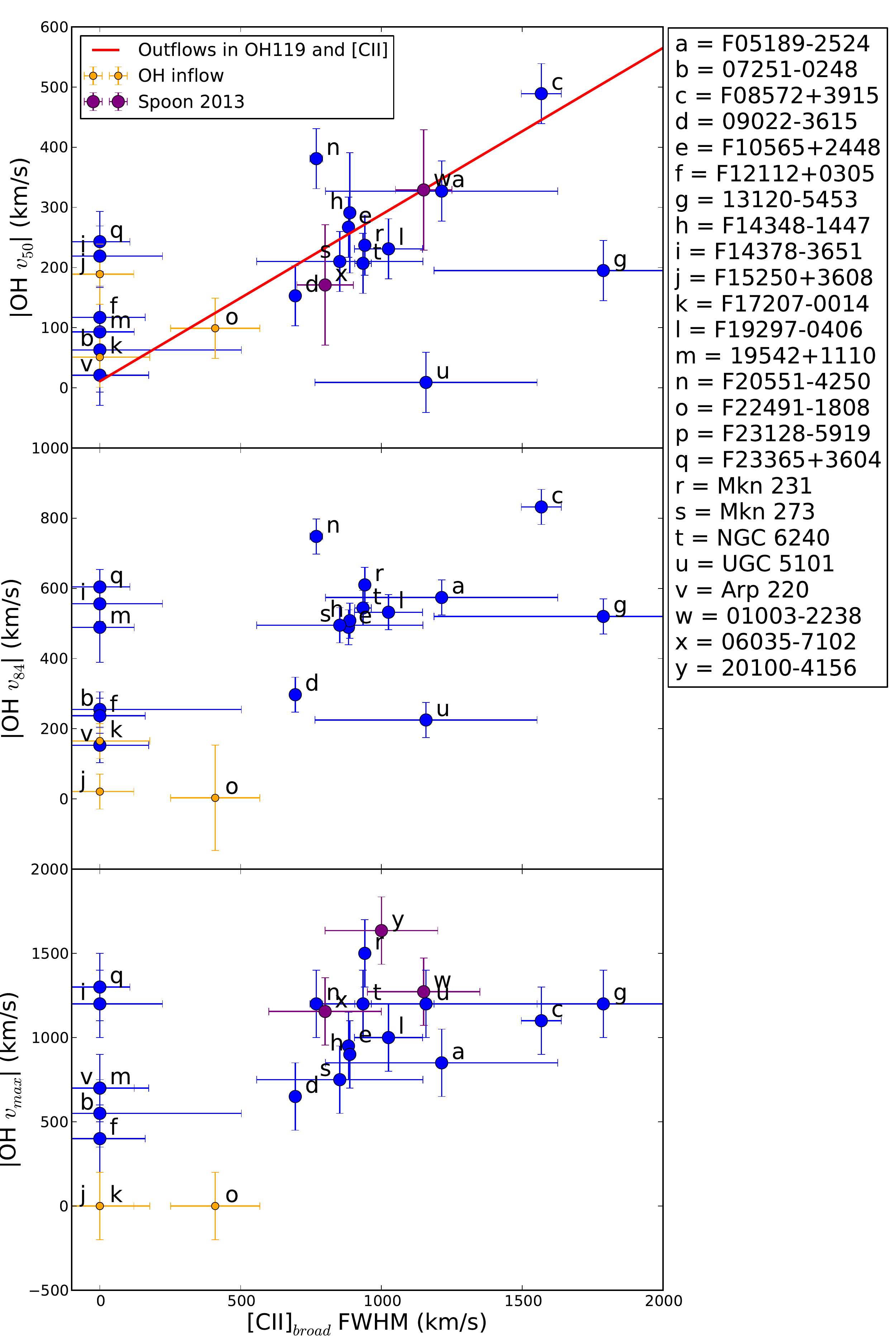}
 \caption{[CII] FWHM of the broad component plotted against  the absolute value of the several OH velocities. Objects without a [CII] broad component are placed at $0\,\text{km}\,\text{s}^{-1}$. The small orange markers represent objects with an inflow in OH119, having a red-shifted absorption ($v_{50} > 50\,\text{km}\,\text{s}^{-1} $). Error bars for OH were adopted from V13, [CII] FWHM errors are listed in Table \ref{tab:cii_obs}. Top panel: the red line is a  weighted fit to all objects with detected outflows both in OH and C$^{+}$.}
 \label{fig:vel_corr}
\end{figure}

A comparison of the OH blue-shifted velocities and broad [CII] FWHM is given in Figure \ref{fig:vel_corr}, which shows the absolute values of the OH119 velocities ($v_{50}, v_{84}, v_{max}$) as function of the broad [CII] FWHM. The absolute value of the OH velocities is taken, because one can not distinguish an outflow from an inflow by observing only the [CII] wings. The three objects with an OH inflow ($v_{50} > 50\,\text{km}\,\text{s}^{-1} $; F15250+3608, F17207-0014 and F22491-1808) are represented by small orange markers. Objects which do not require a broad component for a good fit to the [CII] spectrum are placed at $0\,\text{km s}^{-1}$. We moreover added all three objects from \citet{Spoon-2013} which have an outflow in both [CII] and OH119. \citet{Spoon-2013} give the FWHM of the broad [CII] component and $v_{max}$ of OH119. A blue-shifted and red-shifted component have been fitted to the OH119 spectra of 01003-2238 and 06035-7102, resulting in $v_{50} = -329\,\text{and } -171\,\text{km s}^{-1}$ 
respectively. 20100-4156 has a 
blended and broad absorption component, which we do not attempt to fit, so we have no $v_{50}$ for this object. These objects are represented by purple markers in Figure \ref{fig:vel_corr}, and are taken into account during the line fit (Equation \ref{eq:bestfit}) and the statistical analysis in Table \ref{tab:stats}. 

Sixteen objects in our sample have OH outflows, defined in V13 to be the objects with OH $v_{50} < -50\,\text{km}\,\text{s}^{-1} $. Eleven out of these objects also have [CII] broad wings, while five do not. Of these five objects, IRAS 23365+3604 is a real outlier; it has a high-velocity P-Cygni profile in OH but no trace of broad wings in [CII]. Although there is a small bump on the red side of the line, the width of this bump is smaller than the spectral resolution of PACS, and the feature is not considered to be real. IRAS 19542+1110 and IRAS 17208-0014, have low velocity OH outflows. The corresponding expected [CII] FWHM is too small for any outflow to be separable from the emission from the host galaxy. Finally, 07251-0248 and F14378-3651 may well have broad wings in [CII], but the reduced $\chi^{2}$ value is smaller for the one-component fit.

The objects with a low-velocity OH outflow and no broad [CII] component are consistent with our hypothesis that broad [CII] traces OH outflows, but that a certain minimum velocity is required in order to distinguish emission of the outflow from that of the host galaxy. More qualitatively, out of the eighteen objects which have detected OH119 velocities and no inflow (the blue points in Figure \ref{fig:vel_corr}), fifteen are consistent with this hypothesis. Moreover, it is unlikely that a broad [CII] component traces an inflow: only one of the three objects with an OH inflow has a broad [CII] component, and it has a FWHM of only $409\,\text{km s}^{-1}$, which is low compared to the objects with outflows. These findings suggest that a broad [CII] component with a FWHM of  $> 500\,\text{km s}^{-1}$ is a good indicator for the presence of an OH outflow.  

We go one step further by comparing the actual outflow velocities. If C$^{+}$ traces an isotropic molecular outflow, a correlation is expected between the broad [CII] FHWM and the blue-shifted velocity of the OH absorption. [CII] seems to correlate with the absolute value of all OH velocities (Figure \ref{fig:vel_corr}), in particular with $v_{50}$. A weighted least squares fit to $|\text{OH } v_{50}|$, for all objects that have an outflow detected both in OH and in C$^{+}$, results in: 
\begin{equation}     
 |\text{OH } v_{50}| = 11\, (\pm 96) + 0.28\, (\pm 0.10) \times \text{CII}_{\text{FWHM}} 
 \label{eq:bestfit}
\end{equation}
 where all velocities are given in $\text{km}\,\text{s}^{-1}$. Each observation is weigthed by max[$\sigma_{\text{OH}}$, $\sigma_{\text{[CII]}}$]$^{-2}$, with $\sigma_{\text{OH}}$ and $\sigma_{\text{[CII]}}$ the errors in the OH velocity and broad [CII] FWHM respectively. Relation \ref{eq:bestfit} is represented by the red line in the upper panel of Figure \ref{fig:vel_corr}. The line goes through (0,0) within the uncertainties, as is expected when C$^{+}$ traces the OH outflow.
 
Is the apparent correlation in the upper panel of Figure \ref{fig:vel_corr} significant? We call a correlation statistically significant if the null hypothesis (the data points are randomly selected) can be rejected at a confidence level of 5\%, using the Spearman rank coefficient and the Pearson correlation coefficient. When all objects with outflows in both OH and C$^{+}$ are included, only the Pearson correlation coefficient suggests a correlation. If IRAS 13120-5453 is removed from the sample, the correlation becomes tentative according to the Spearman rank coefficient and significant according to the Pearson correlation coefficient. Table \ref{tab:stats} lists the correlation coefficients and confidence levels. The confidence levels for a two-tailed test have been retrieved from a table for low-number statistics, and are applicable to sample sizes of 4-30 data points \citep{Wall-2003}.

\begin{deluxetable}{lcccccc}
 \tablecaption{Correlation coefficients}
\tablehead{ \colhead{Parameter} & \colhead{N} & \colhead{$\rho_{s}$} & \colhead{P$_{\rho}$} & \colhead{r} & \colhead{P$_{r}$} & \colhead{P$_{r}$($N < 30$)}  \\
\colhead{(1)} & \colhead{(2)} & \colhead{(3)} & \colhead{(4)} & \colhead{(5)} & \colhead{(6)} & \colhead{(7)}
}

\startdata
CII$_{broad}$ - $|\text{OH} v_{50}|$ ($^{*}$) & 13 & 0.34 & ... & 0.36 & 0.23 & 0.05 \\
CII$_{broad}$ - $|\text{OH} v_{84}|$ ($^{*}$) & 11 & 0.47 & 0.2 & 0.39 & 0.24 & ... \\
\hline
CII$_{broad}$ - $|\text{OH} v_{50}|$ ($^{**}$) & 12 & 0.55 & 0.1 & 0.75 & 0.01 & 0.02 \\ 
CII$_{broad}$ - $|\text{OH} v_{84}|$ ($^{**}$) & 10 & 0.59 & 0.1 & 0.65 & 0.04 & 0.05 \\
\hline
CII$_{broad}$ - $\alpha_{AGN}$ & 14 & 0.4 & 0.2 & 0.27 & 0.35 &  ... \\
CII$_{broad}$ - $L_{AGN}$  & 14 & 0.06 & ... & -0.04 & 0.89 &  ... \\
\hline
 $|\text{OH} v_{50}|$- $\alpha_{AGN}$ & 16 & 0.45 & 0.1 & 0.54 & 0.03 & 0.05 \\
$|\text{OH} v_{50}|$ - $L_{AGN}$  & 16 & 0.29 & ... & 0.29 & 0.28 &  ... \\
\hline
CII$_{broad}$ - $|\text{Na} v_{50}|$ & 8 & 0.48 & ... & 0.74 & 0.04 &  0.05 \\
$\text{OH} v_{50}$ - $\text{Na} v_{50}$ & 10 & 0.7 & 0.05 & 0.74 & 0.01 &  0.02 \\
OH mass - [CII] mass & 8 & 0.61 & 0.2 & 0.9 & 0.01 & 0.01 \\
\enddata
\tablecomments{(1) quantities considered for the test, (2) number of objects, (3) Spearman rank coefficient $\rho_{s}$, (4) confidence level for rejection of the null-hypothesis (there is no correlation between the 2 parameters) P$_{\rho}$, (5) Pearson correlation coefficient, (6) two-sided area after a Fisher transformation P$_{r}$,  (7) confidence level for rejection of the null-hypothesis, using the t-value from the Pearson correlation coefficient. Note that P$_{\rho}$ and P$_{r}$($N < 30$) have been retrieved from a table for low-number statistics, and are applicable to sample sizes of 4-30 data points. If no value is given, the confidence levels are $> 0.2$ and not included in the tables for low-numbers statistics. $^{*}$ including all objects with a detected outflow in both OH and C$^{+}$. $^{**}$ like $^{*}$ but excluding IRAS 13120-5453.}
\label{tab:stats}
\end{deluxetable}

Although part of the lack of a strong correlation between the OH and C$^{+}$ velocities can be explained by the small sample, there is also intrinsic scatter in the relation.  We assumed that each object has a large opening angle outflow, which is not necessarily the case. Moreover, there are uncertainties in the fits, in particular for objects with faint wings. And finally, C$^{+}$ and OH may trace different parts of the outflow. So although C$^{+}$ is in general a good tracer of the molecular outflow, significant deviations from the best fit in Equation \ref{eq:bestfit} can remain in individual objects.

\subsection{Correlation with AGN fraction and luminosity}

In a previous analysis of 28 local ULIRGs, we found the OH $v_{50}$ to correlate with the AGN fraction $\alpha_{agn}$ and AGN luminosity, although with significant scatter (V13).  Part of the scatter is caused by uncertainty in the AGN luminosity, part again by our assumption of a spherical outflow with a large opening angle, and a constant coupling efficiency. Even in the absence of these uncertainties, an intrinsic scatter is expected due to different driving mechanisms and variable AGN luminosities. Nevertheless, we compare the AGN luminosity and AGN fraction with the broad [CII] FHWM and OH $v_{50}$, in order to see whether our subsample of V13 is large enough to reveal the same correlation as found earlier. Table \ref{tab:sample} lists the AGN luminosity and fraction for our sources. $\alpha_{agn}$ has been calculated from $f_{30}/f_{15}$ because these numbers are known for every object in the sample, and because this method shows the tightest relation with other methods to determine the AGN 
fraction \citep{Veilleux-2009}. The error on the AGN fraction is typically 20\%. 

\begin{figure}
 \centering
 \includegraphics[scale=0.4]{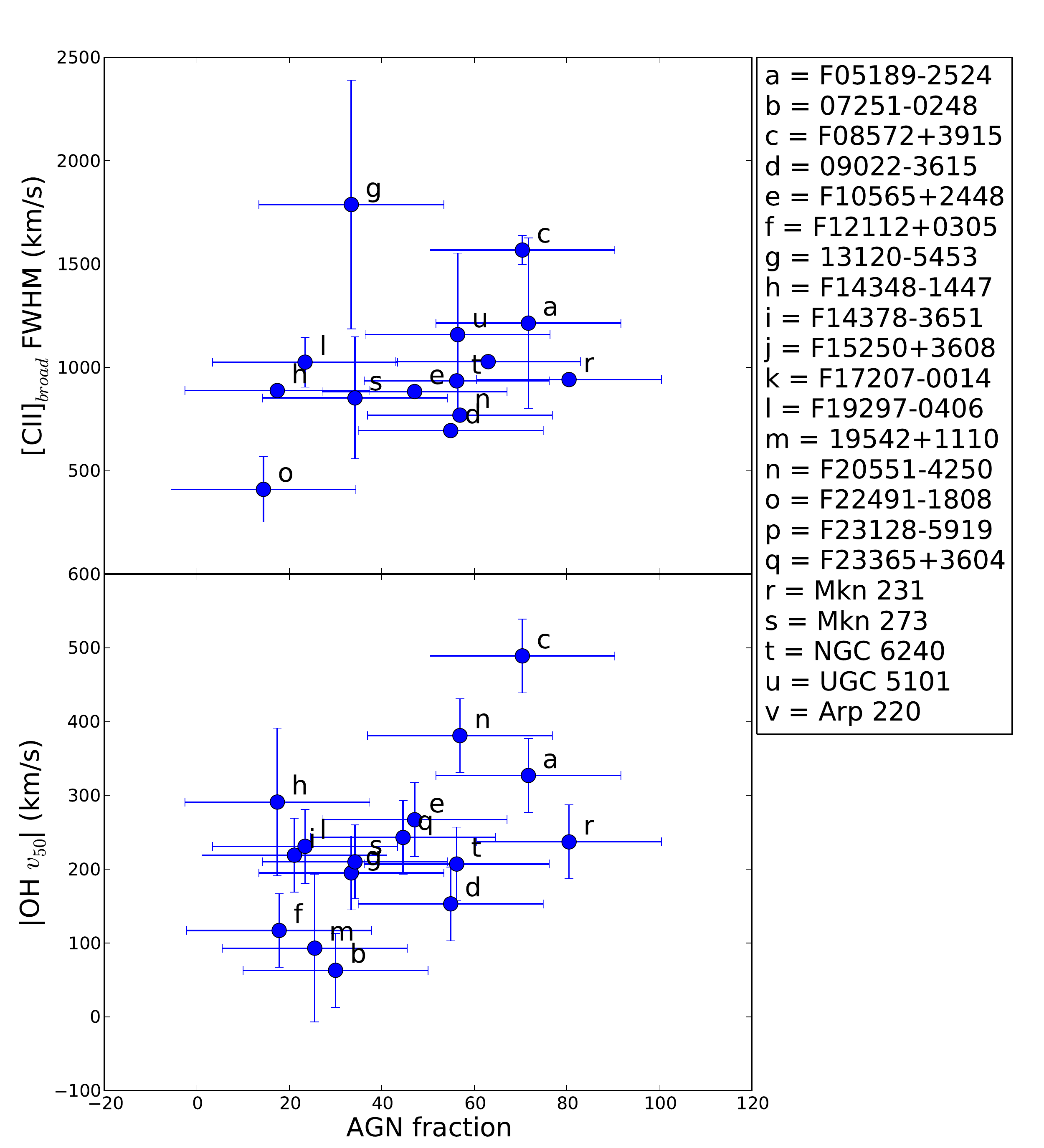}
  \caption{Upper panel: broad [CII] FWHM versus AGN fraction as listed in Table \ref{tab:sample}. Lower panel: OH $v_{50}$ versus AGN fraction.  }
 \label{fig:agn}
\end{figure}

\begin{figure}
 \centering
 \includegraphics[scale=0.4]{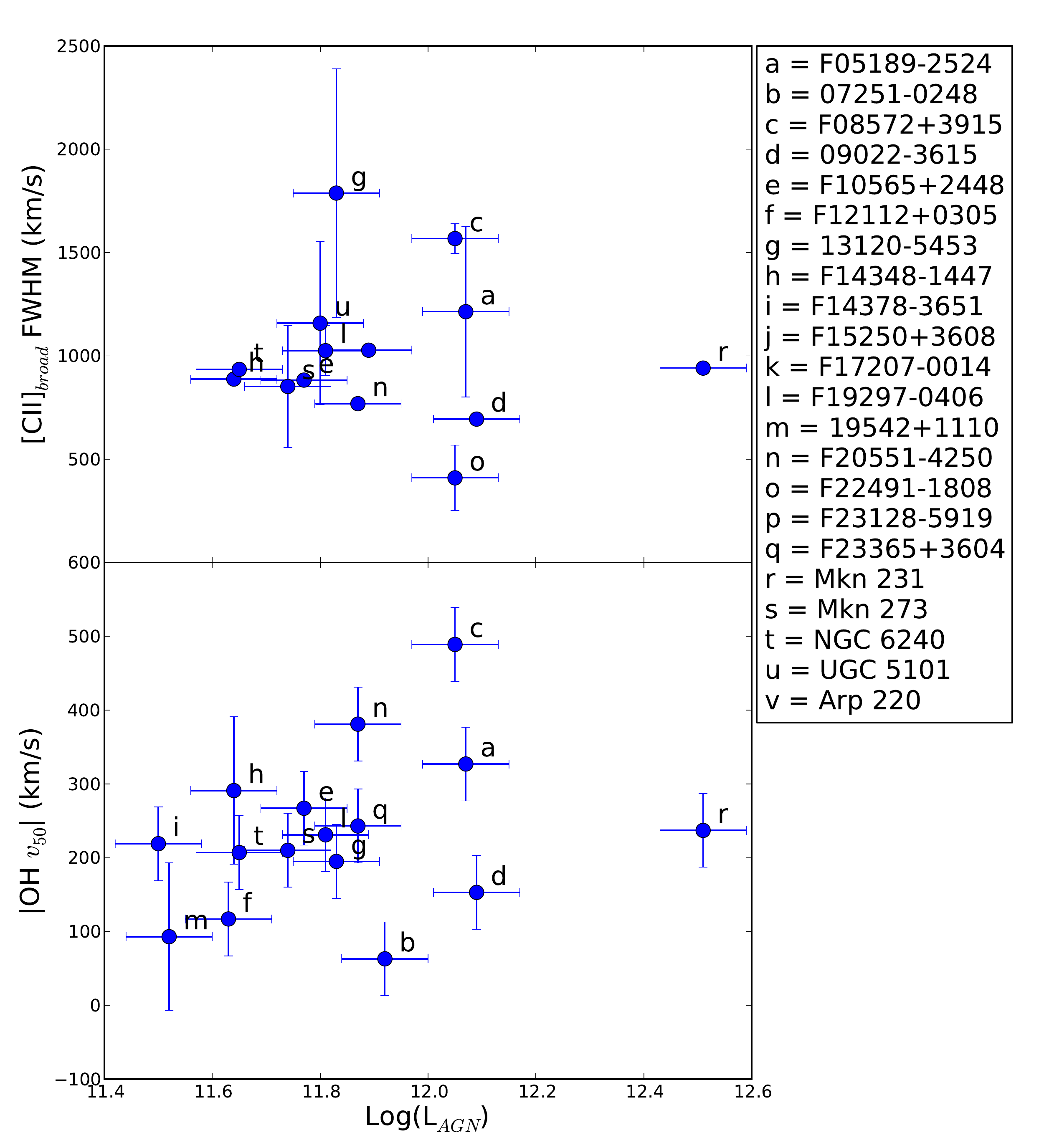}
  \caption{Upper panel: [CII] FWHM versus AGN luminosity as listed in Table \ref{tab:sample}. Lower panel: OH $v_{50}$ versus AGN luminosity.  }
 \label{fig:agnlum}
\end{figure}

Figure \ref{fig:agn} shows the relation of both the broad [CII] FHWM and OH $v_{50}$ with   $\alpha_{agn}$ for all objects with a detected outflow. There is no statistically significant correlation in any of the two plots, see Table \ref{tab:stats}. Figure \ref{fig:agnlum} is similar to Figure \ref{fig:agn}, but now with the logarithm of the AGN luminosity on the x-axis. The uncertainties are taken to be 20\% of the total luminosity. Also here, no statistically significant correlation is found with either the broad [CII] FWHM or  OH $v_{50}$. We conclude that the current sample is too small, and covers too small a range in $L_{AGN}$, to find the same correlation as in V13.

\subsection{Additional observations: Na I D}
\begin{figure}
 \centering
  \includegraphics[scale=0.4]{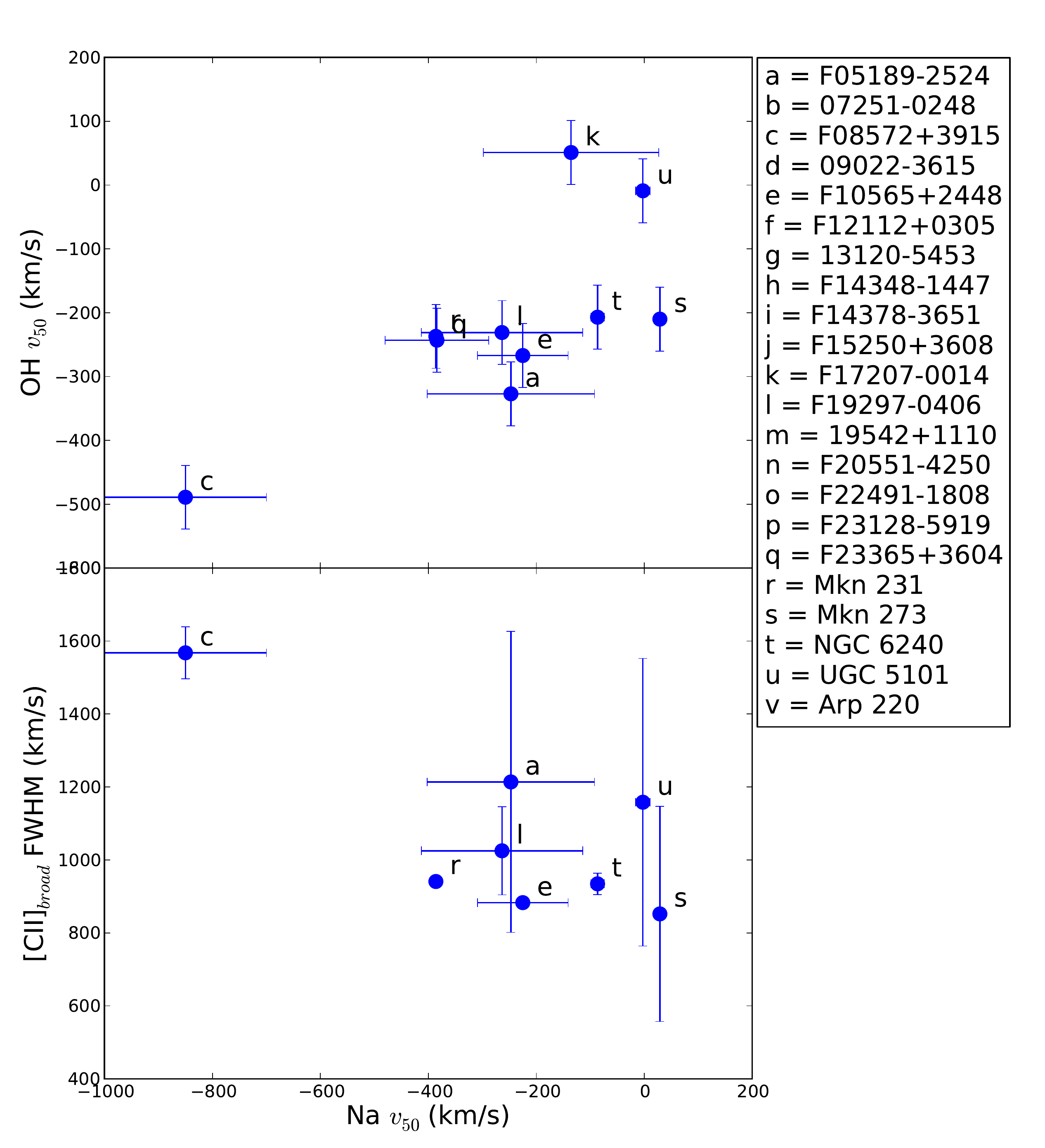}
 \caption{Na I D $v_{50}$ compared to OH $v_{50}$ (upper panel) and [CII] FHWM  (lower panel). The upper panel shows objects with an OH inflow or outflow (we do not plot the absolute values here, in order to differentiate between inflows and outflows), the lower panel shows only objects with a detected [CII] outflow. }
 \label{fig:na}
\end{figure}

Ten objects in our sample have additional observations of blue-shifted Na I D $\lambda \lambda 5890,5896$ absorption, observed by \citet{Rupke-2005a}, \citet{Rupke-2005b},  \citet{Rupke-2013b} and \citet{Martin-2005}. Because Na I D observations are often spatially resolved, several velocities are available for each object. Table \ref{tab:oh_observations} lists the largest and smallest $v_{50}$ values for the outflowing neutral gas, which were averaged for our analysis here.  Figure \ref{fig:na} compares Na I D $v_{50}$ to OH $v_{50}$ and the broad [CII] FWHM. F08572+3915 has by far the highest velocity in OH and Na, and the second highest velocity in [CII]. Also for the other objects,  Na I D $v_{50}$ seems to correlate well with OH $v_{50}$ (for all objects with a detection in both tracers), and with [CII] (for objects with a detection in Na I D and broad wings in [CII]). We give the Spearman rank coefficients and Pearson correlation coefficients in Table \ref{tab:stats}. Despite the small sample, the 
rejection levels suggest a significant correlation of the broad [CII] FWHM and OH $v_{50}$ with Na $v_{50}$, which is however driven mostly by F08572+3915.

\subsection{Additional observations: CO}
\label{co}
Seven objects in our sample have CO(1-0) or CO(2-1) observations with sufficient S/N and velocity coverage to reveal broad wings. A comparison with [CII] is facilitated by  CO and [CII] both appearing in emission and having similar critical densities.  The critical density for the fine structure transition [CII] is $3.2 \times 10^{3}\,\text{cm}^{-3}$ (using collision rates with atomic hydrogen at $100\,\text{K}$, \citet{Barinovs-2005}). For CO(1-0) and CO(2-1), $n_{crit} = 2.1 \times 10^{3}\,\text{cm}^{-3}$ and $n_{crit} = 2.2 \times 10^{4}\,\text{cm}^{-3}$ respectively (using collision rates with para-H$_{2}$ at $100\,\text{K}$,  \cite{Yang-2010}).

\citet{Feruglio-2010} present CO(1-0) observations of Mrk 231. The line profile has been fitted with two components: a narrow one and a broad one with a FWHM of $870\,\text{km}\,\text{s}^{-1}$ (compared to $971\,\text{km}\,\text{s}^{-1}$ in [CII]). At higher spatial resolution, the source is slightly resolved at 1.1-1.3 kpc scale \citep{Cicone-2012}. NGC 6240 has been observed in CO(1-0) by \citet{Feruglio-2013}, who find a line profile extending from  $−600\,\text{km}\,\text{s}^{-1}$ to $+800\,\text{km}\,\text{s}^{-1}$ (compared to $-1000\,\text{km}\,\text{s}^{-1}$ to $+1500\,\text{km}\,\text{s}^{-1}$ in [CII]). 

\citet{Cicone-2014} present CO(1-0) observations of F23365+3604,  F10565+2448, Mrk 273 and F08572+3915 from our sample. F23365+3604 -- with detected outflows in Na I D and OH but not in [CII] --  clearly has broad CO wings, with flux densities of only 1\% of the the peak emission. If the outflow is equally faint in the neutral gas traced by C$^{+}$, it would explain why no outflow was detected in [CII]: our PACS observations of [CII] do not reach the required sensitivity to detect such a faint component. The CO(1-0) profile for F10565+2448 has very bright wings, which is also the case in our [CII] spectra.  F08572+3915 has the fastest CO outflow of the sample, with wings peaking around $-800\,\text{km}\,\text{s}^{-1}$ and $+900\,\text{km}\,\text{s}^{-1}$. In our sample, this object also has the highest outflow velocity in  OH and Na I D, and the second highest velocity in [CII].   

Recently, \citet{Garcia-Burillo-2015} published CO(2-1) spectra of IRAS F17207-0014, which revealed a complex line profile. After fitting it with 3 narrow line components, a residue is left between $-700\,\text{km}\,\text{s}^{-1}$ and  $-800\,\text{km}\,\text{s}^{-1}$. There is however no obvious broad component,  again in agreement with our findings from [CII]. 

The low-J CO and [CII] line profiles show many similarities, and we plan to provide a quantitative comparison in a future paper.

\subsection{Uncertainties In The Fits And Alternative Velocity Determination}
Some spectra in our sample have a [CII] line profile that can be fitted in different ways, resulting in large errors in the FWHM of the broad component. For example, 13120-5453 could either have an extremely broad component, or no broad component at all. The  NGC 6240 line profile can be fitted with two or three components, which results in different values for the broad component FWHM. We therefore investigated alternatives. A more objective method  is to take the velocities that enclose a certain fraction of the total (host galaxy + outflow) flux (the area-method). We applied this method at both 90\% and 95\% of the total flux, and found that 90\% is a good compromise between being sensitive to faint emission and being dominated by noise.  However, the velocities derived from this method do not correlate at all with the broad [CII] FWHM, the OH velocities, the Na I D velocities or the AGN fraction. 

This discrepancy arises because the area-method depends strongly on the emission from the host galaxy, but the luminosity of the host galaxy does not correlate with that of the outflow. An outflow may therefore not be caught when the central component is narrow, or its velocity may be overestimated if the central component is broad. Therefore, separating the host emission from the outflow emission seems necessary, even if this introduces large errors. 

Another uncertainty lies in the decision to extract the spectrum from the central or the $3 \times 3$ spaxels. If all spectra are extracted from the central spaxel, the correlation between [CII]$_{broad}$ FWHM and the OH velocities weakens a bit. Most importantly, the [CII]$_{broad}$ FWHM of F08572+3915 decreases from $1586\,\text{km}\,\text{s}^{-1}$ to $898\,\text{km}\,\text{s}^{-1}$, which weakens the correlation between [CII]$_{broad}$ FWHM and Na I D $v_{50}$. Otherwise, the results remain the same.

\section{DISCUSSION}
\label{discussion}

Studying a similar sample of 24 ULIRGs in OH119 and [CII], \citet{Spoon-2013} and \citet{Farrah-2013} only found 4 objects with broad wings in [CII]. In 3 objects the [CII] line width matches well with the line width of the blue-shifted OH absorption.  The reason for finding only few [CII] broad components probably is, as the authors note themselves, that the observations do not have the required sensitivity to detect faint wings. The ULIRGs in their sample are fainter and at higher redshift ($z < 0.262$) than in our sample ($z<0.1$).

\subsection{Outflow Mass}
\label{mass}
The OH absorption only traces the molecular outflow along our line of sight to the FIR continuum. Calculation of the total outflow mass thus requires an assumption on the opening angle and geometry of the outflow. Moreover, the absorbed flux depends on the OH column density towards the source, on the covering factor of OH onto the FIR continuum, and on the FIR continuum intensity and spectral shape. A single OH transition is therefore not sufficient for mass determinations. We did however detect three or more OH transitions in some sources, and calculated the outflow mass for the 9 sources which 1) have detections of at least three out of the four transitions: OH119, OH79 ($^{2}\Pi_{1/2}\, 1/2\, \rightarrow\, ^{2}\Pi_{3/2}\, 3/2 $), OH84 ($^{2}\Pi_{3/2}\, 7/2\, \rightarrow\, 5/2 $) and OH65 ($^{2}\Pi_{3/2}\, 9/2\, \rightarrow \, 7/2 $), 2) with clear P-Cygni profiles or broad wings, and 3)  for which the spectra and fits look normal. 

\begin{figure}
\centering
\includegraphics[scale=0.5]{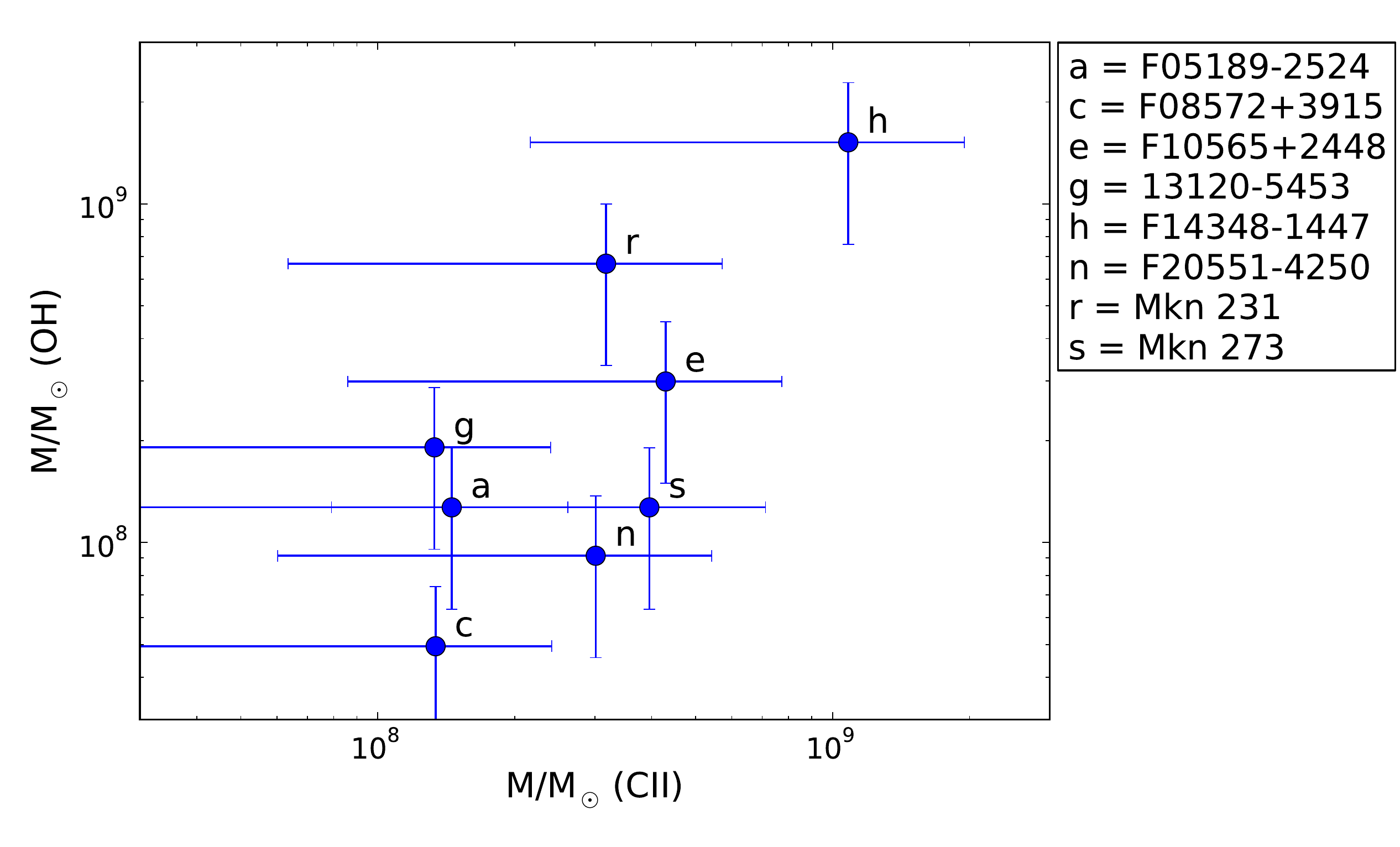}
 \caption{Total molecular outflow mass based on OH line modelling by Gonz\'{a}lez-Alfonso (in preparation) on the y-axis, and atomic outflow mass based on the luminosity of the [CII] broad component on the x-axis. }
 \label{fig:oh_mass}
\end{figure}

The model is described in \citet{Gonzalez-2014}, and final results will be published in Gonz\'{a}lez-Alfonso (in preparation). A short description of the model is given here. Each object is described by the sum of three components: two outflow components (with different velocity, spatial extents and far-IR radiation sources) and one quiescent component. Free parameters of the model are: the inner- and outer- radius of the component, the dust temperature, the continuum optical depth at $100\,\mu$m, the gas velocity at $R_{in}$, the gas velocity at $R_{out}$, the OH column density, the impact parameter (giving the on-sky spatial extend of the outflow) and the scaling factor (related to the covering factor of OH with respect to FIR continuum). Each component is divided in at least 20 concentric shells. The statistical-equilibrium populations are calculated in each shell, and the emerging line shapes are calculated and convolved with the PACS spectral resolution. These modelled line profiles are then compared to 
the three or more available spectra in OH, and the best fit is used. The mass is derived from the outflow components only. Figure \ref{fig:oh_mass} compares these outflow masses based on OH absorption with those based on [CII] emission. 

The mass of the neutral atomic outflow was derived from the [CII] broad component using the cooling law as described in the appendix of \citet{Tielens-1985}, and assuming that [CII] is optically thin:
\begin{equation}
 n^{2}\Lambda = \frac{ g_{u}/g_{l}\, \text{exp}[-h\nu_{ul}/kT] }{1 + n_{cr}/n + g_{u}/g_{l}\, \text{exp}[-h\nu_{ul}/kT]} \, \chi_{C^{+}}nA_{ul}h\nu_{ul} 
\end{equation}
This equation describes the cooling of the interstellar medium (ISM) per unit volume due to line emission ($[n^{2}\Lambda] = \text{J}\,\text{s}^{-1}\,\text{cm}^{-3}$). The ratio of the statistical weights of the upper and lower level $g_{u}/g_{l} = 2$,  $h\nu_{ul}/k = 91\,\text{K}$, the critical density $n_{cr} = 3.2\times 10^{3}\,\text{cm}^{-3}$ in neutral medium  \citep{Barinovs-2005}, the abundance $\chi_{C^{+}} = 1.4\times 10^{-4}$, the Einstein coefficient for spontaneous emission $A_{ul} = 2.4 \times 10^{-6}\,\text{s}^{-1}$, and the frequency $\nu_{ul}=1.9 \times 10^{12}\,\text{Hz}$. Filling in these constants results in the following equation:
\begin{equation}
 \frac{M_{N}}{M_{\odot}} = 1.34 \frac{L_{\text{CII}}}{L_{\odot}} \times \frac{1 + n_{cr}/n + 2\,\text{exp}[-91\text{K}/T]}{2\,\text{exp}[-91\text{K}/T]}
 \label{eq:mass}
\end{equation}
where we assumed that 25\% of the mass is in helium, and that all broad [CII] emission arises from atomic gas.

In order to derive the density and temperature we need either a CO line SED or FIR SED plus density tracers, but these observations are not available for the outflows.  Instead, we will use $n = 10^{5}\,\text{cm}^{-3}$ and $T = 100\,\text{K}$, which should be typical conditions in ULIRGs  \citep{Sanders-1996, Veilleux-2009, Mashian-2015}. The fraction in Equation \ref{eq:mass} becomes 2.28. This number is quite conservative, as it is less than twice as large as the factor retrieved  for infinitely large density and temperature (1.5).
With an uncertainty of 50\% due to unknown temperature and density and a 30\% flux calibration error, we adopt a total error of 80\%.
Table \ref{tab:cii_obs} lists the luminosities of the broad components and the derived total neutral gas masses. These masses are probably too high for NGC 6240, F14348-1447 and 09022-3615 ($M > 10^{9} M_{\odot}$), because the outflow emission could not be separated well from the host galaxy emission.  

 There appears to be a tentative 1:1 relation between the derived outflow masses from OH and [CII], albeit within the large uncertainties (Figure \ref{fig:oh_mass}). Together with the observed trend between the broad [CII] FWHM and OH $v50$, this supports our hypothesis that 
 a broad [CII] component  is a plausible tracer of molecular outflows in galaxies.
 The mass has been calculated from OH by adopting an abundance of $2.5 \times 10^{-6}$ relative to H nuclei and including a factor of 1.4 due to Helium. Both derivations should therefore result in similar masses if the same amount of particles is involved. An exact 1:1 relation is not necessarily expected. OH absorption arises predominantly from the molecular gas while  C$^{+}$ emission arises from atomic and molecular gas. Some overlap between the two tracers is expected, but it is not known how much. Yet, within these uncertainties, the mass derived from the luminosity of the broad [CII] component seems to agree well with the outflow mass derived from OH. The relation is significant according to the Pearson correlation coefficient, but not if the Spearman rank coefficient is used (Table \ref{tab:stats}). However, we stress here that the OH masses are preliminary results, and that the correlation is driven by the large mass in IRAS F14348-1447 ($1-2 \times 10^{9}\,M_{\odot}$).

\subsection{Origin of the [CII] broad wings}
The P-Cygni profiles and blue-shifted absorptions in OH are clear signs of outflowing gas. Broad wings in [CII] on the other hand, can be caused by outflows, inflows, turbulence, or a high velocity rotation disk. However, the OH spectra have shown that the ULIRGs in our sample contain molecular outflows more often than inflows (16 outflows versus 3 inflows). Furthermore, all outflow masses derived from OH and [CII] lie along a 1:1 relation within the uncertainties. Adding to that the trend of the broad [CII] FWHM with OH $v_{50}$, leads us to the conclusion that [CII] traces and is likely entrained in the OH outflow.

This does not mean that [CII] broad wings exclusively trace molecular outflows: the broad [CII] FWHM also shows a trend with Na I D blue-shifted velocity (Figure \ref{fig:na}, Table \ref{tab:stats}), suggesting an origin in atomic gas. V13 already found a correlation between Na I D $v_{50}$ and OH $v_{50}$, showing that the atomic and molecular outflow are mixed or that OH absorption arises mostly from the atomic gas. 

Multi-phase outflows have been observed before, although the connection between the different phases is not always clear \citep{Shih-2010, Westmoquette-2013, Kreckel-2014, Morganti-2015}. Gas phases with similar outflow velocities can be explained by gas mixing. For example, \citet{Contursi-2013} explain the similar velocities of [OIII] $\lambda\,88 \mu$m, [OI] $\lambda\, 63\,\mu$m and $\lambda\, 145\,\mu$m, [CII], and CO(1-0) in M82, by assuming that the outflow consists of small molecular cloudlets which are ionized on the outside, but remain molecular inside. In this way, the different phases do not only have the same velocity, but are also cospatial.  Spatially resolved observations of the outflows would therefore help  to explain the connection between the gas phases.

\citet{Rupke-2013a} present spatially resolved IFU observations of Na I D and H $\alpha$ of the following ULIRGs in our sample: F08572+3915, F10565+2448, Mrk 231, Mrk 273 and F17207-0014 (F17207-0014 only has Na I D detected). All of them have extended  outflows with a radius $\geq 1\,\text{kpc}$. These kpc-scales may be hard to unite with the scale of the OH absorption. PACS did not resolve the OH absorption, but there are several indications that the absorption arises from scales $\leq 1 $ kpc:
\begin{itemize}
 \item Using the model as described in Section \ref{mass}, \citet{Sturm-2011b} find inner radii of 100-200 pc for the OH outflow for five objects in our sample, and outer radii of a few hundred pc.
 \item \citet{Spoon-2013} use the anti-correlation between OH119 equivalent width and the silicate strength to argue that the OH119 absorption arises in the buried nucleus. 
 \item \citet{Rupke-2013b} present spatially resolved observations of the H$_{2}$ outflow in F08572+3915 (NW) at 400 pc. The outflow velocities correspond well with those observed in OH. 
 \end{itemize}
There is however a bias in the determination of the spatial scale at which OH absorption takes place, because a bright FIR continuum is needed. If this continuum has a smaller spatial extent than the OH outflow, a significant part of the OH outflow will not be illuminated. \citet{Lutz-2015} recently published the Herschel/PACS FIR radii at 70 $\mu$m for our objects, and provided us with the radii at 100 $\mu$m (private communication). These radii range from below 0.5 kpc to 2.8 kpc, and are typically around 1 kpc for our sample. The OH outflow could therefore be more extended than what we infer from the PACS observations.

If the outflow is located at a kpc from the nucleus, would it be able to escape the gravitational potential of the galaxy? Based on the widths of the narrow components of the [CII] line profile (deconvolved with the spectral PACS PSF), we derive a typical dynamical mass of $10^{10}\,M_{\odot}$. We do not take any inclination effects into account for the disk-dominated objects. For a central mass of $10^{10}\,M_{\odot}$, the escape velocity is $\sim 300\,\text{km s}^{-1}$, which is reached by most C$^{+}$ outflows. Part of the gas could therefore escape the central few  kiloparsec of the galaxy, although it may not escape the halo. 

\section{SUMMARY AND CONCLUSION}
\label{conclusion}

We address the question of whether [CII] at $158\, \mu\text{m}$ is a good tracer of outflows in general, and molecular outflows in particular. The sample consists of 22 local and bright ULIRGs, of which 16 objects have molecular outflows as traced by OH $119\,\mu$m absorption at velocities $< -50\,\text{km}\,\text{s}^{-1}$. Because [CII] appears in emission, any isotropic and spatially unresolved outflow becomes visible as broad wings in the line profile. We fitted the [CII] line profiles with two Gaussians, in order to separate host emission from outflow emission, and use the FWHM of the broad component to estimate the outflow velocity. The main results are:
\begin{enumerate}
 \item Fourteen objects in our sample show broad [CII] wings. Eleven of these objects contain an OH outflow, one contains an inflow, one object has a noisy OH spectrum and is removed from the analysis, and one has a faint outflow which is not traced by OH $v_{50}$ (the median velocity of the blue-shifted absorption). If all objects with an outflow in [CII] and OH119 are taken into account,  the FWHM of the broad [CII] component shows a tentative correlation with OH $v_{50}$ and $v_{84}$.
 
 \item Observations of Na I D are available for 10 sources in the sample. The FWHM of the broad [CII] component shows a trend with the Na I D $v_{50}$. This is not surprising, since Na I D $v_{50}$ was already found to correlate with the OH $v_{50}$ in V13.  [CII] may thus trace both the molecular and atomic gas in the neutral outflow, and part of the OH absorption could arise from the atomic gas as well.  
 
 \item The broad [CII] FWHM does not show a correlation with either the AGN luminosity or the AGN fraction. This might be caused by a combination of large uncertainties in the AGN luminosity and fraction, intrinsic scatter, and the small sample: no correlation is found between OH $v_{50}$ and AGN fraction or luminosity either, although such a correlation was found when using a larger sample. 

 \item Preliminary estimates of the molecular outflow mass, based on at least 3 OH line profiles, are given for 9 objects. These masses lie roughly on a 1:1 relation with the outflow mass derived from the broad [CII] component, and can be up to $10^{9}\,M_{\odot}$. This shows that both OH and C$^{+}$ trace the same (or similar fractions of the) massive outflows observed in ULIRGs.
 
 \item We conclude that the broad [CII] wings are a good tracer of both the molecular and atomic phases of the outflow, and that the outflow mass based on [CII] is a good estimate of the molecular outflow mass.   
 
\end{enumerate}

We wish to thank Steve Hailey-Dunsheath for discussing equation \ref{eq:mass} with us, and Henrik Spoon for providing us with the OH119 spectra of objects in his sample.  We also thank the referee for his/her helpful comments. 
J.F. acknowledges support from the NHSC/ JPL subcontracts 139807 and 1456609; Basic research in IR astronomy at NRL is funded by the US-ONR. E.G-A is a Research Associate at the Harvard-Smithsonian
Center for Astrophysics, and thanks the Spanish Ministerio de Econom\'{\i}a y Competitividad for support under project FIS2012-39162-C06-01, and NASA grant ADAP NNX15AE56G. A.S. thanks the DFG for support via German-Israeli Project Cooperation grant STE1869/1-1.GE625/15-1. S.V. acknowledges support by NASA through Herschel contracts 1427277 and 1454738. A.V. acknowledges support from the Leverhulme Trust in the form of a Research Fellowship.
PACS has been developed by a consortium of institutes led by MPE (Germany) and including UVIE (Austria); KU Leuven, CSL, IMEC (Belgium); CEA, LAM
(France); MPIA (Germany); INAF-IFSI/OAA/OAP/OAT, LENS, SISSA (Italy); IAC (Spain). This development has been supported by the funding agencies
BMVIT (Austria), ESA-PRODEX (Belgium), CEA/CNES (France), DLR (Germany), ASI/INAF (Italy), and CICYT/MCYT (Spain).

\clearpage

\bibliographystyle{apj}
\bibliography{reference}

\section{APPENDIX A: GAUSSIAN FITS}

\begin{figure} 
 \centering
\includegraphics[scale=0.3]{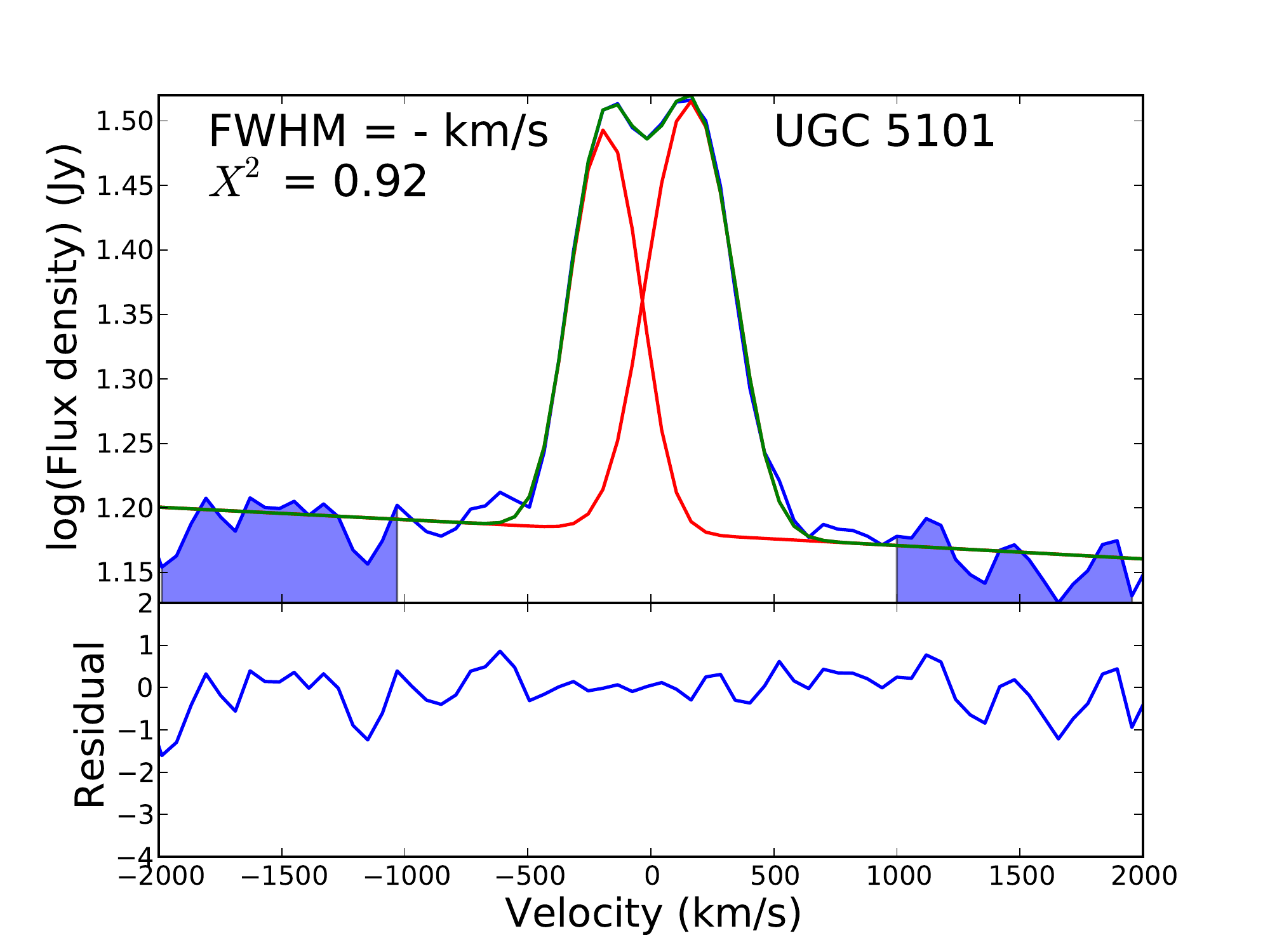}
\includegraphics[scale=0.3]{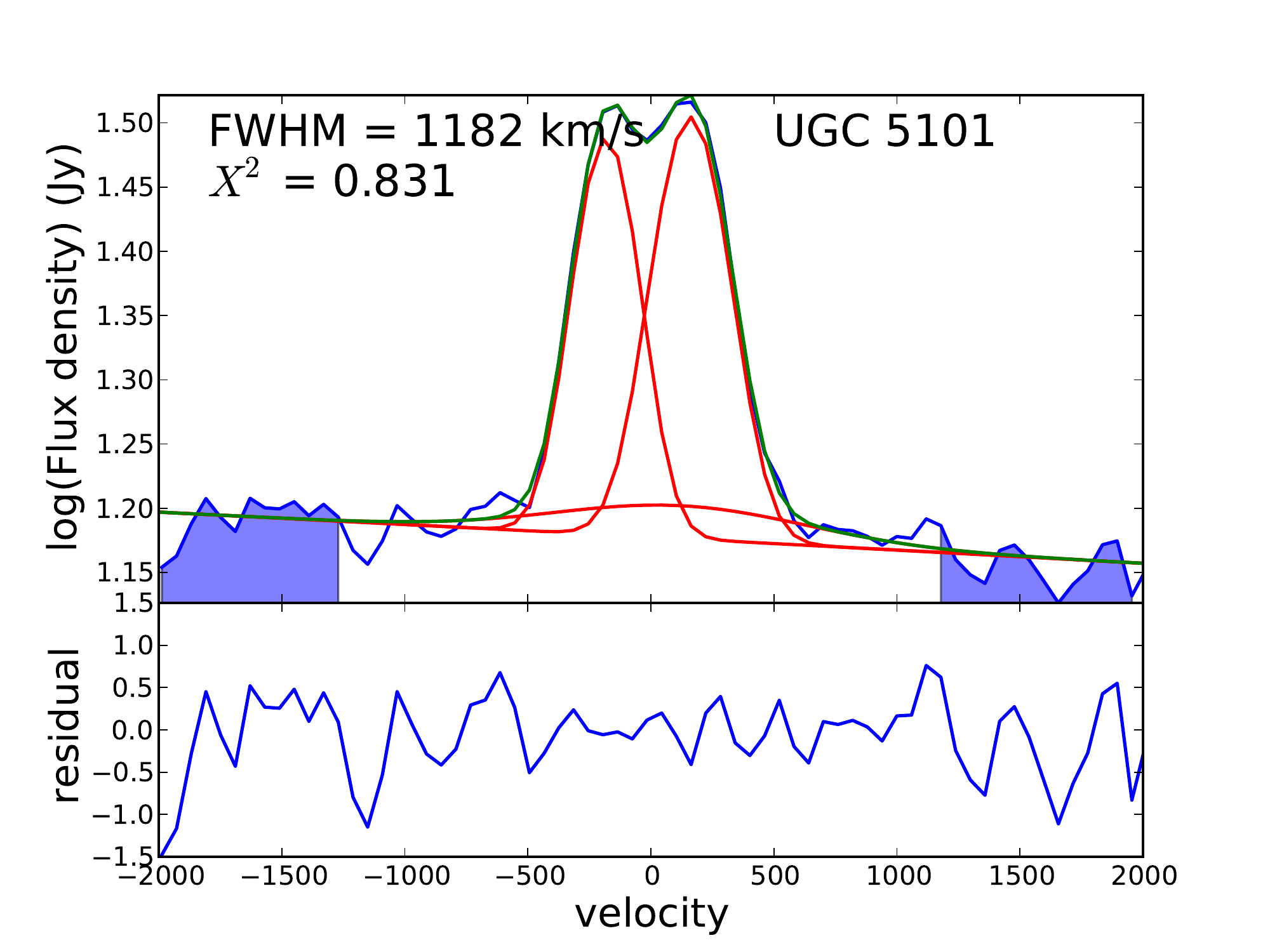}
\includegraphics[scale=0.3]{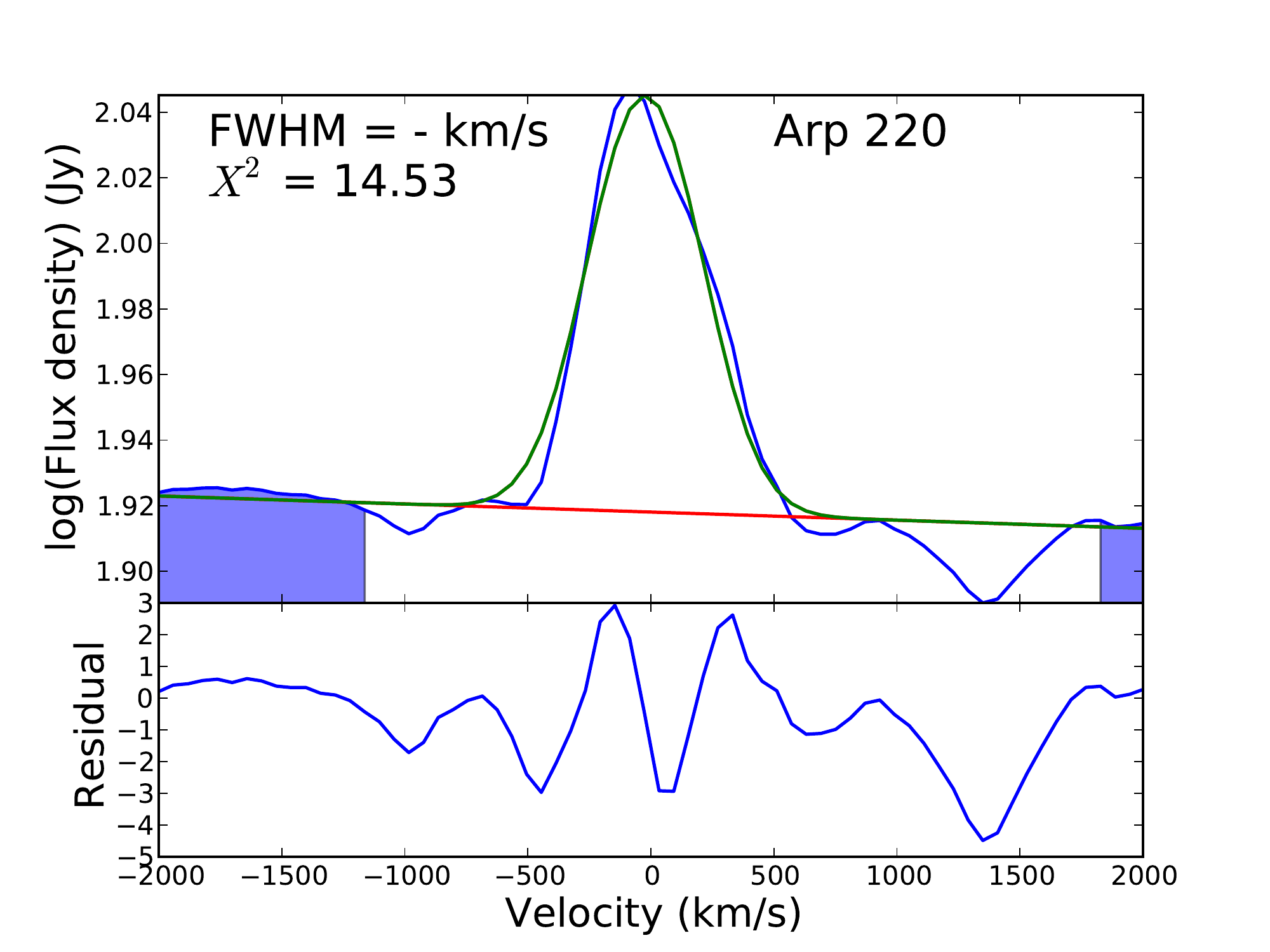}
\includegraphics[scale=0.3]{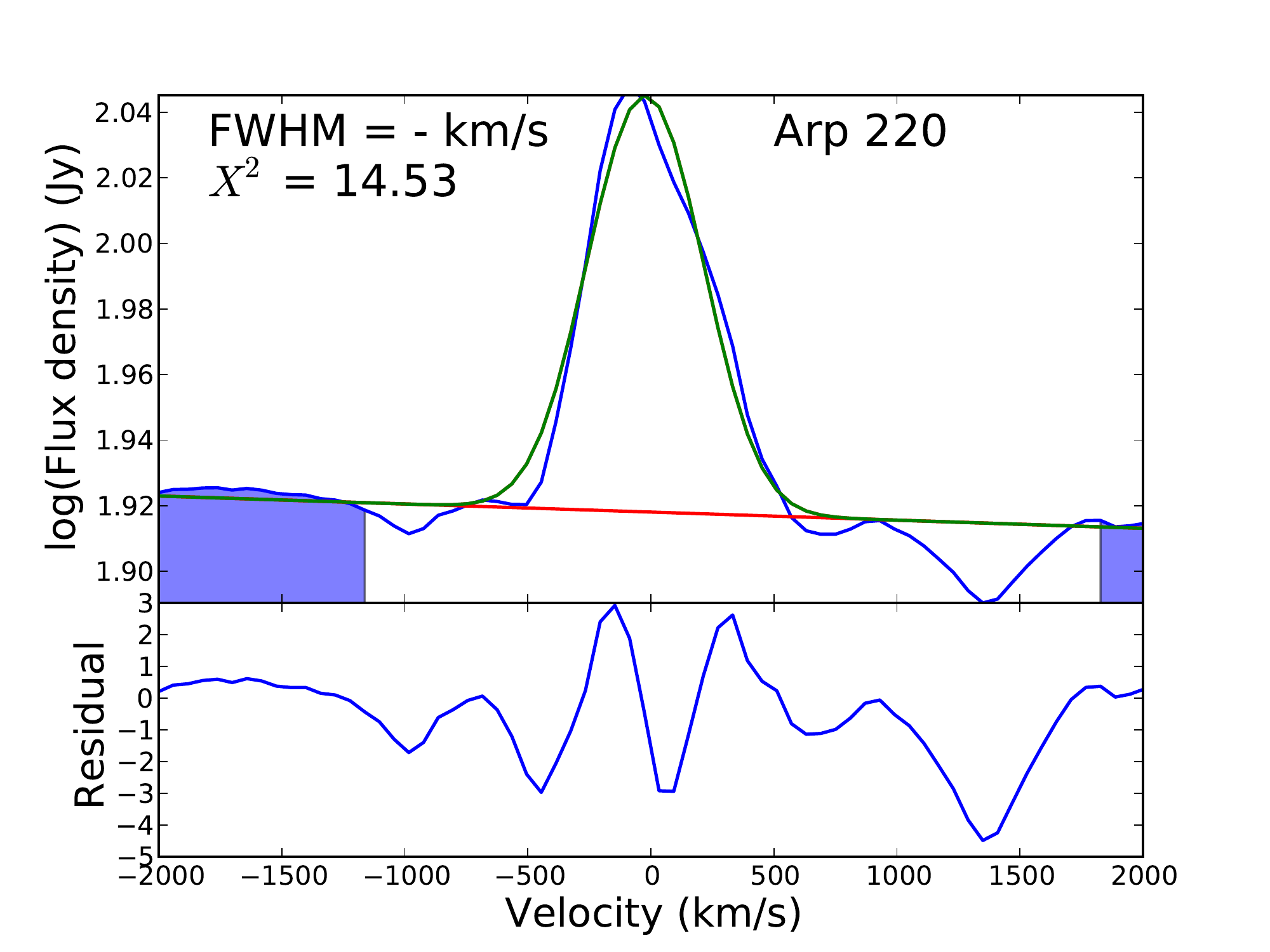}

\caption{Gaussian fits to [CII] spectra: the flux density in the upper panel is shown in logarithmic scale, to make the faint wings stand out. The residual in the lower panel is given in linear scale. The FWHM of the broad component and the reduced $\chi^{2}$ value is given in the figure. The PACS spectral resolution is 240 km s$^{-1}$ FWHM. Each source has been fitted twice with different placements of the continuum, as shown by the shaded area. }
\label{fig:appA}
\end{figure}

Figure \ref{fig:appA} shows the [CII] spectra for each source in our sample. Each subfigure lists the FHWM of the broad component and the reduced $\chi^{2}$ value for the fit. The fit has been done to the normal (linearly scaled) flux density, but here the y-axis is shown in a logarithmic scale, to increase the visibility of the low signal wings. The residual plot below each figure is shown in linear scale. Shaded areas mark the range that was chosen for the continuum fit. The [CII] spectra for each source in our sample are plotted twice, in order estimate the error on the FWHM of the broad component. In the first subfigure, the continuum was fitted close to the line, at good signal to noise, but potentially removing broad lines and  minimizing the FWHM of the broad component. In the second figure, the continuum was fitted far away from the line, at the edges where the signal to noise ratio was low, but leaving plenty of room for a broad component. Because the spectra are only fully covered from $-1300\,
\text{km s}^{-1}$ to $+1300\,\text{km s}^{-1}$, and the part outside this region has been scanned less frequently, the noise steadily increases at velocities $< -1300\,\text{km s}^{-1}$ and $> 1300\,\text{km s}^{-1}$. We therefore avoid fitting the continuum in high-velocity regions with sudden peaks and dips in the spectrum. After continuum subtraction, we attempted to fit two Gaussians (and in exceptional cases three Gaussians) to the line profile.  Table \ref{tab:appA} contains all the parameters of the Gaussian fits.

 \begin{figure} \centering
\includegraphics[scale=0.3]{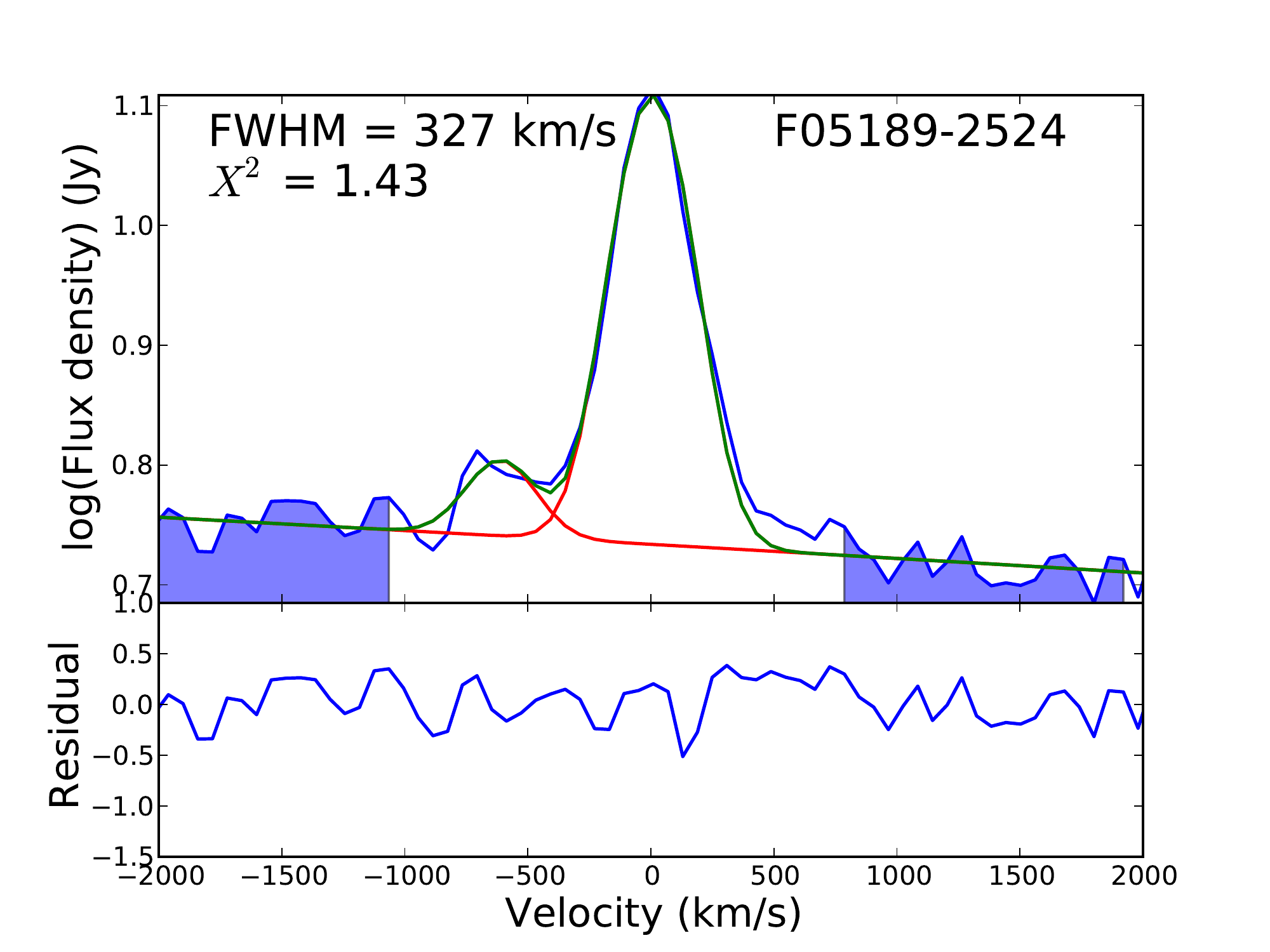}
\includegraphics[scale=0.3]{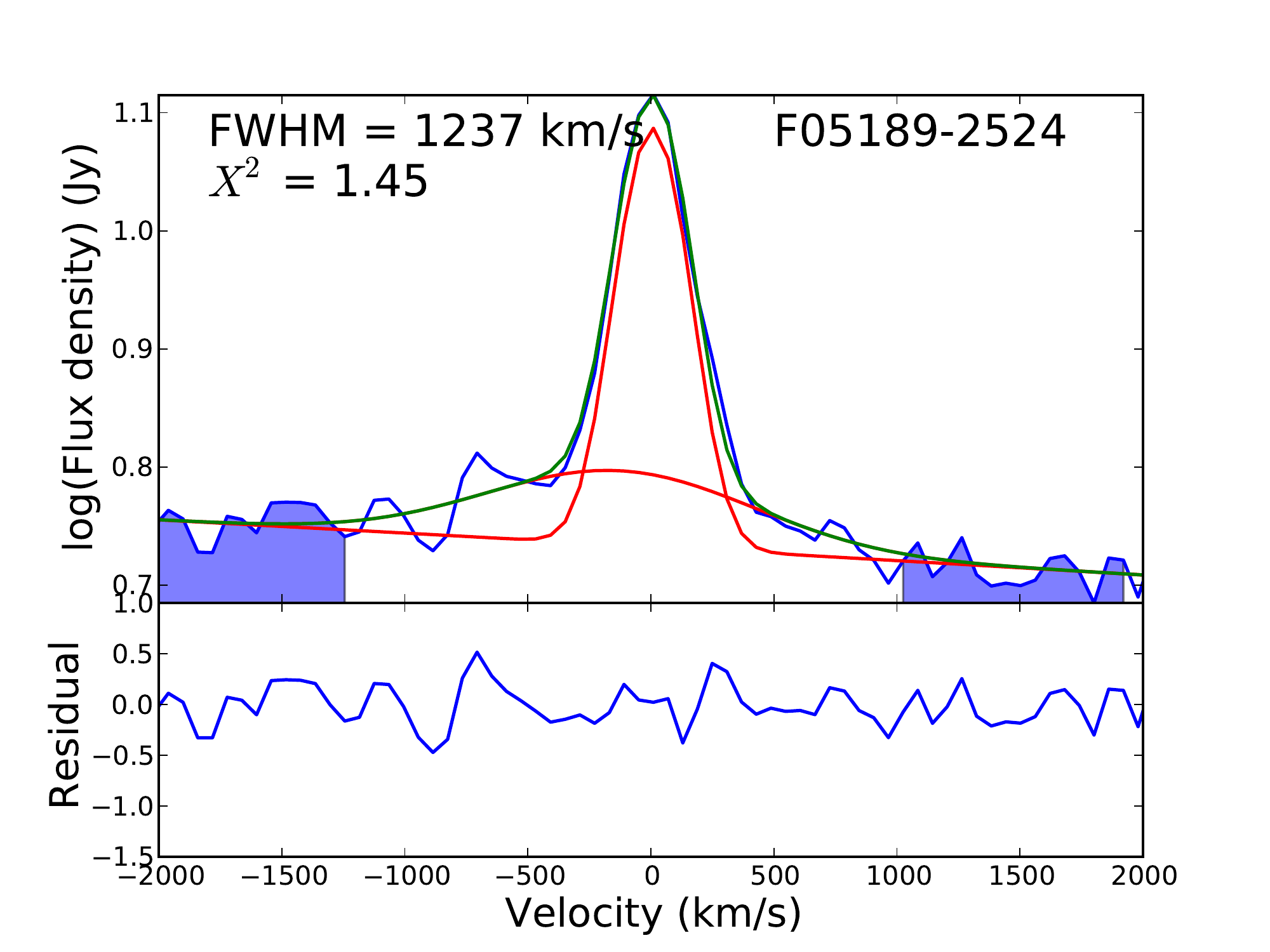}
\includegraphics[scale=0.3]{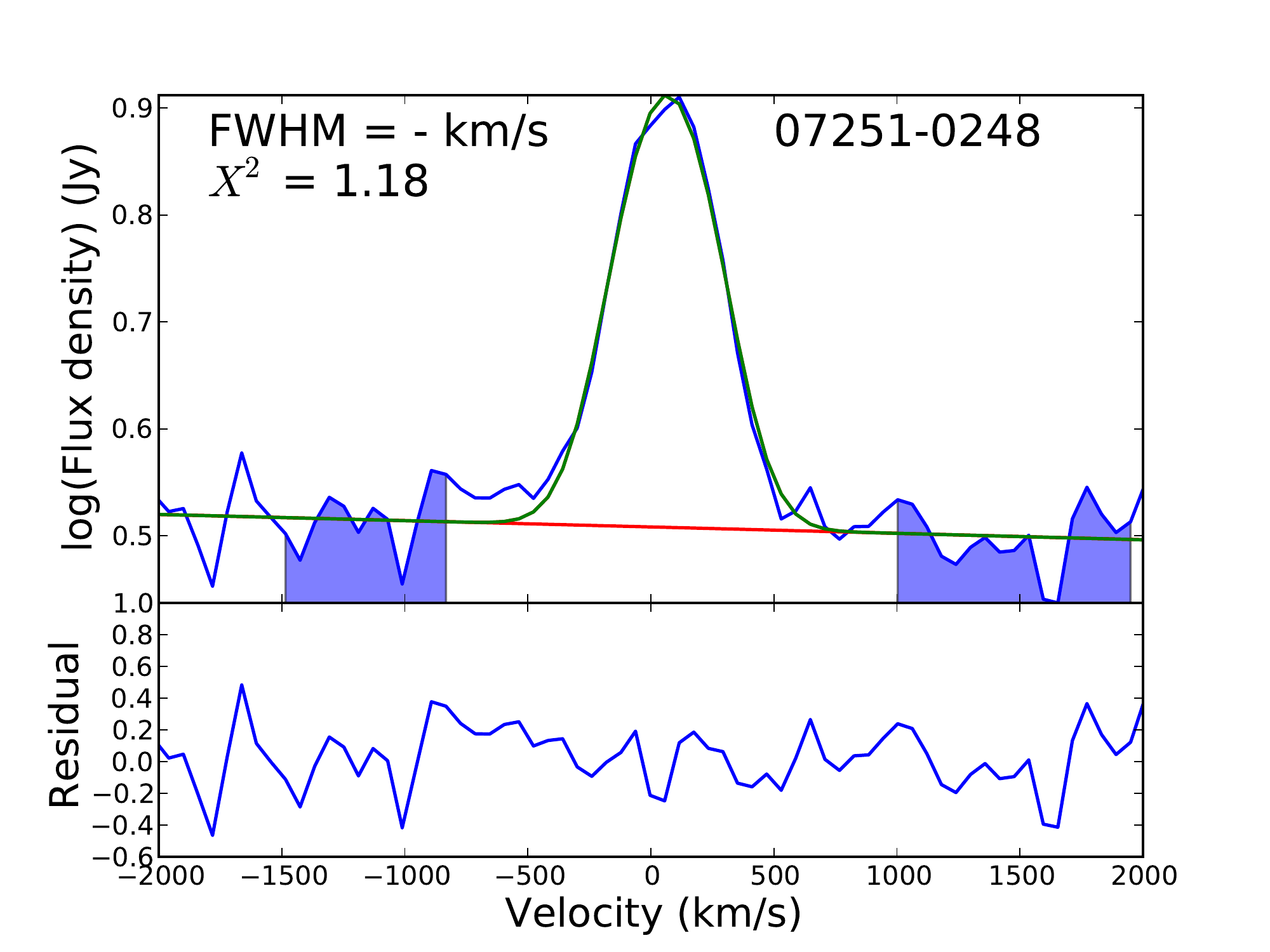}
\includegraphics[scale=0.3]{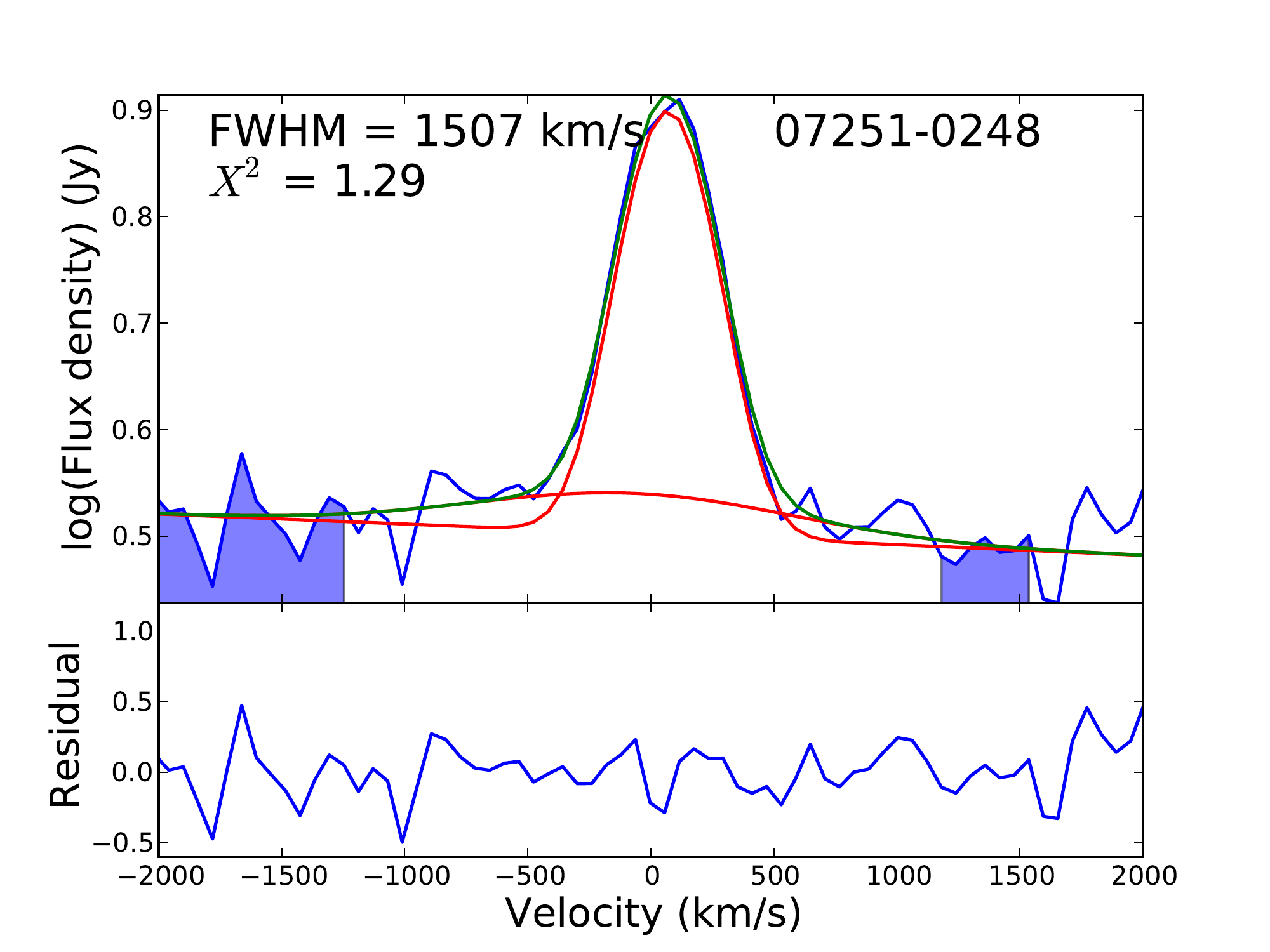}
\includegraphics[scale=0.3]{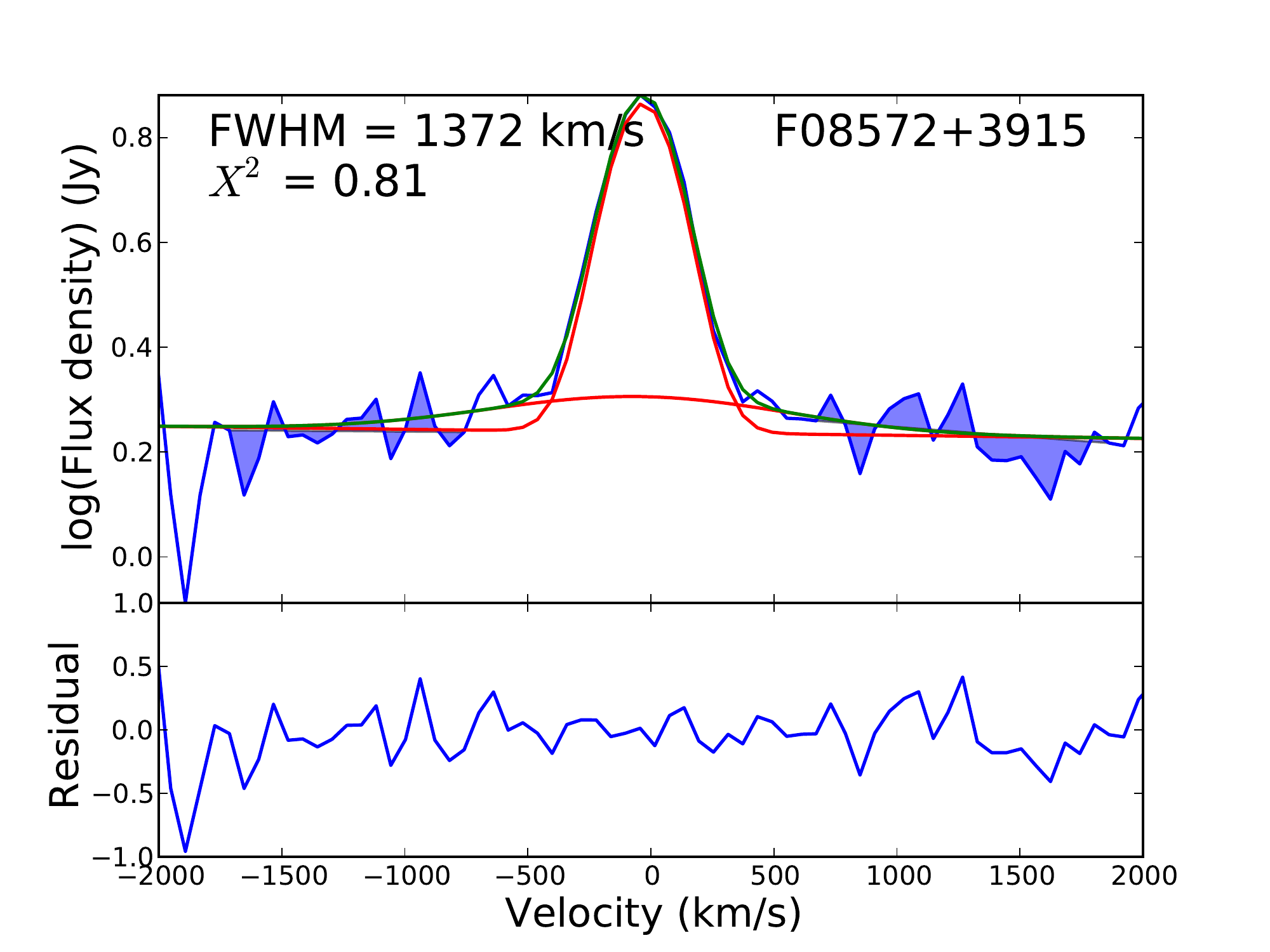}
\includegraphics[scale=0.3]{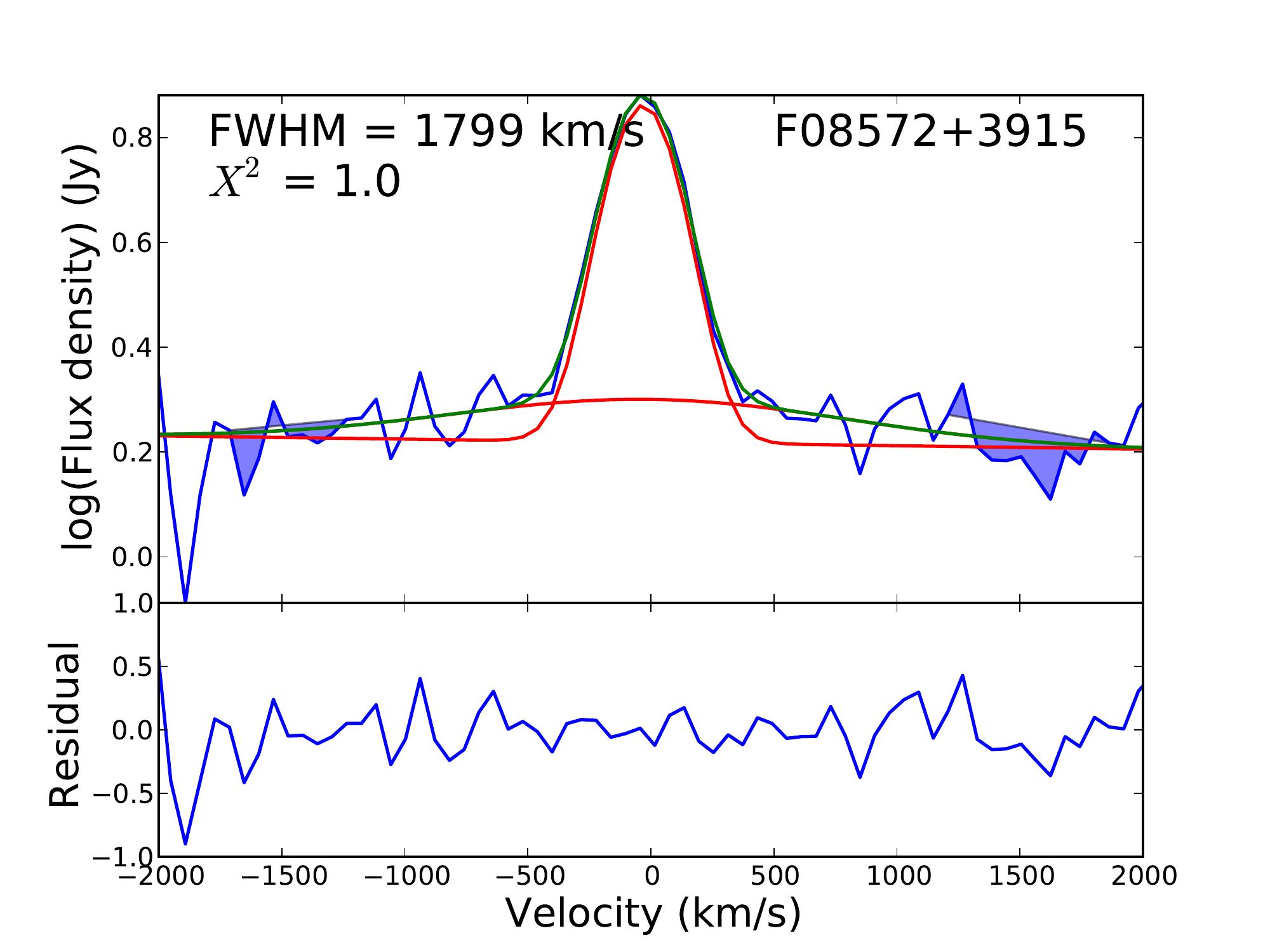}
\includegraphics[scale=0.3]{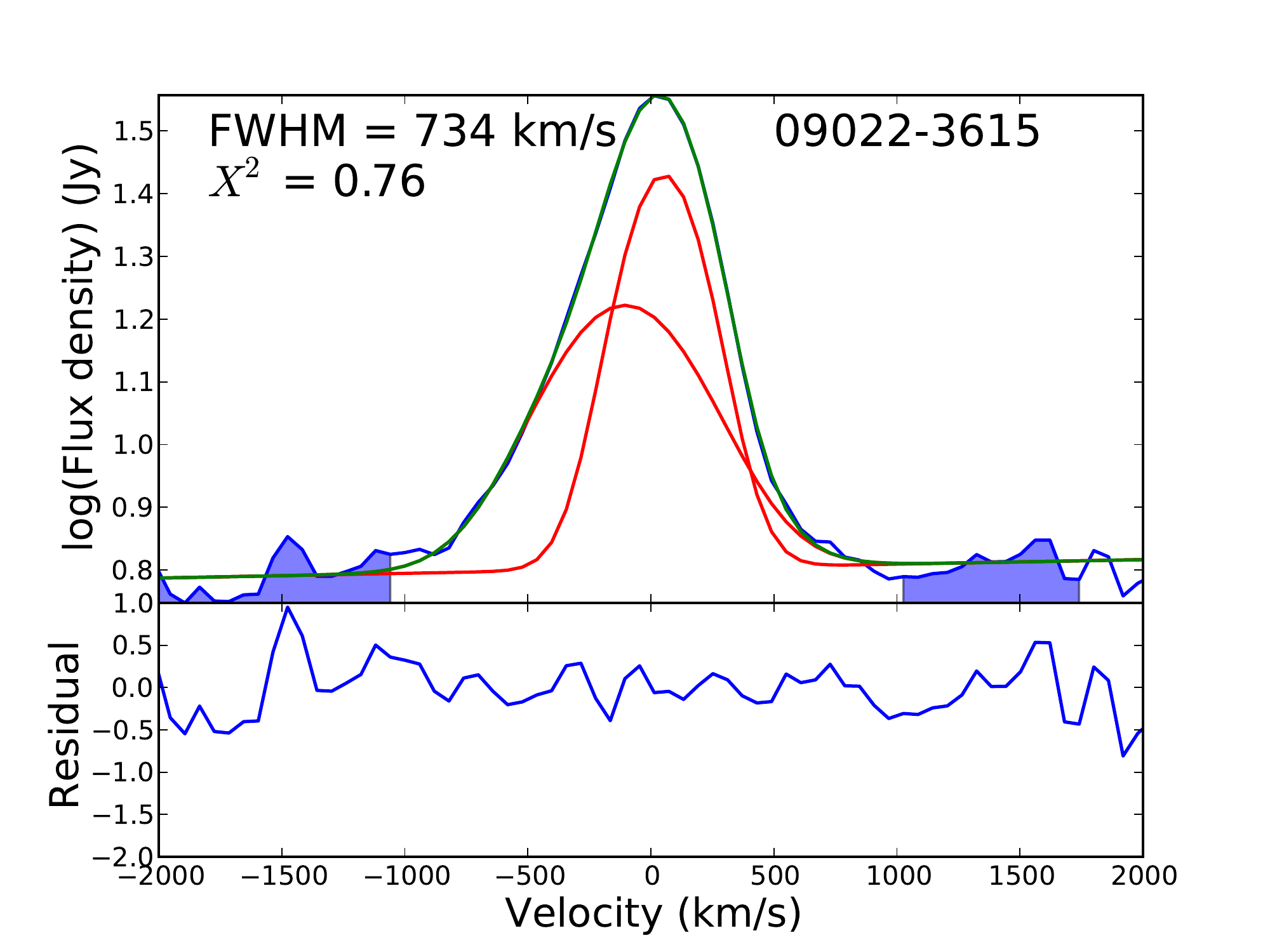}
\includegraphics[scale=0.3]{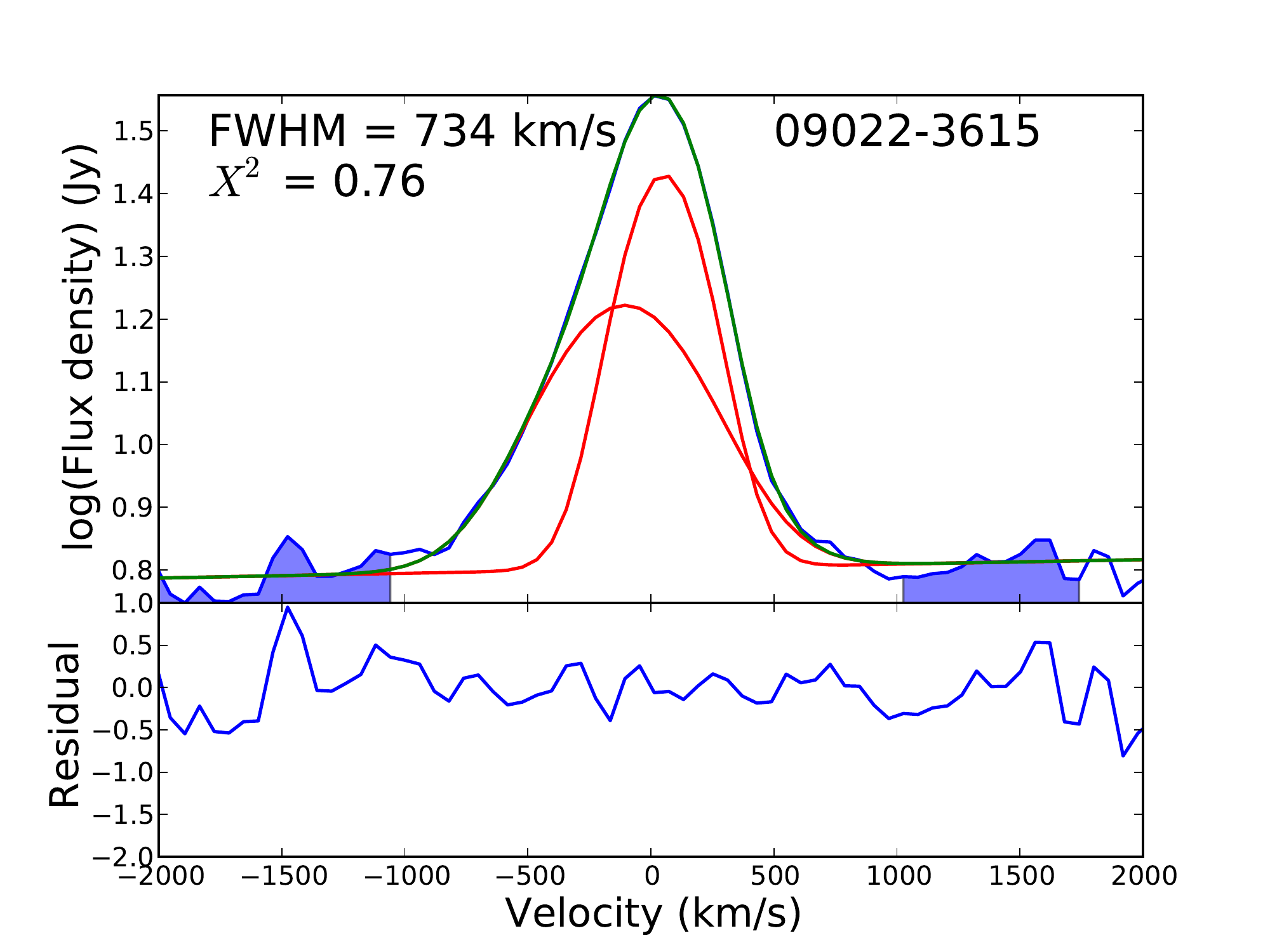}
\includegraphics[scale=0.3]{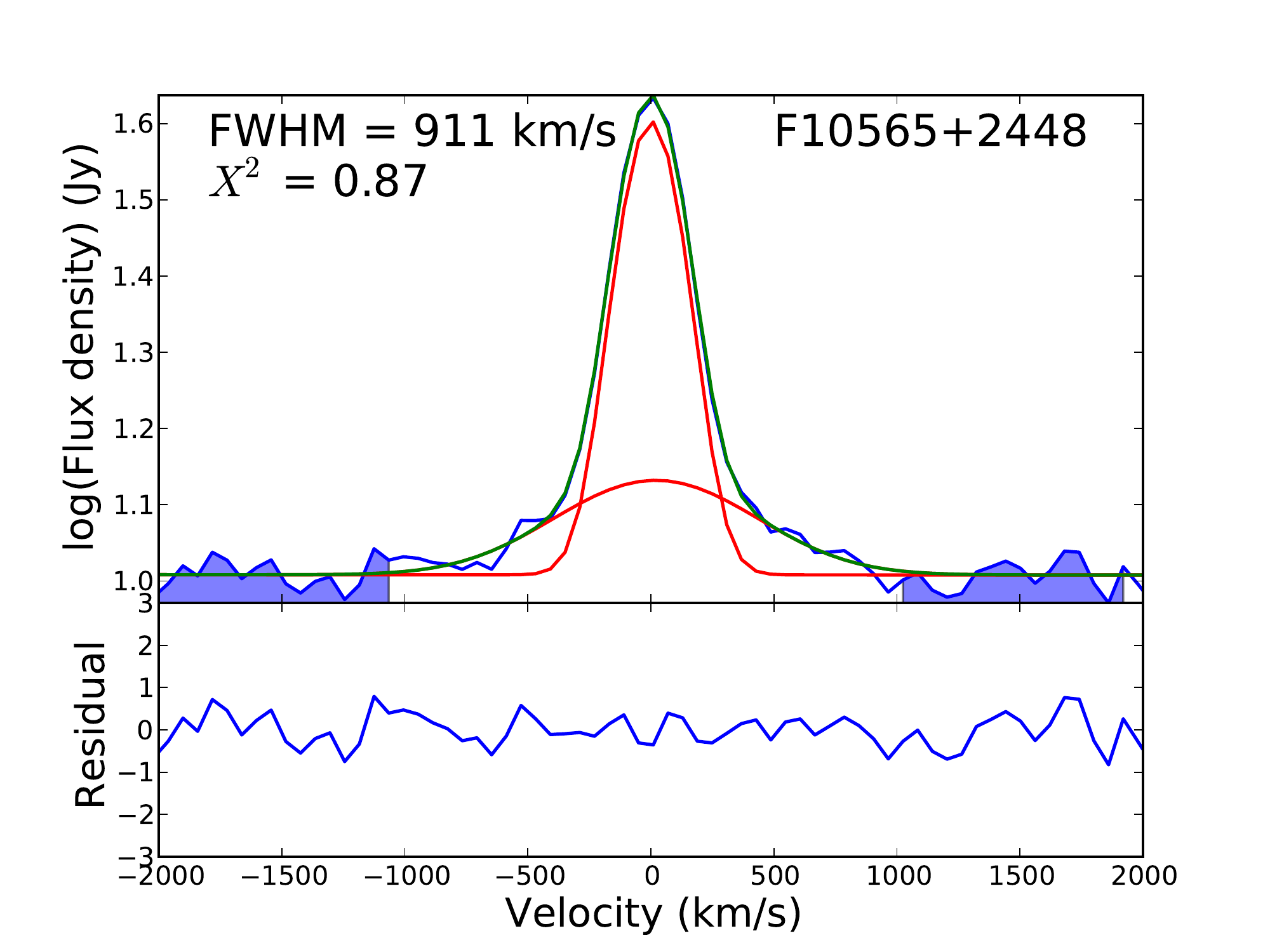}
\includegraphics[scale=0.3]{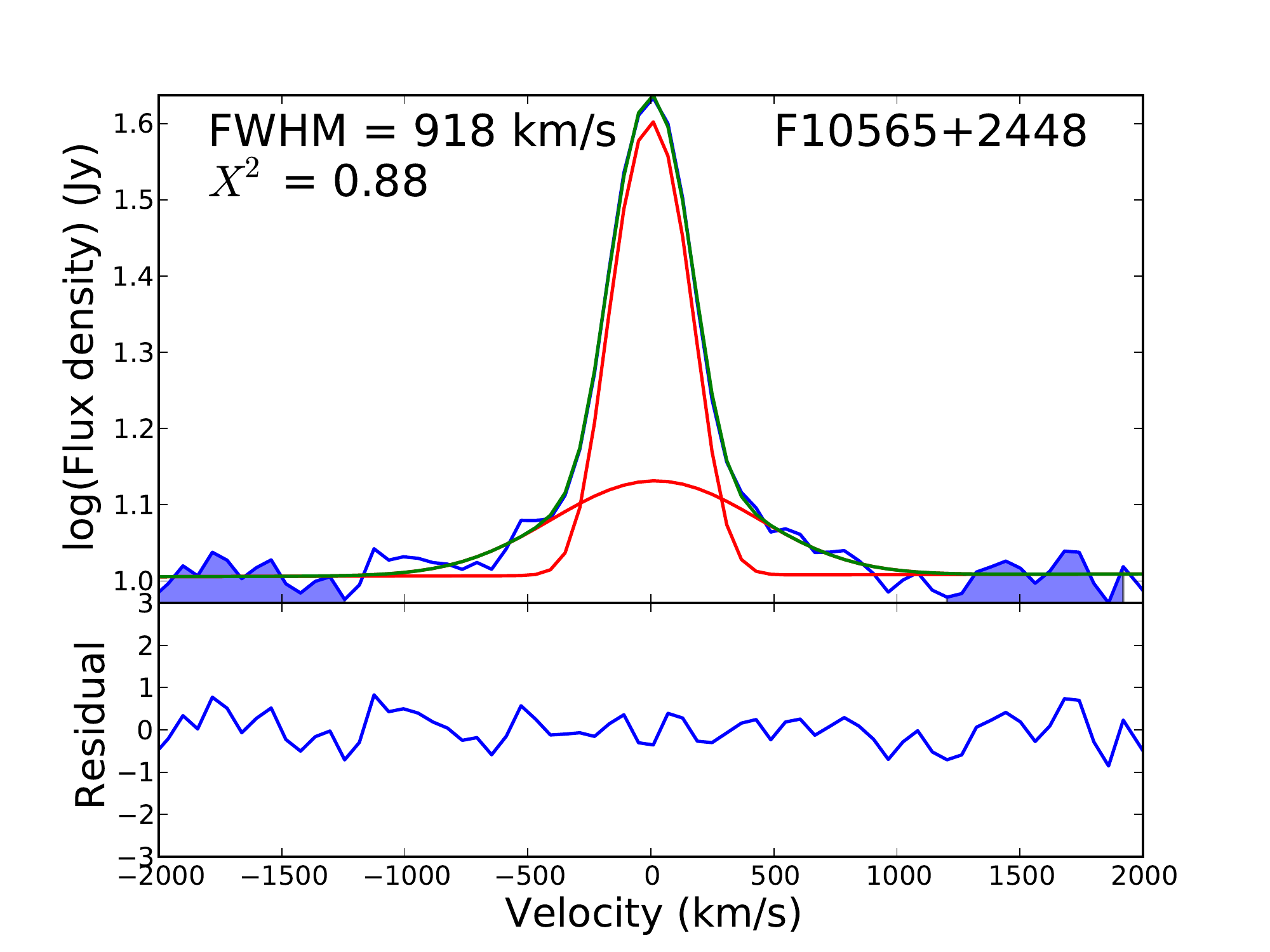}
\end{figure} 
 \begin{figure} 
 \centering
\includegraphics[scale=0.3]{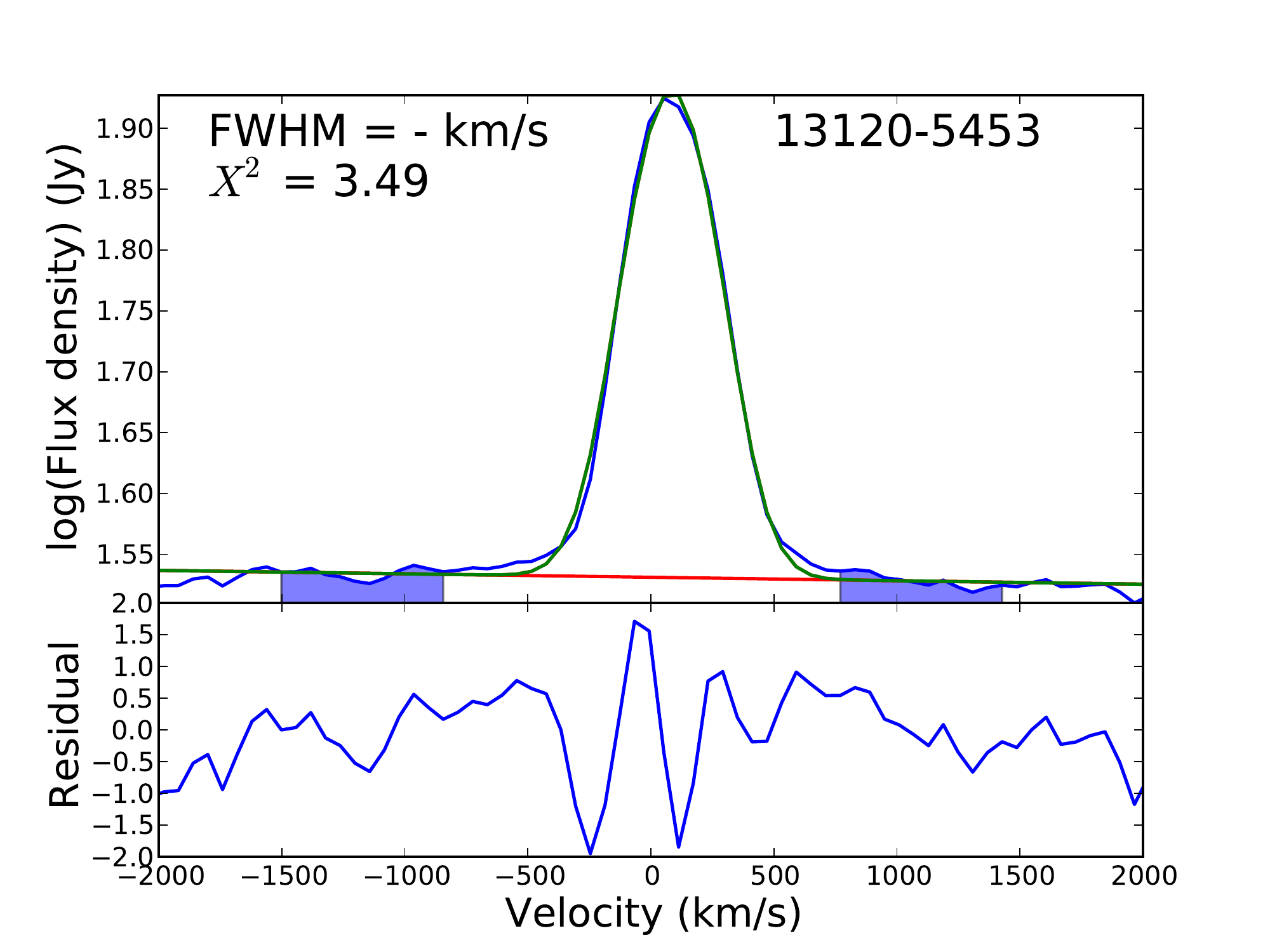}
\includegraphics[scale=0.3]{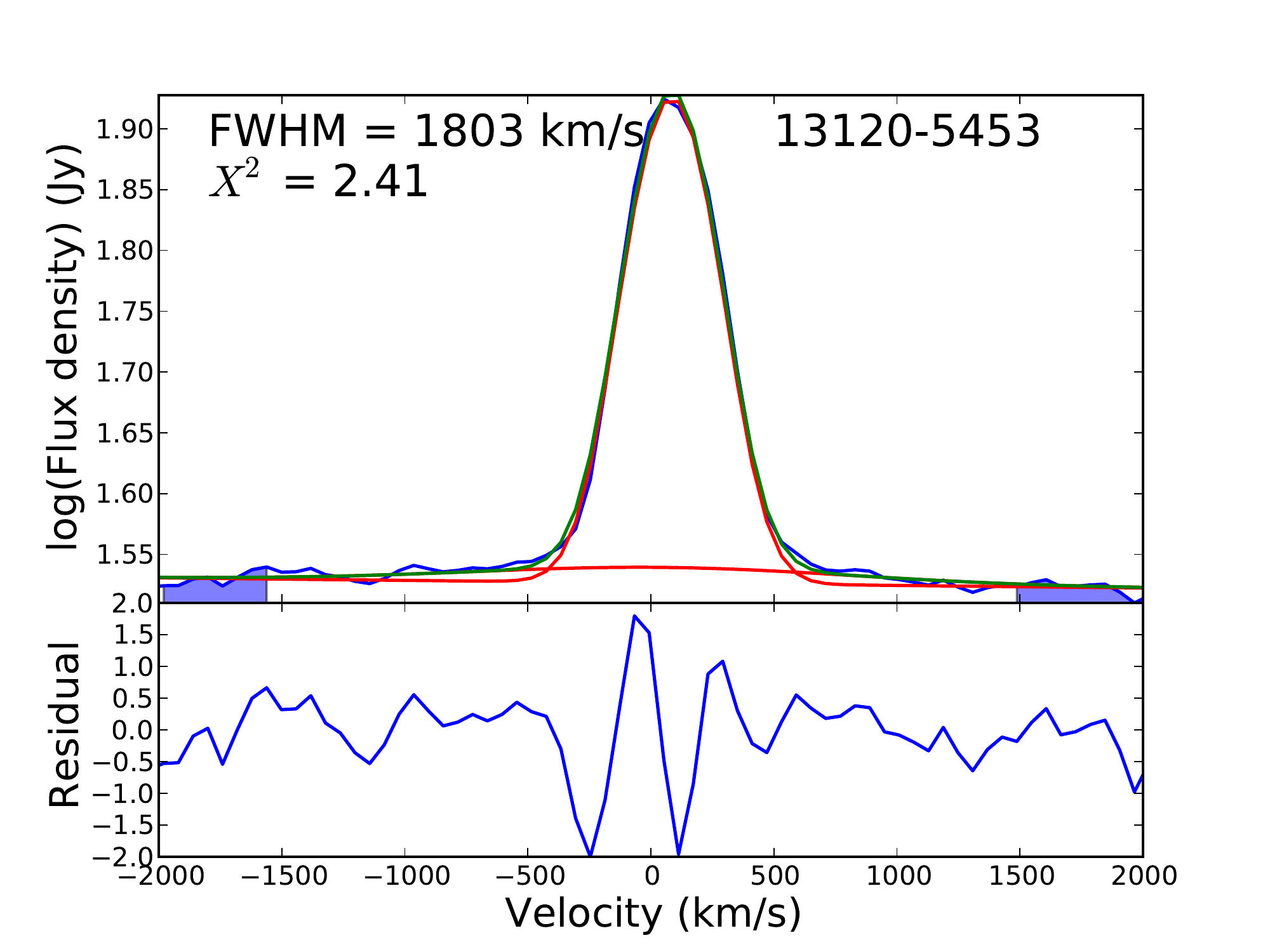}
\includegraphics[scale=0.3]{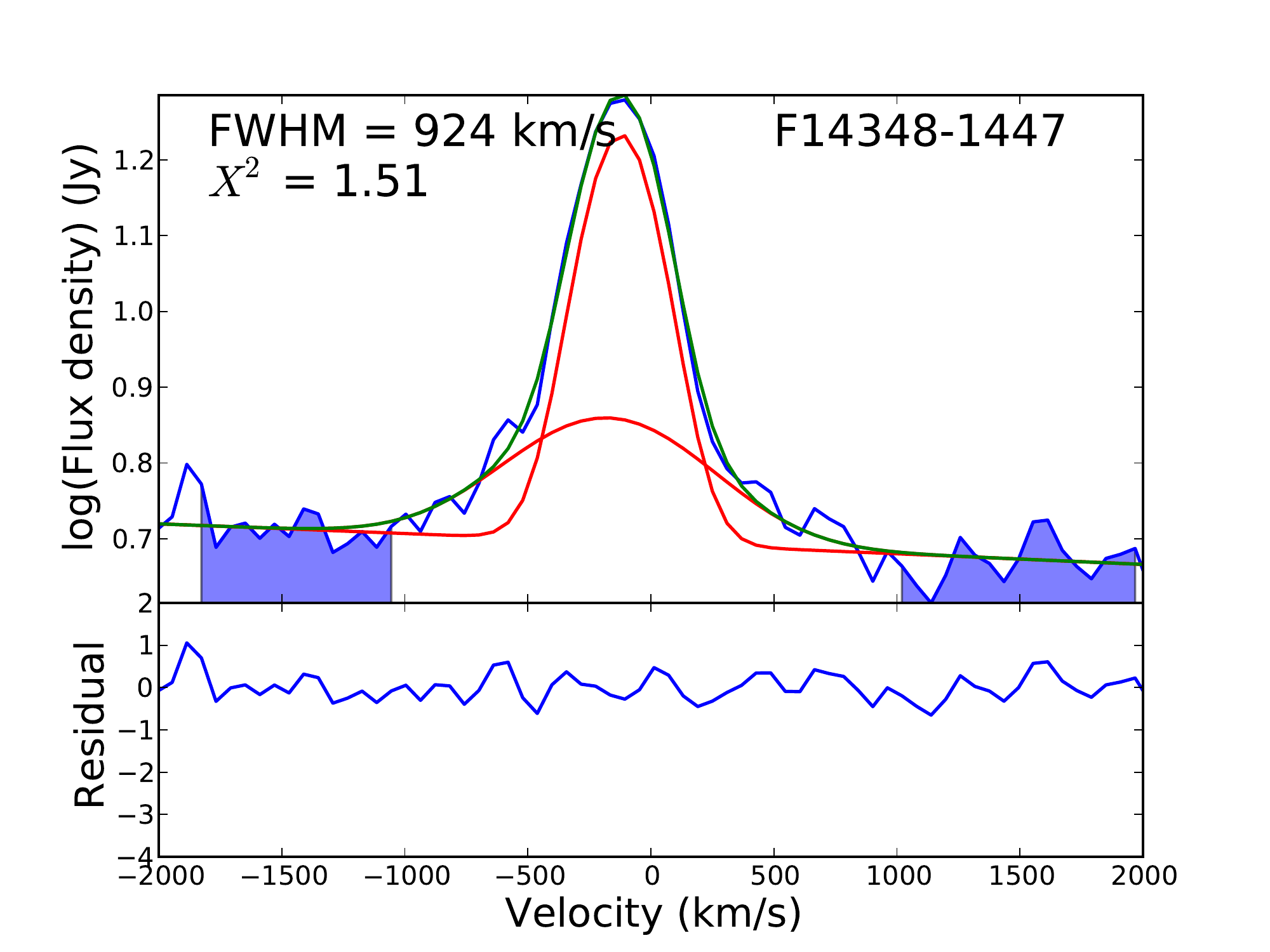}
\includegraphics[scale=0.3]{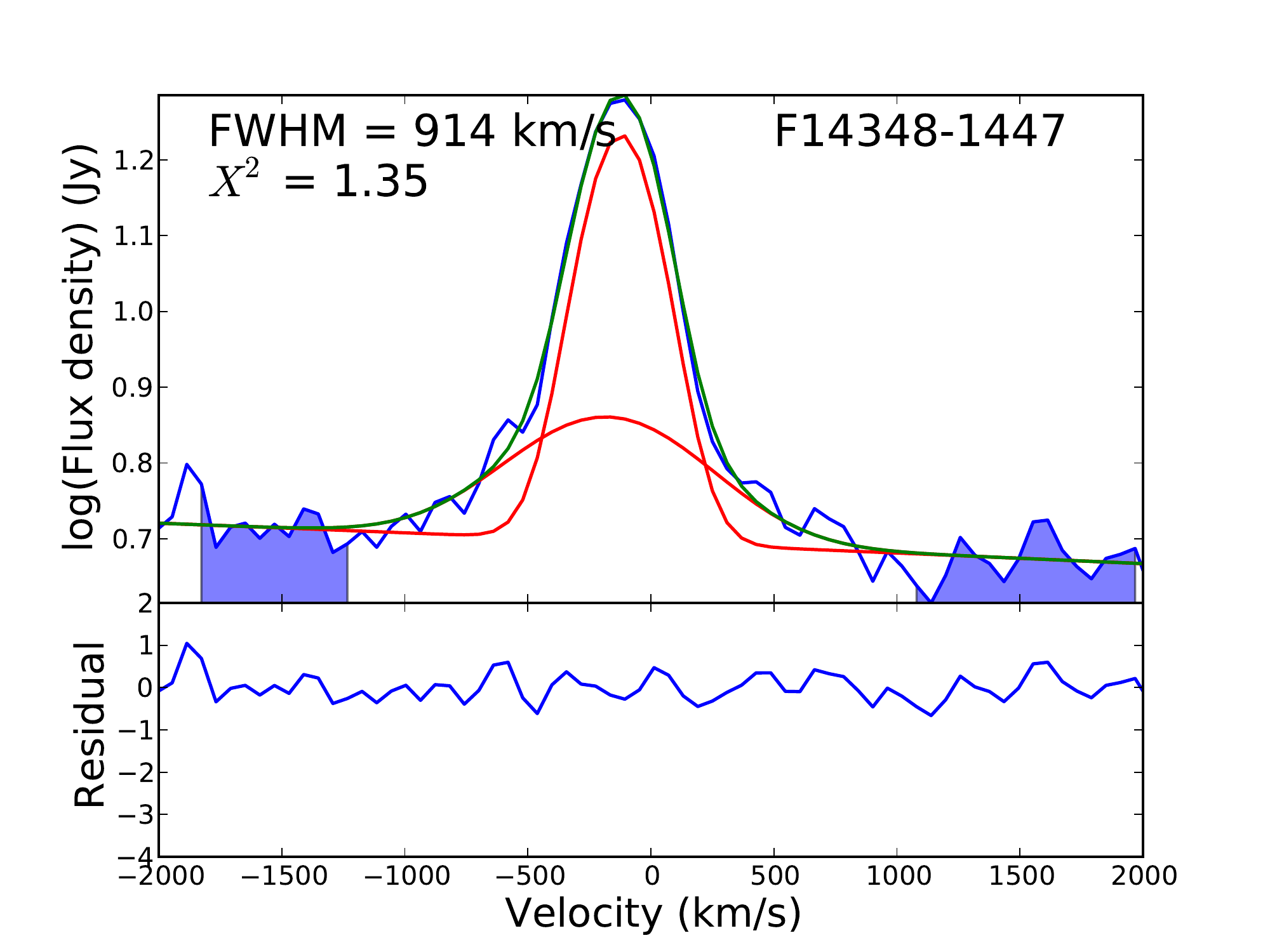}
\includegraphics[scale=0.3]{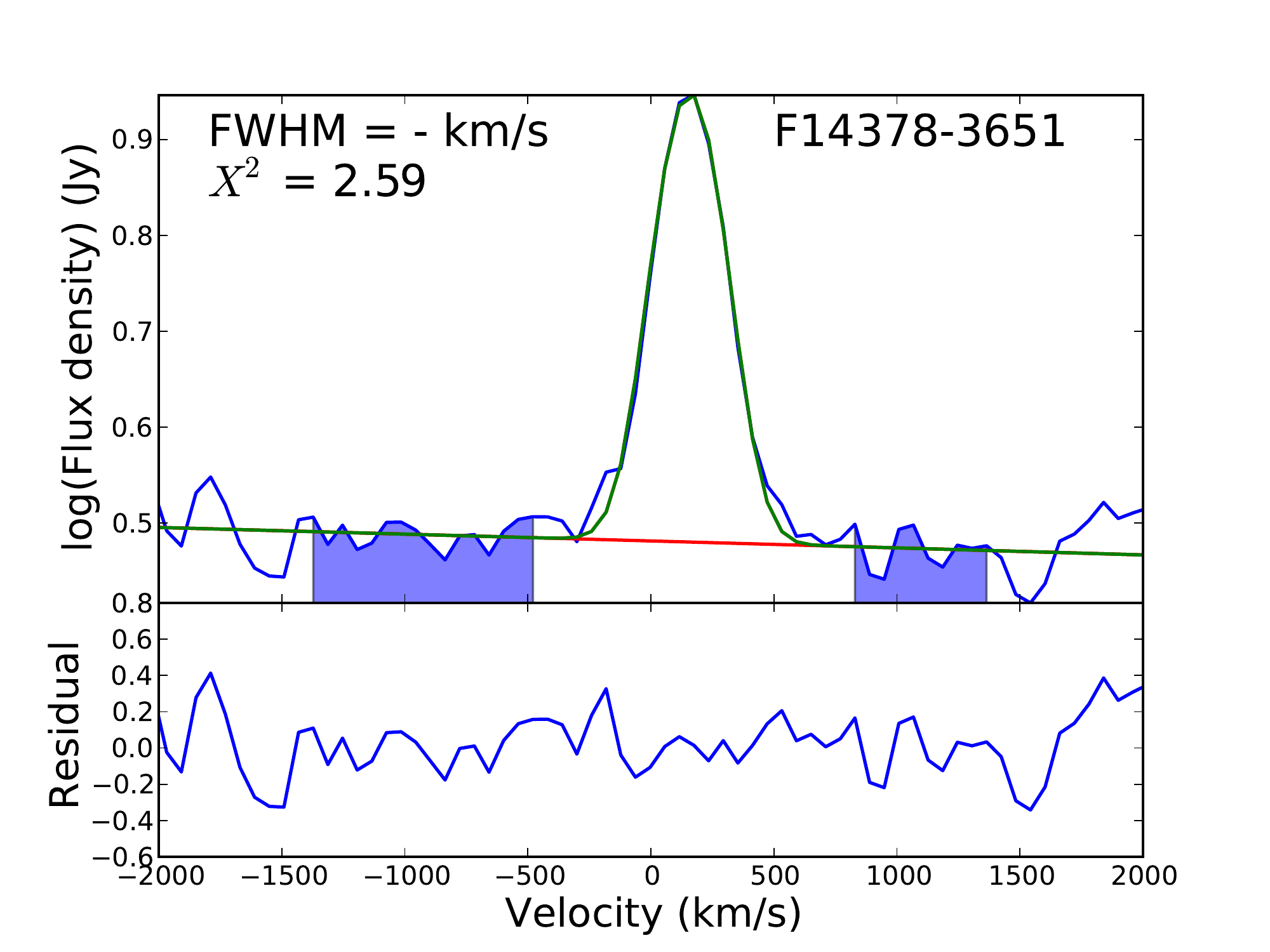}
\includegraphics[scale=0.3]{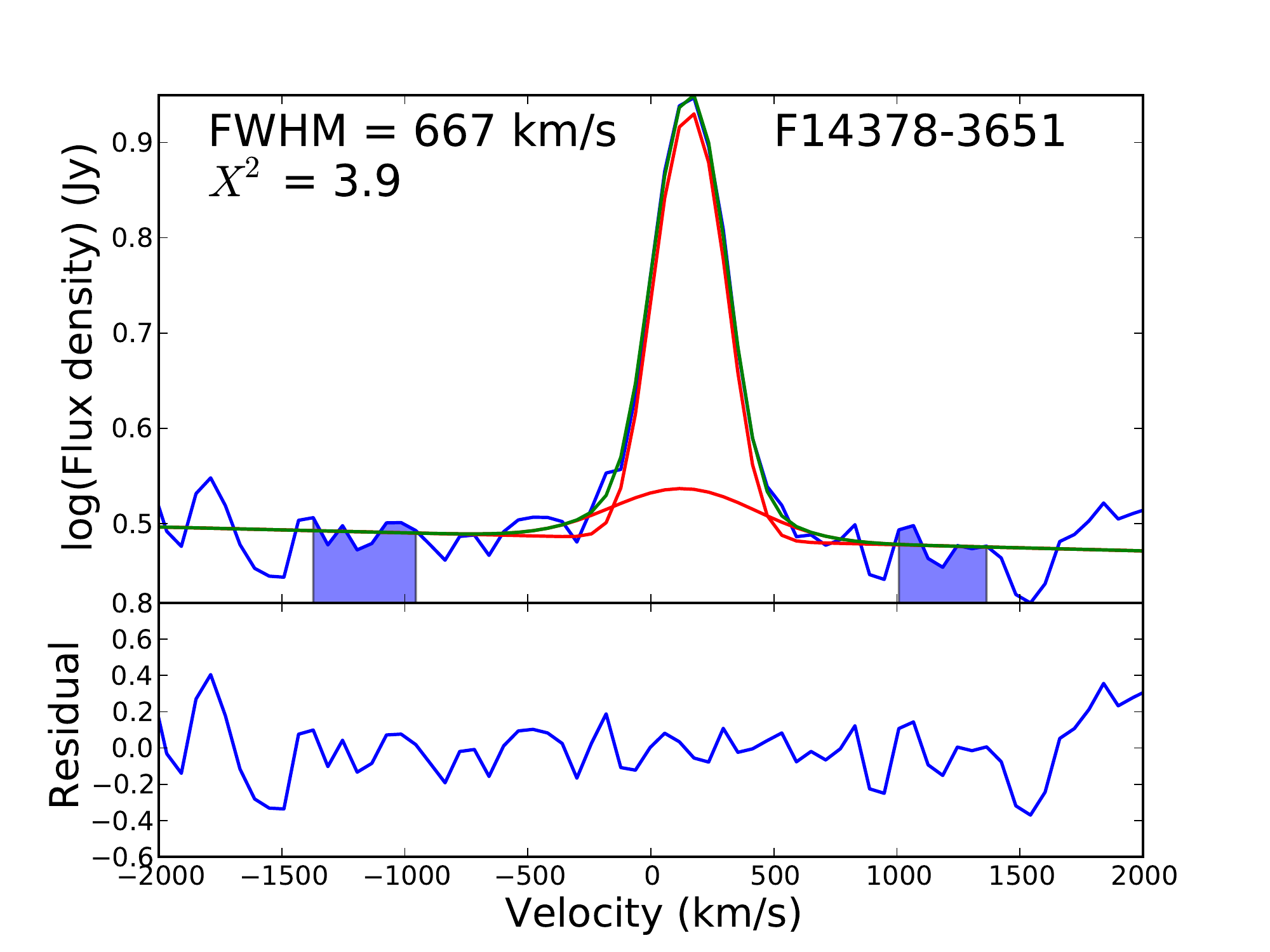}
\includegraphics[scale=0.3]{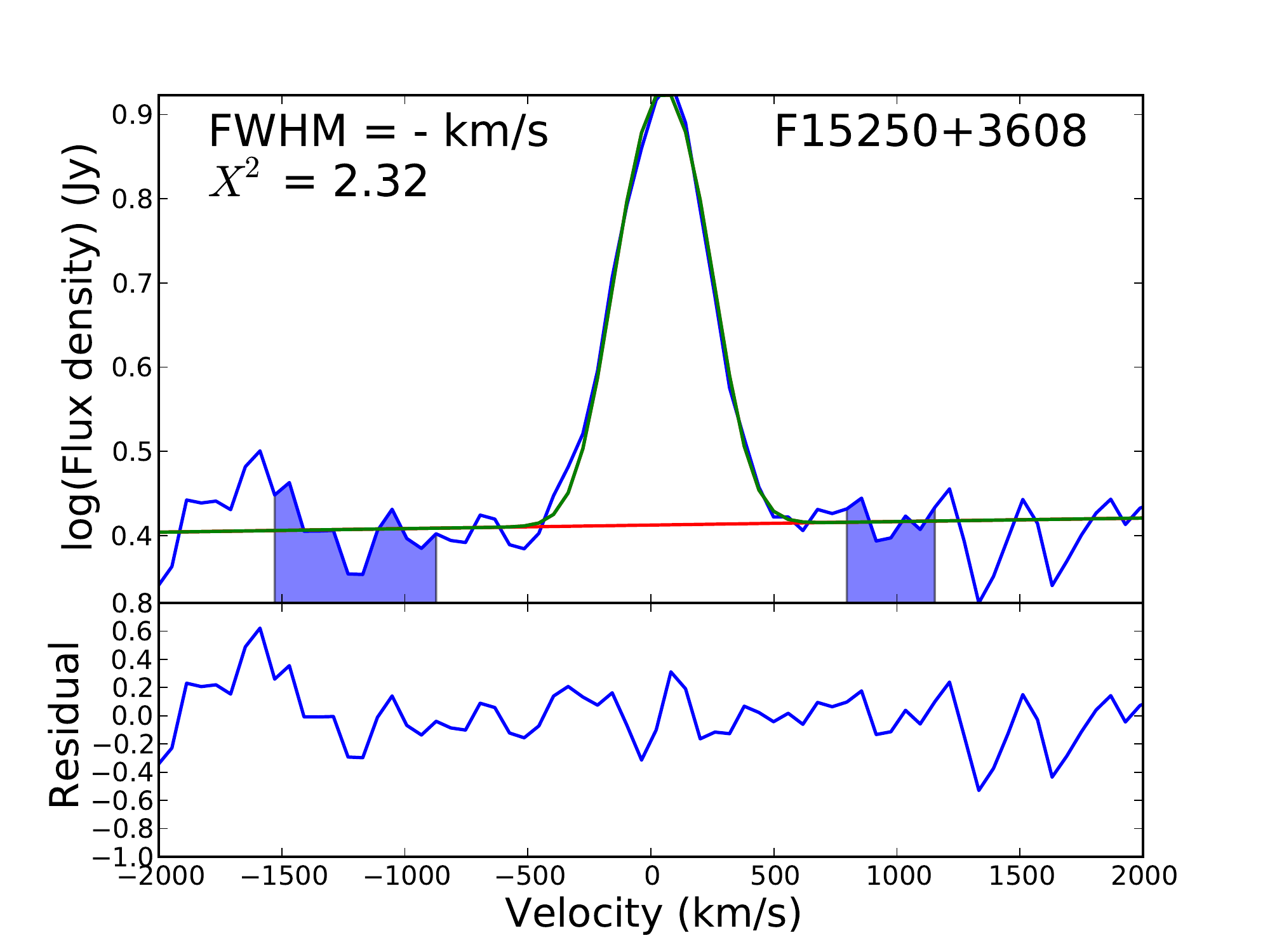}
\includegraphics[scale=0.3]{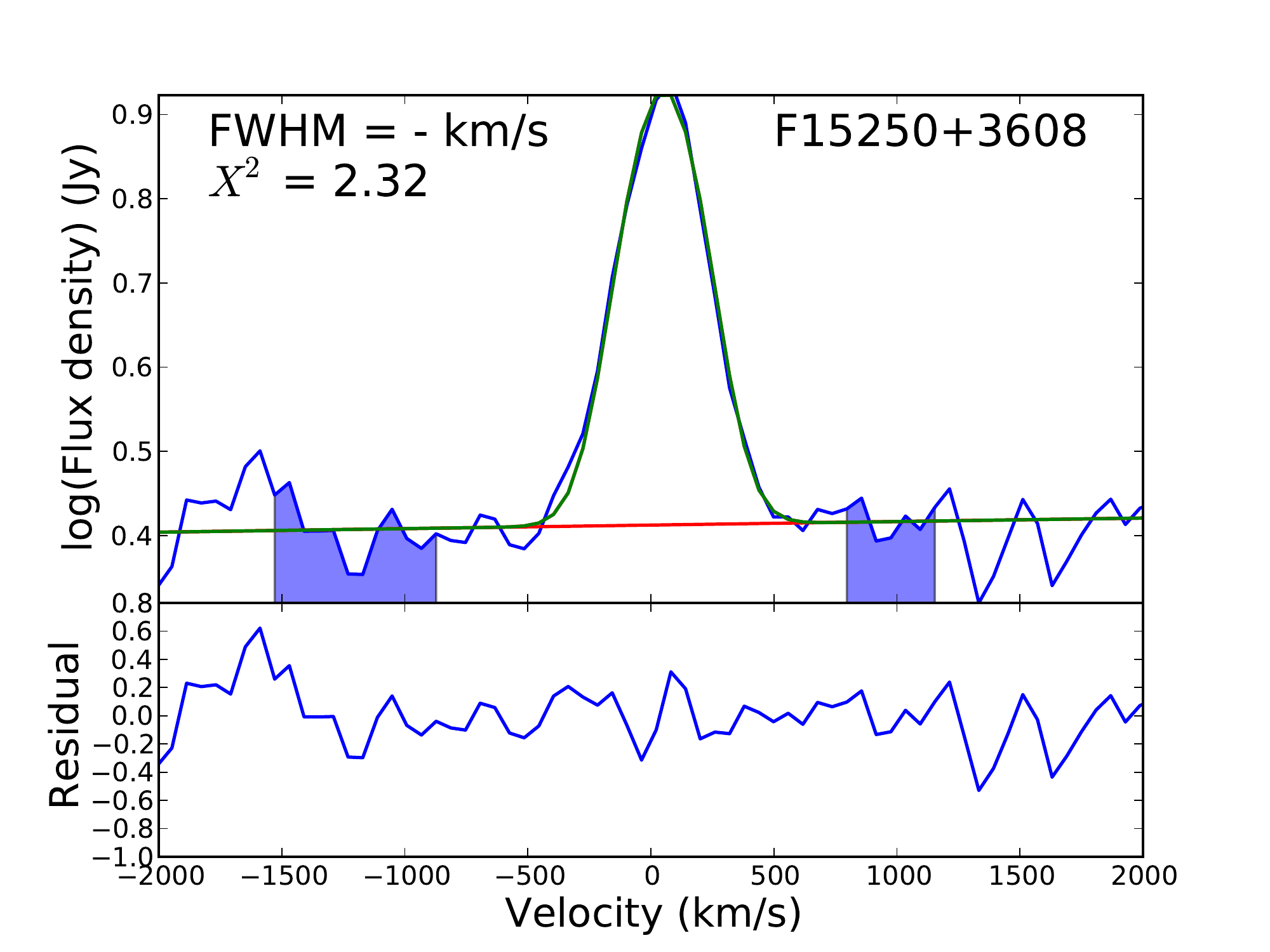}
\includegraphics[scale=0.3]{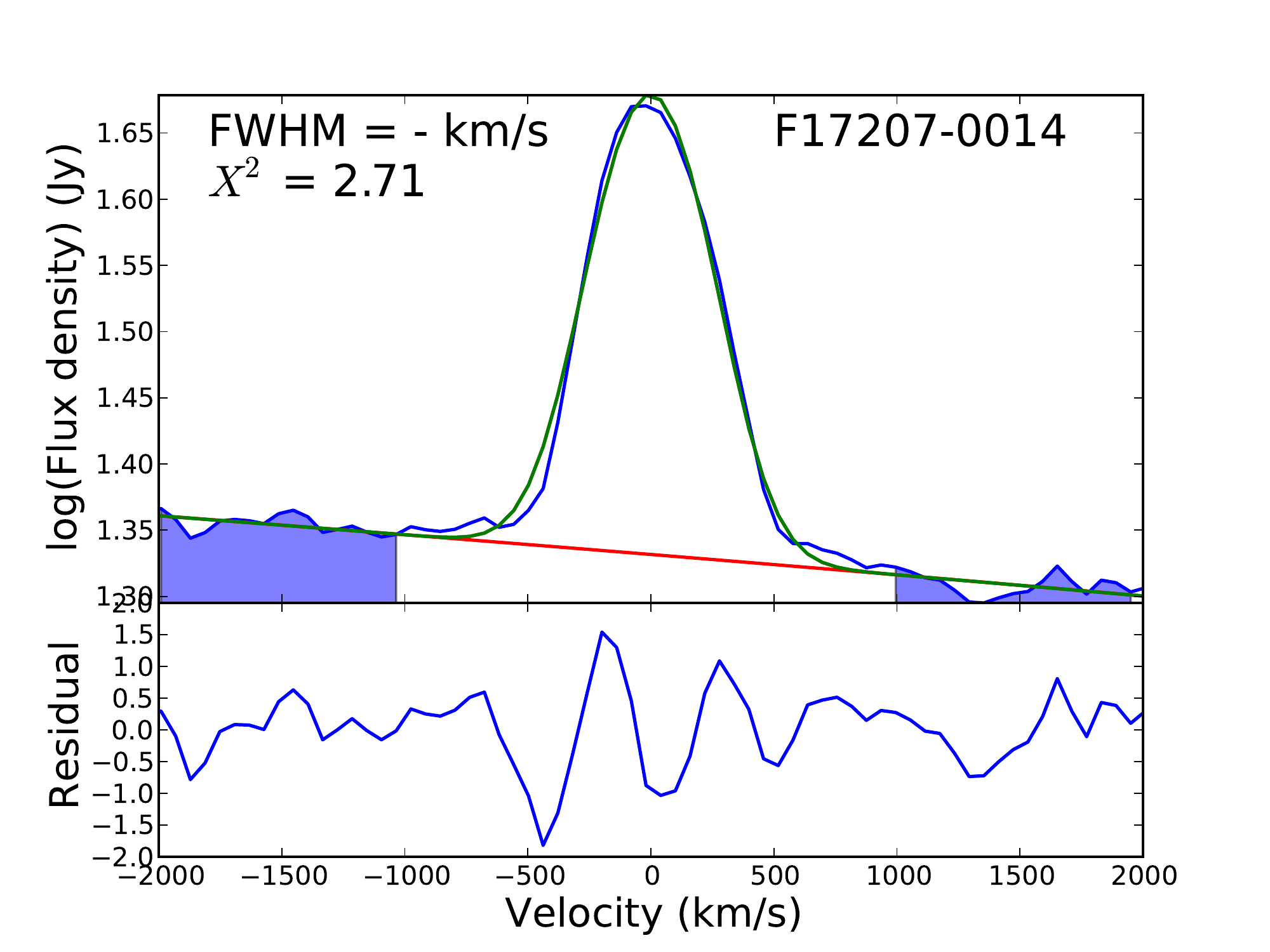}
\includegraphics[scale=0.3]{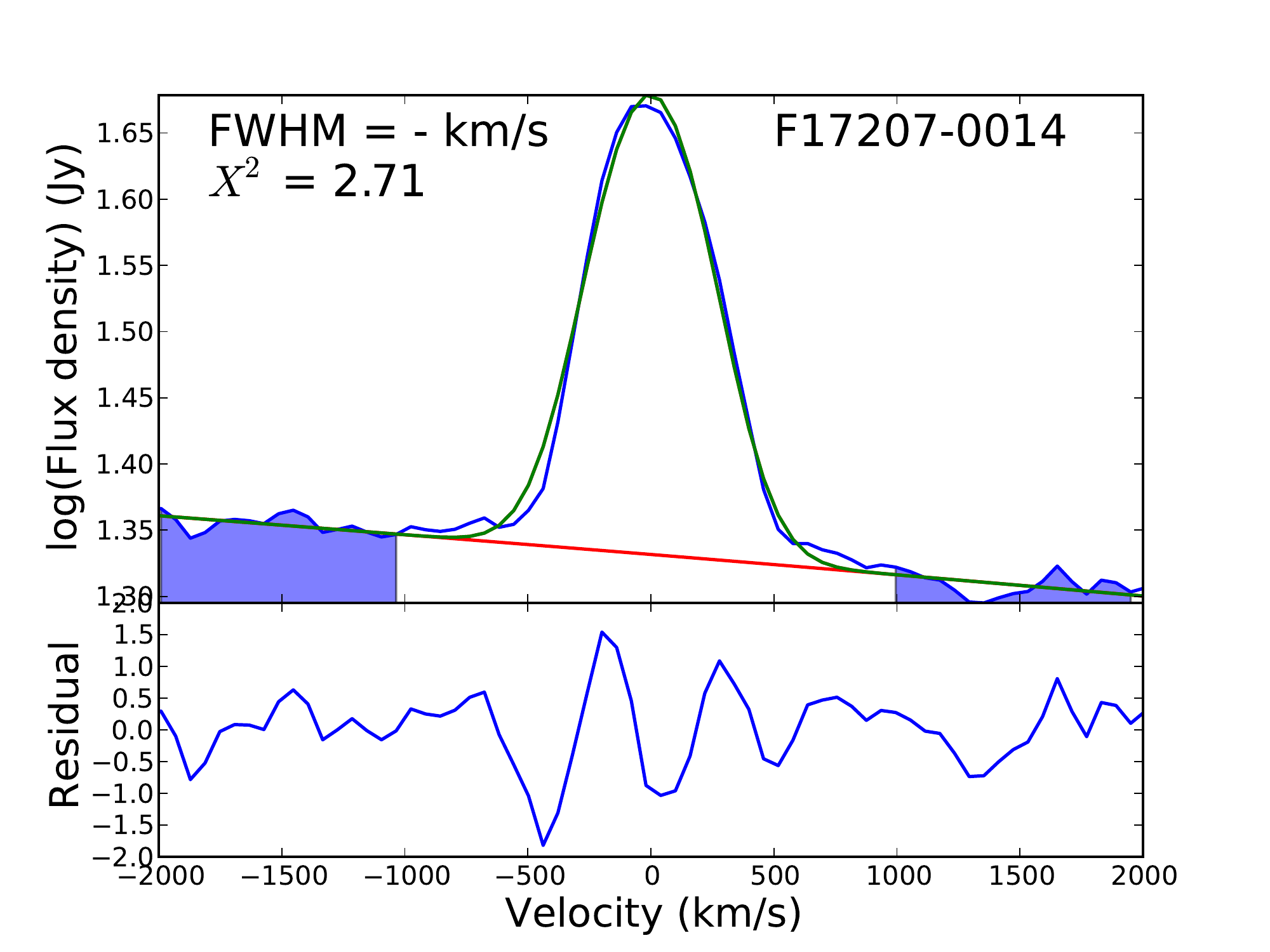}\end{figure} 
 \begin{figure} 
  \centering
\includegraphics[scale=0.3]{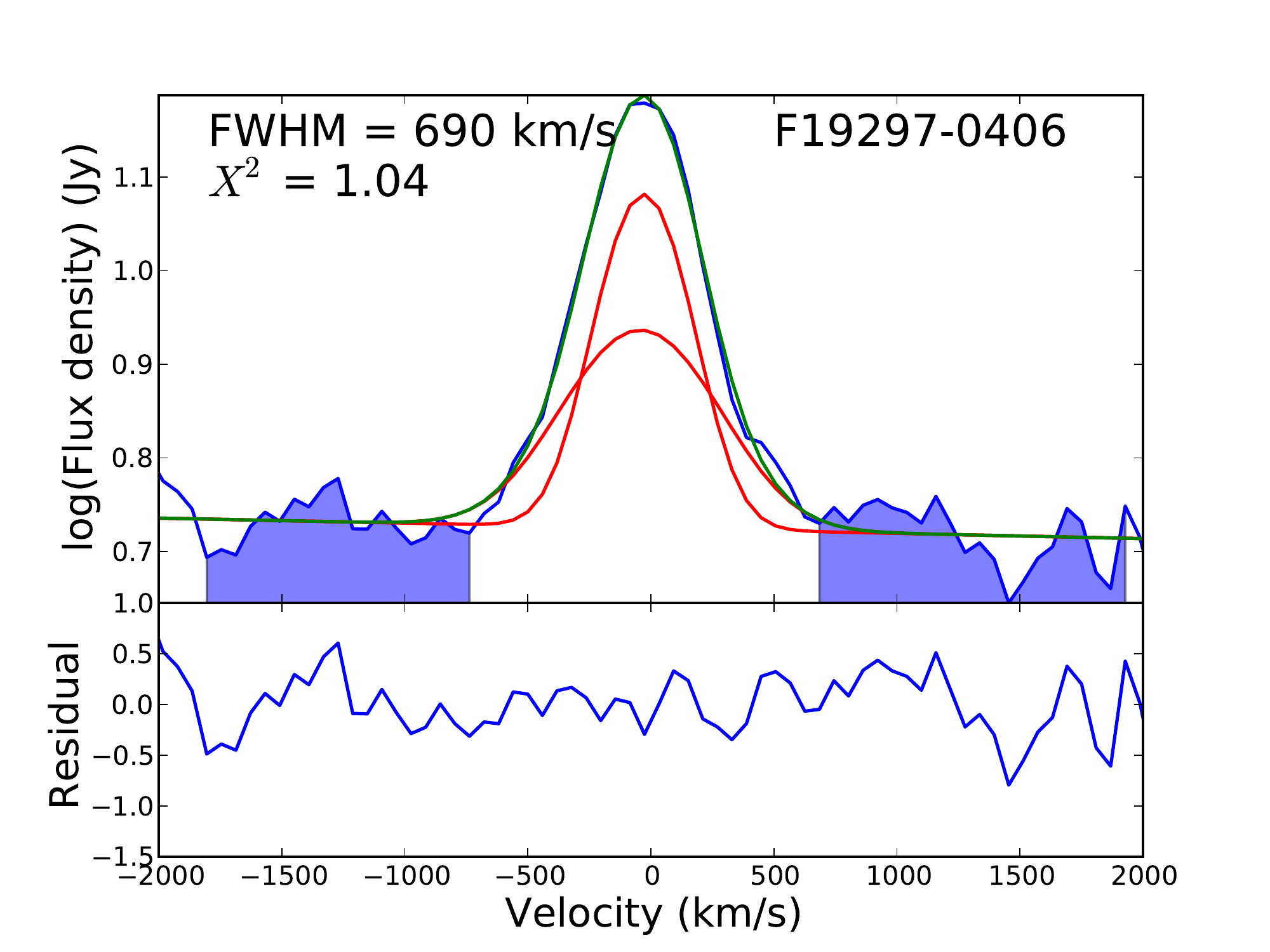}
\includegraphics[scale=0.3]{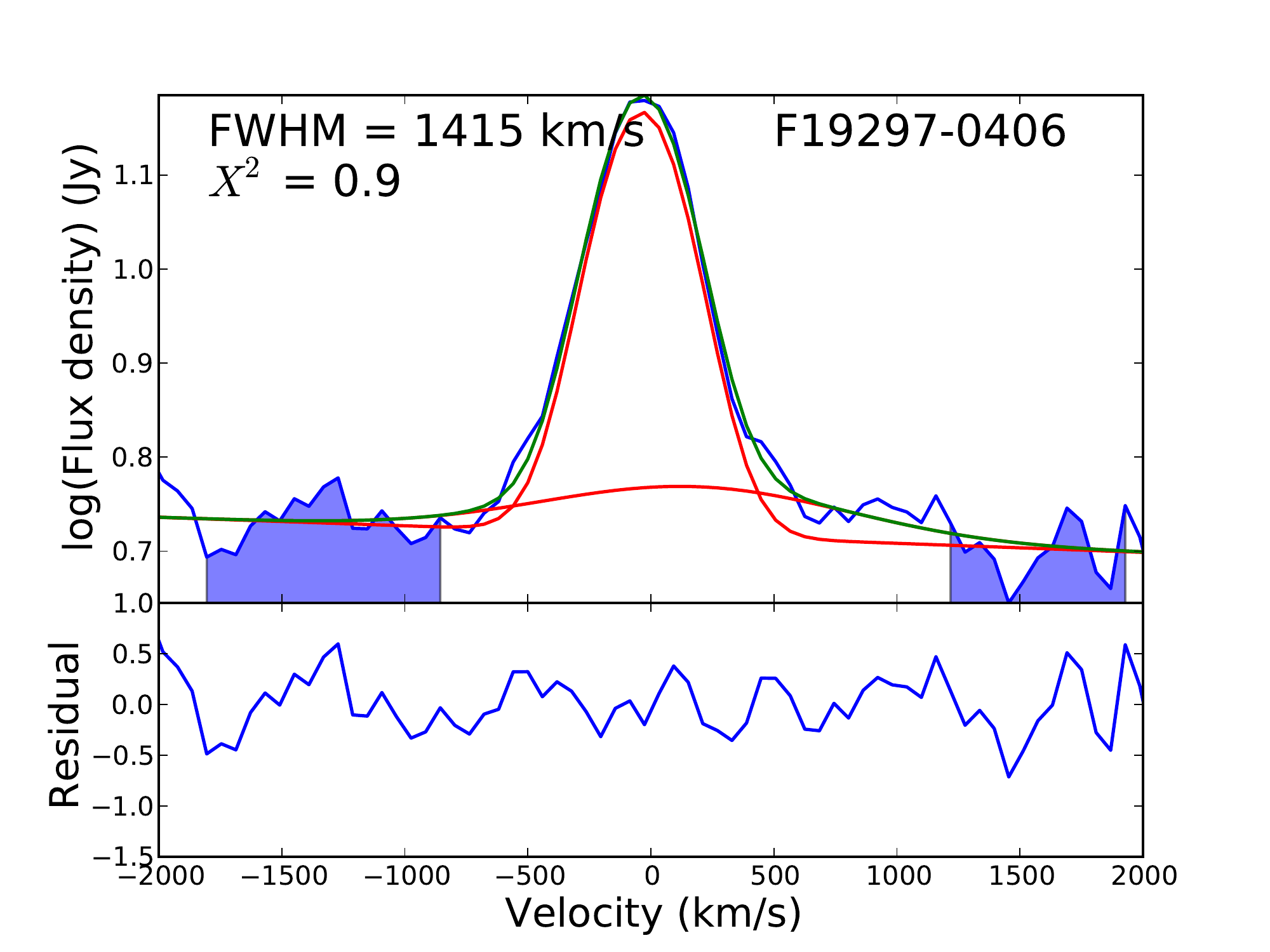}
\includegraphics[scale=0.3]{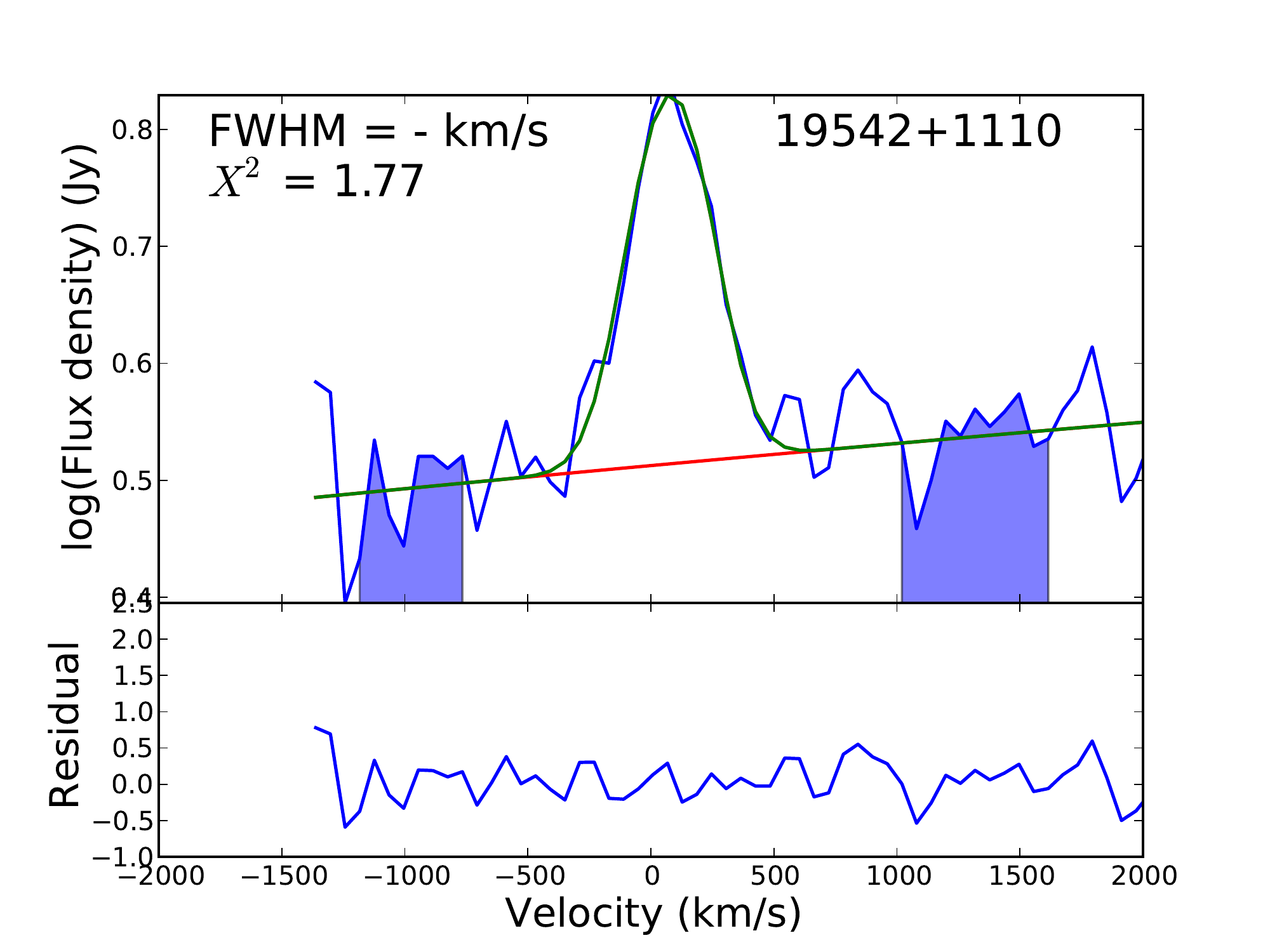}
\includegraphics[scale=0.3]{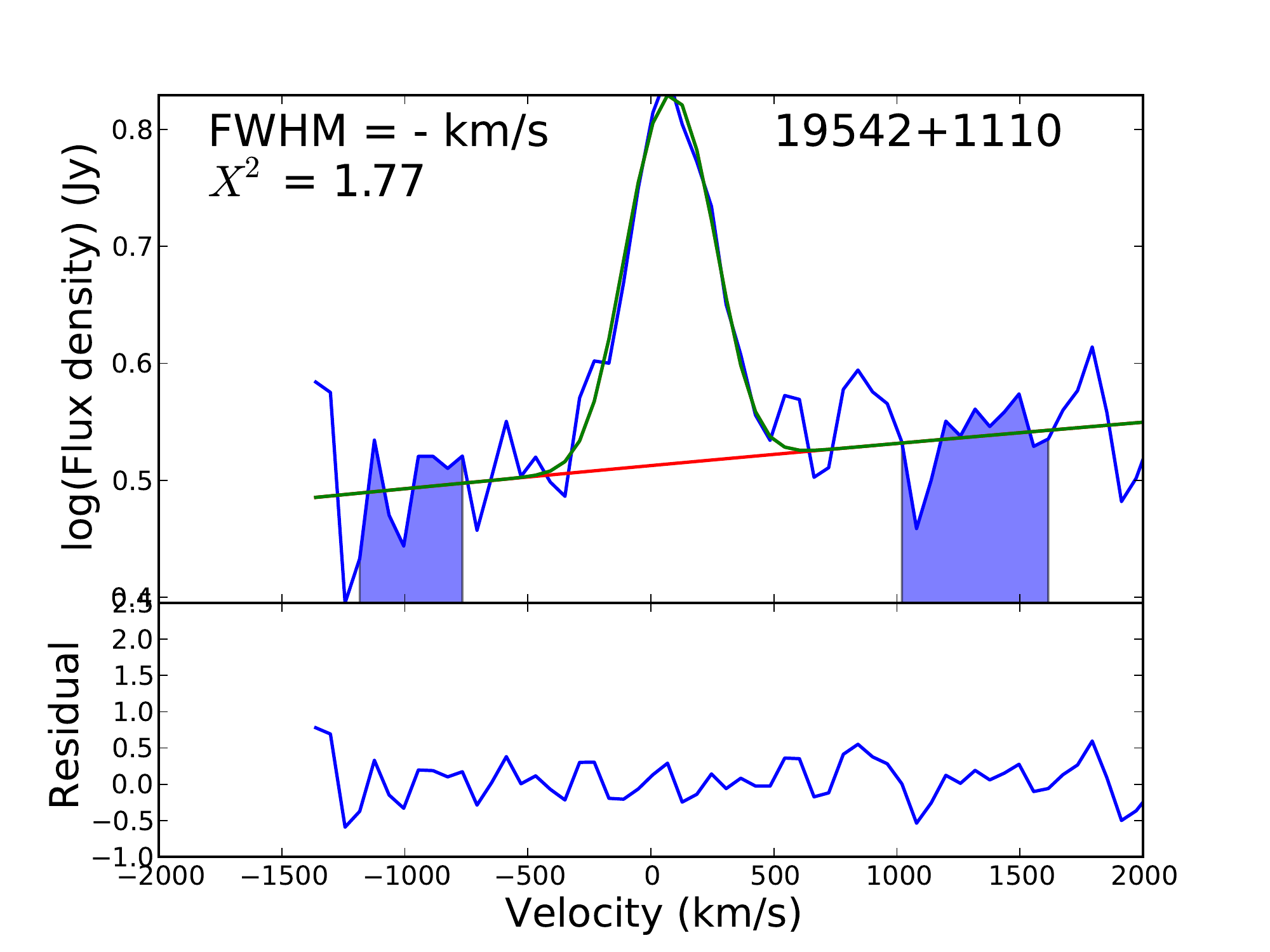}
\includegraphics[scale=0.3]{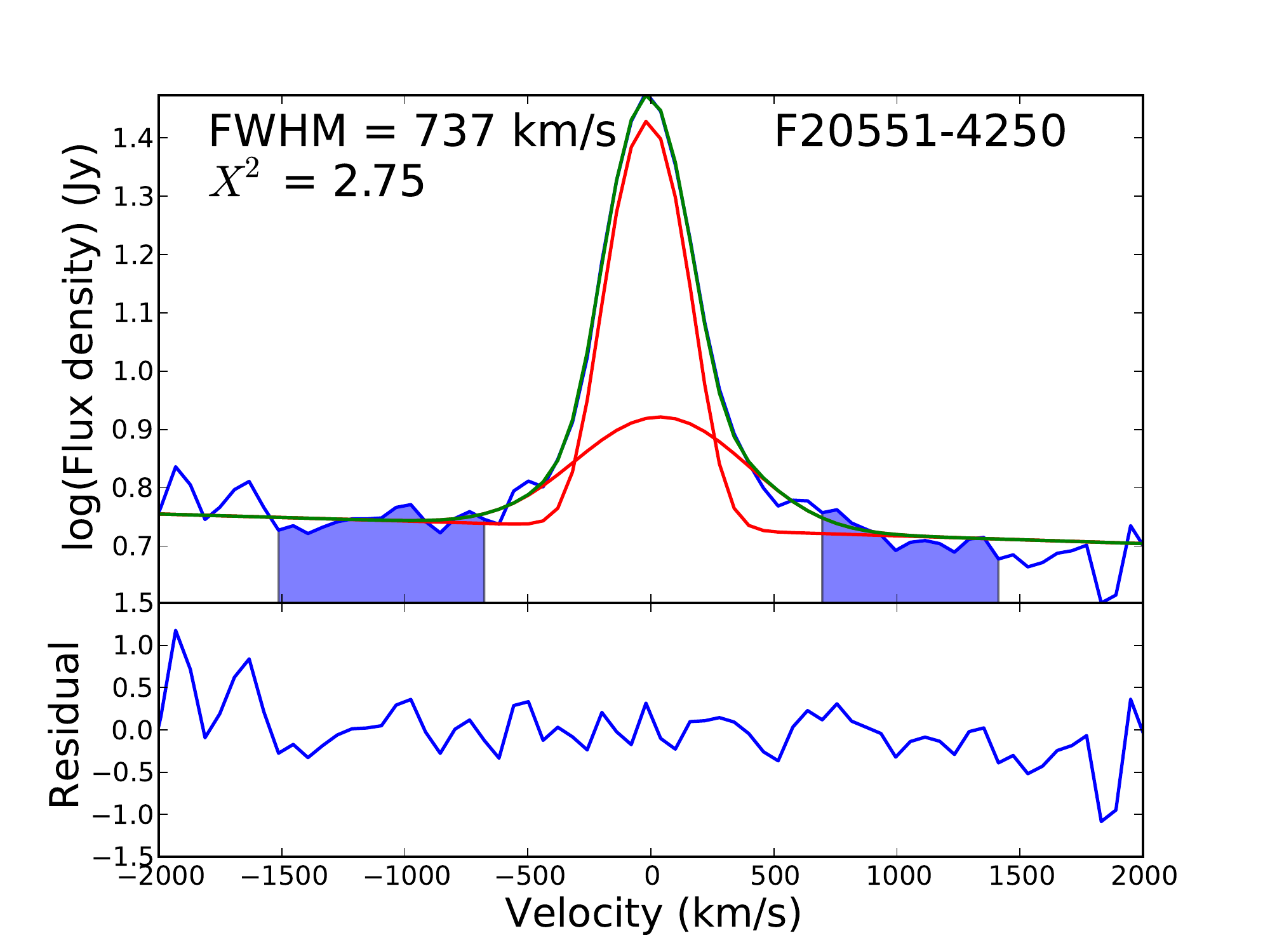}
\includegraphics[scale=0.3]{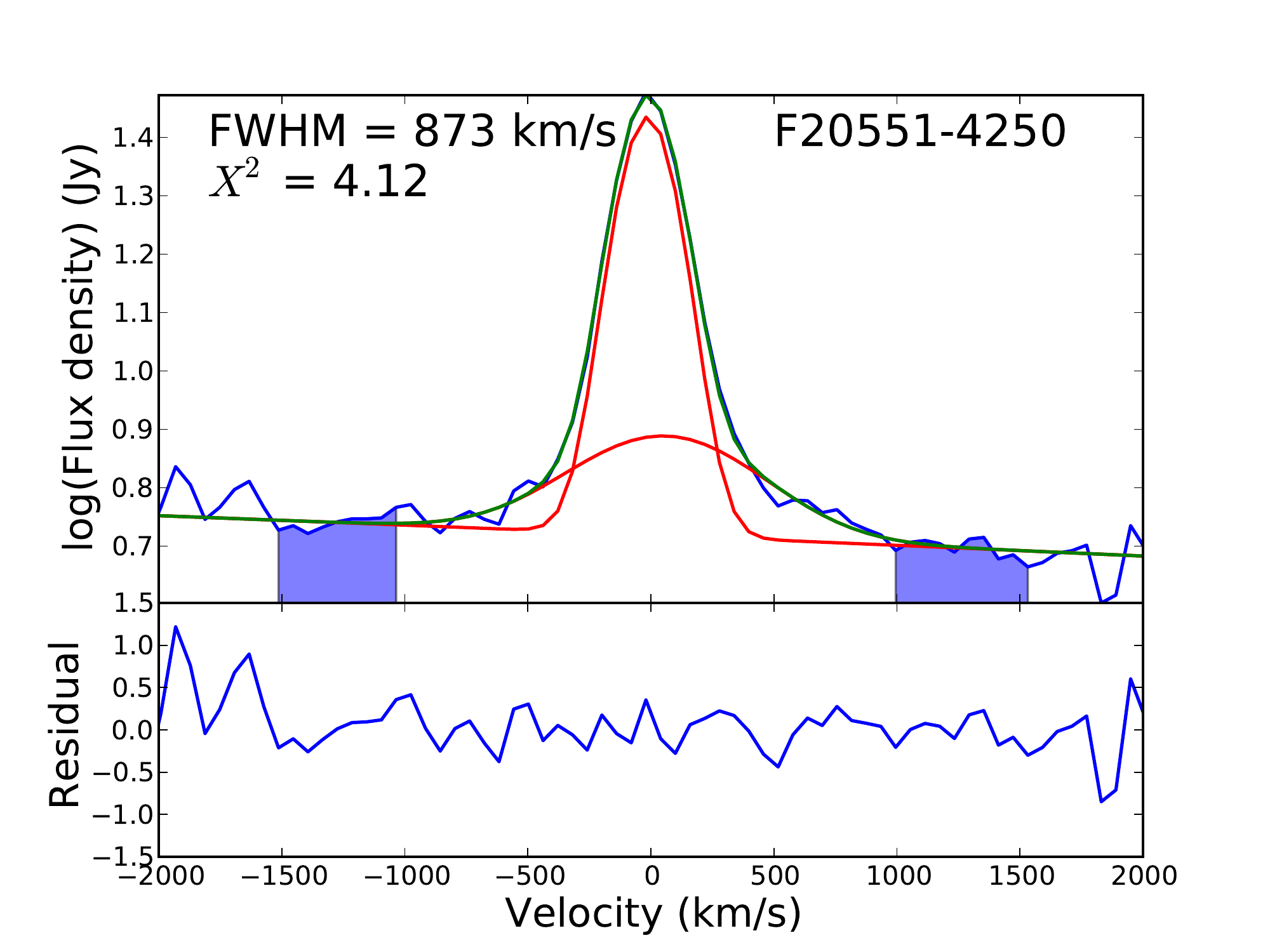}
\includegraphics[scale=0.3]{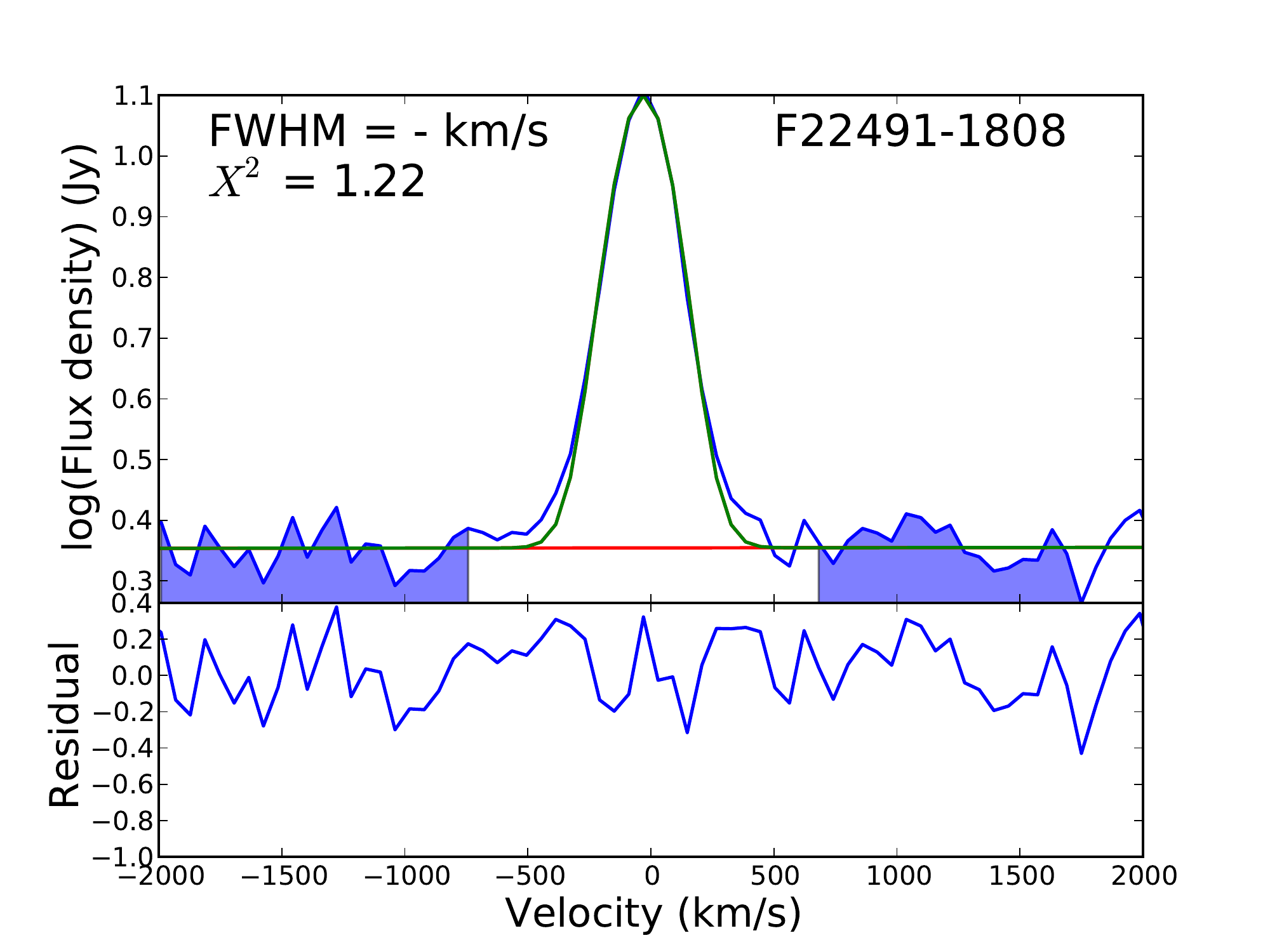}
\includegraphics[scale=0.3]{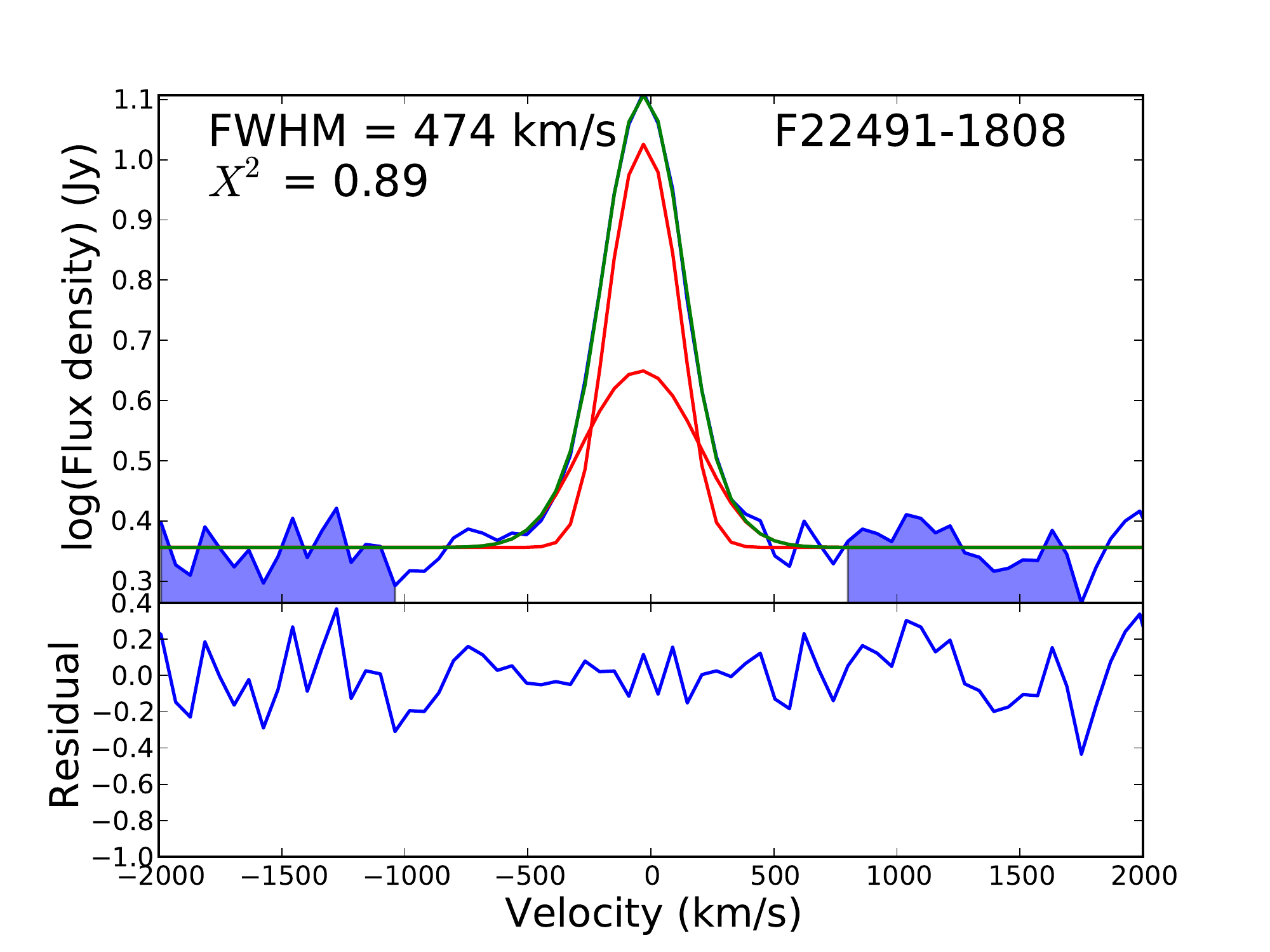}
\includegraphics[scale=0.3]{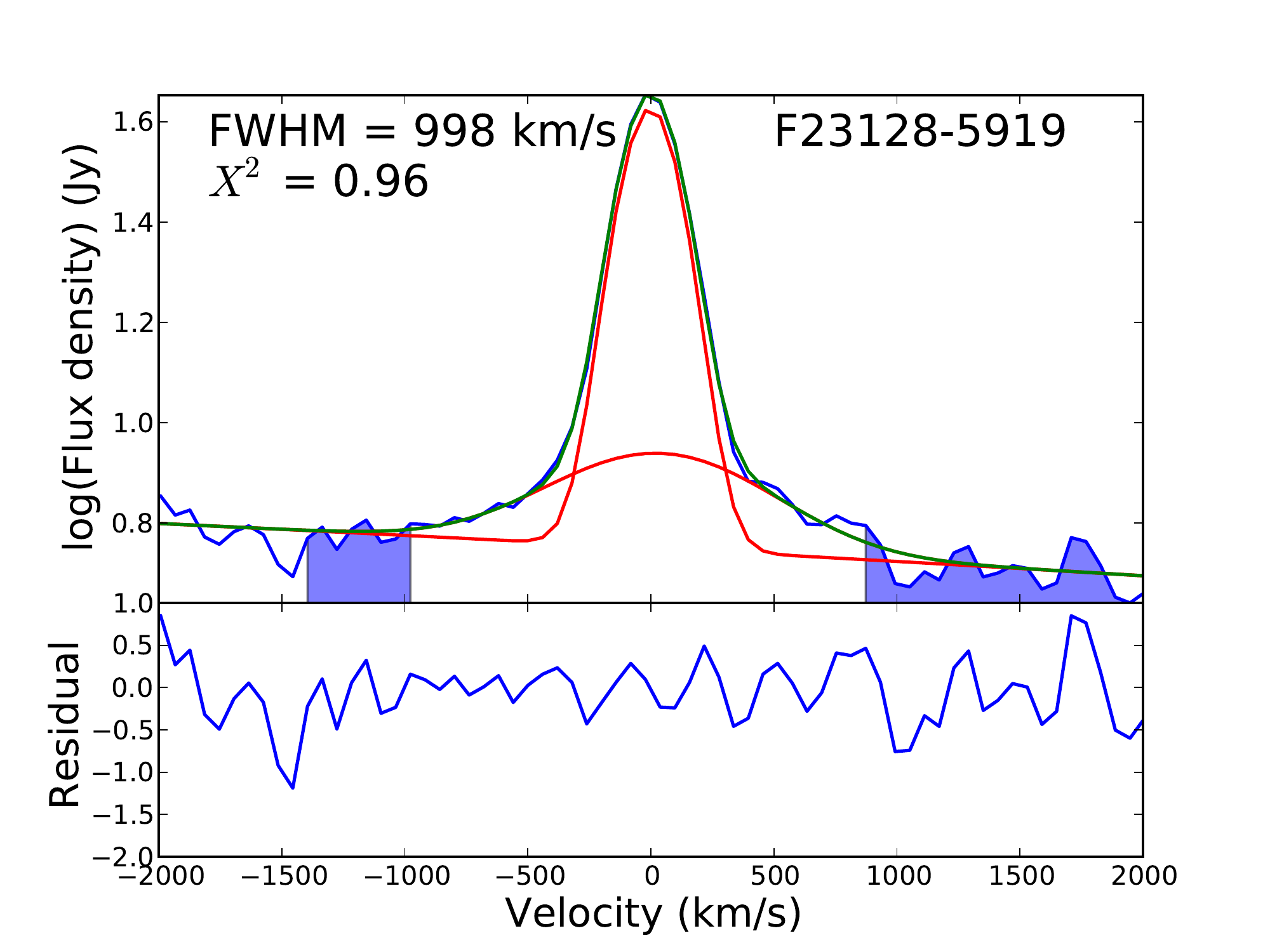}
\includegraphics[scale=0.3]{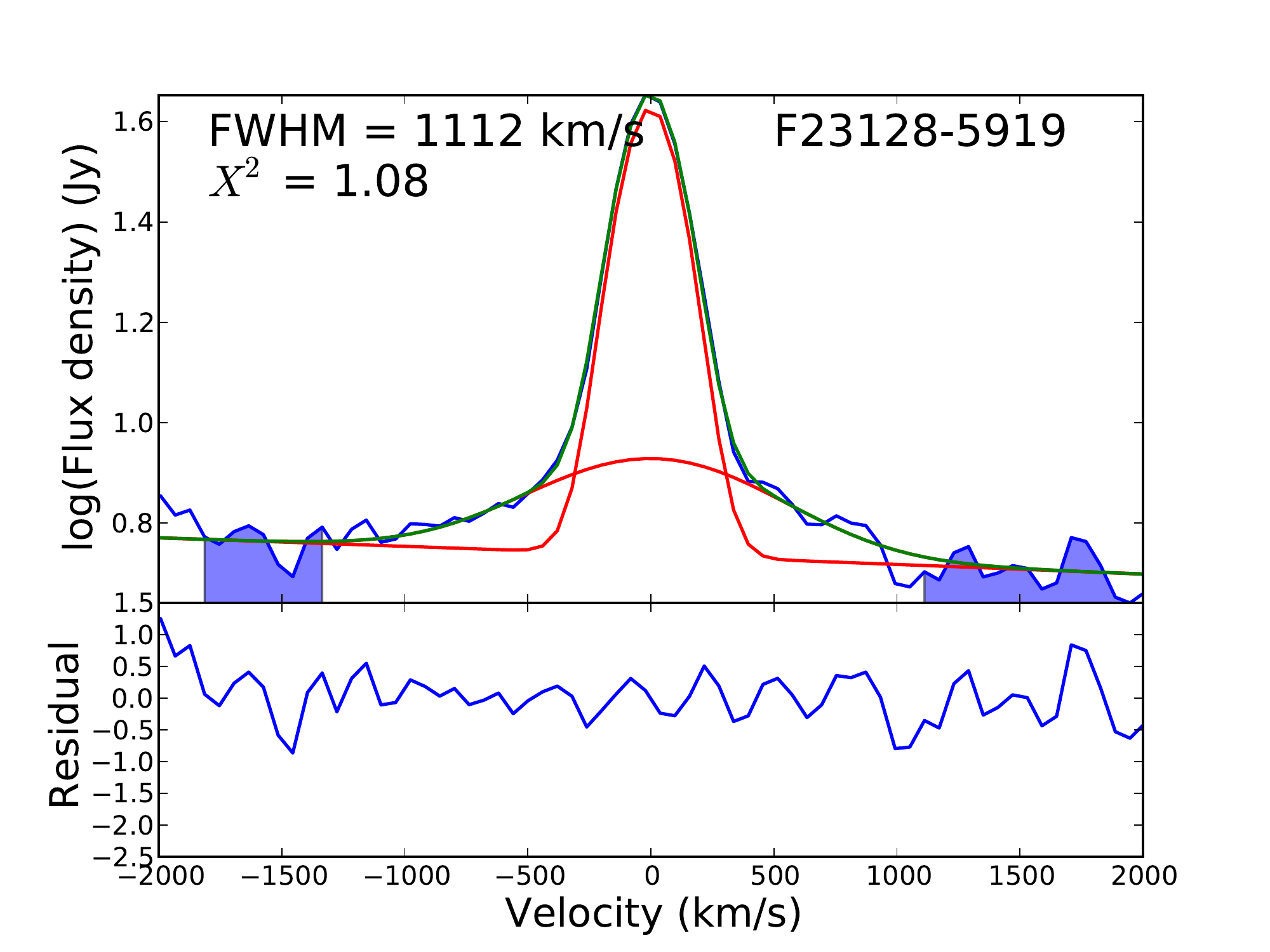}\end{figure} 
 \begin{figure} 
  \centering
\includegraphics[scale=0.3]{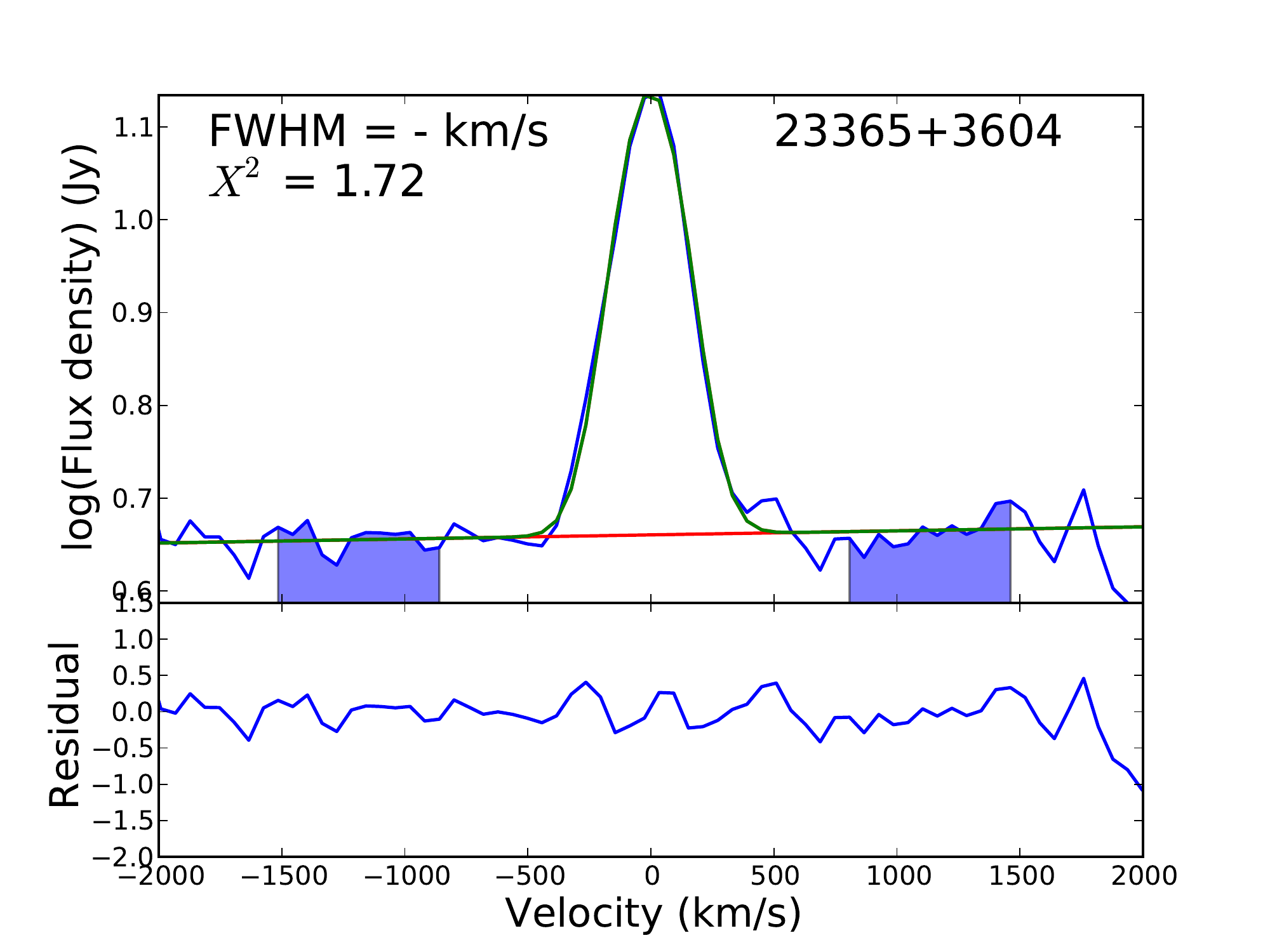}
\includegraphics[scale=0.3]{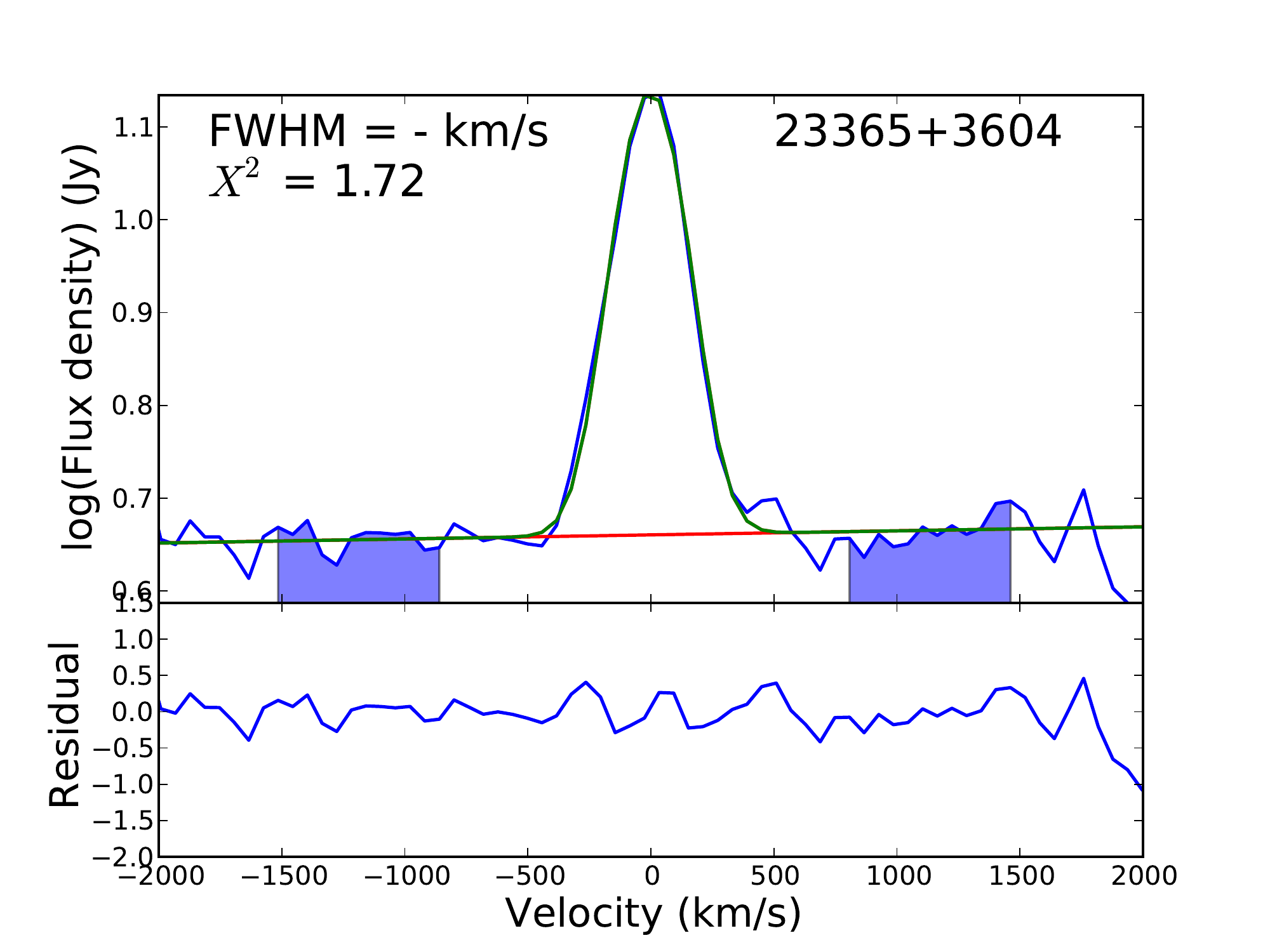}
\includegraphics[scale=0.3]{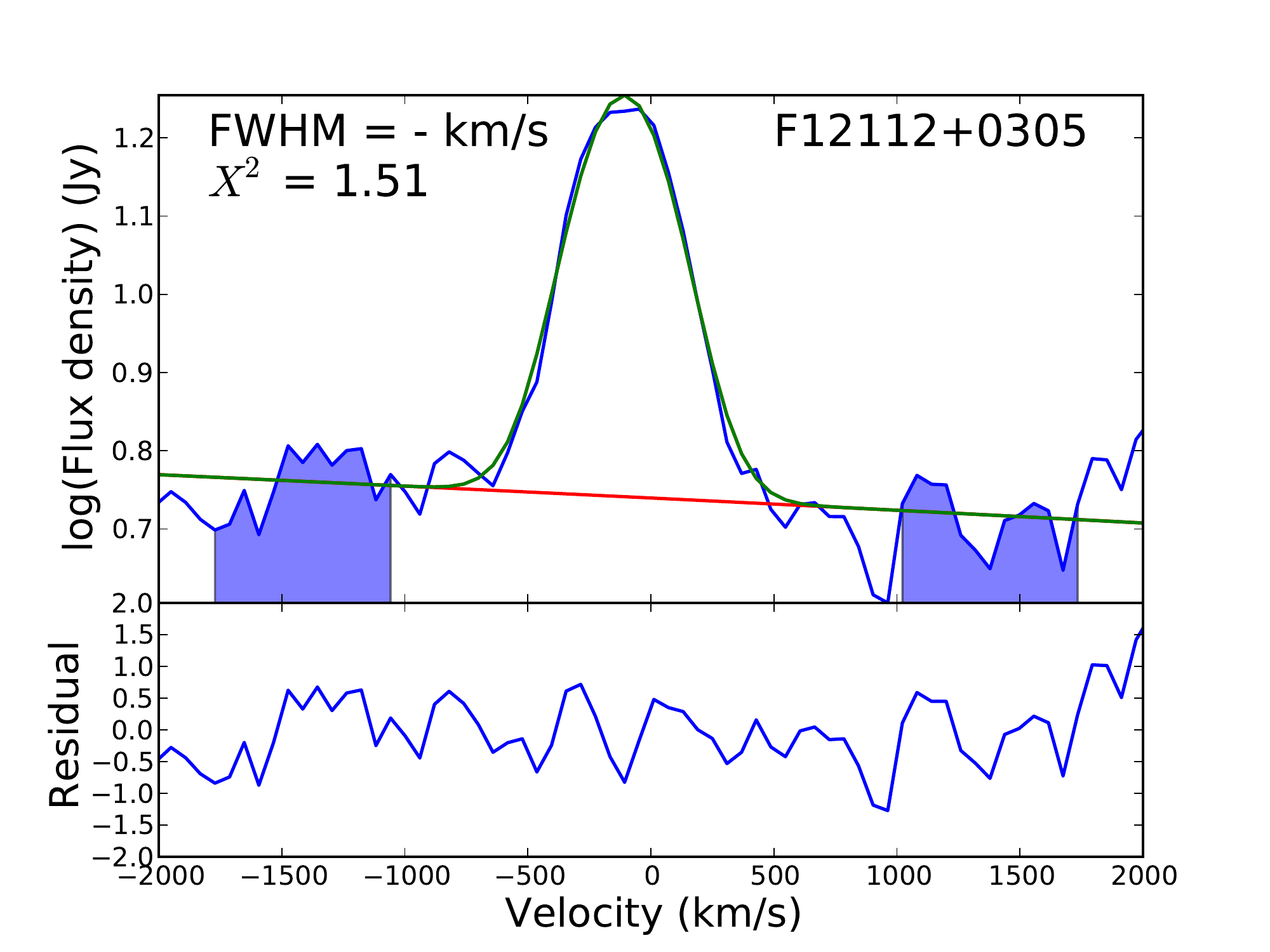}
\includegraphics[scale=0.3]{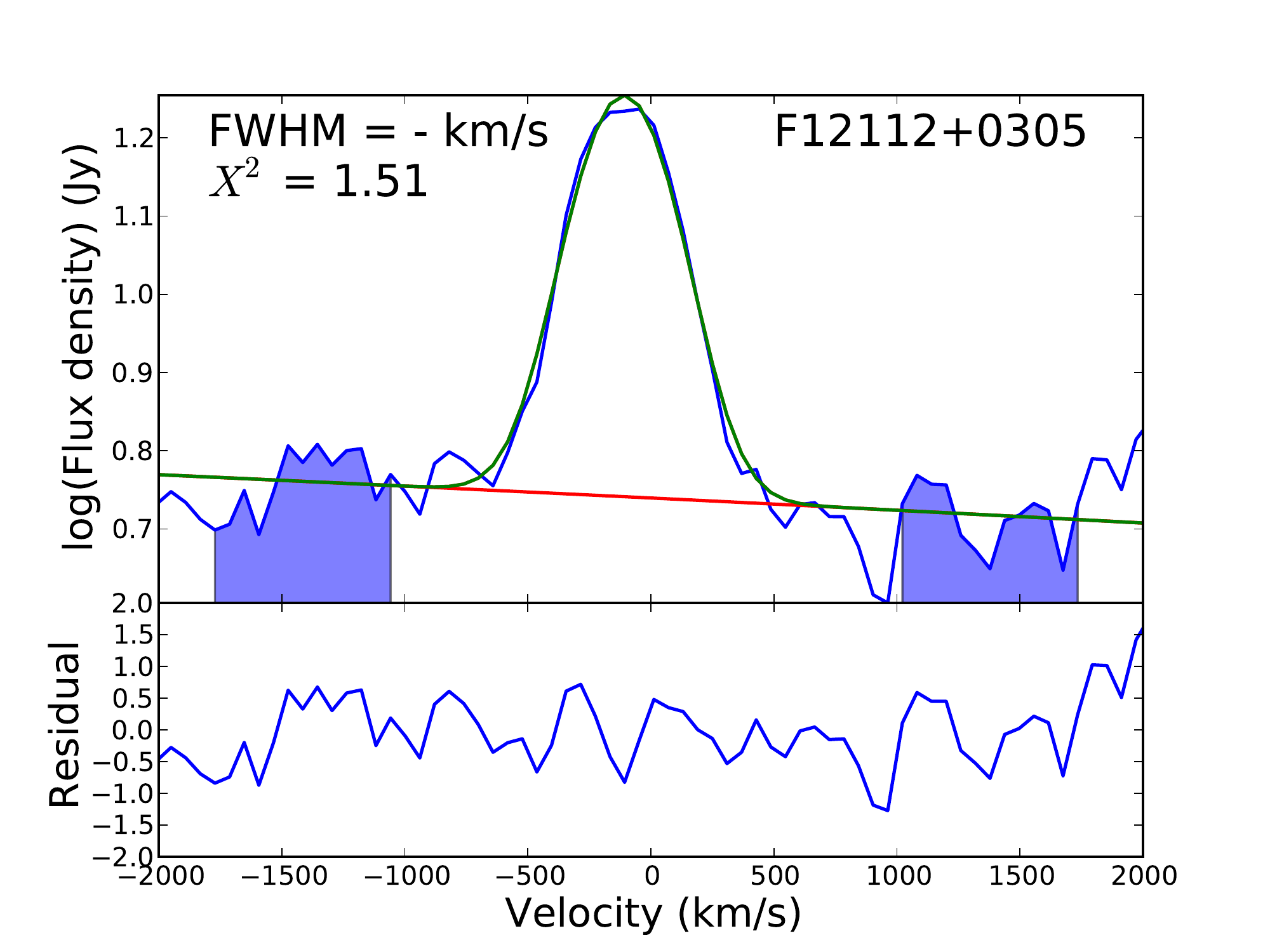}
\includegraphics[scale=0.3]{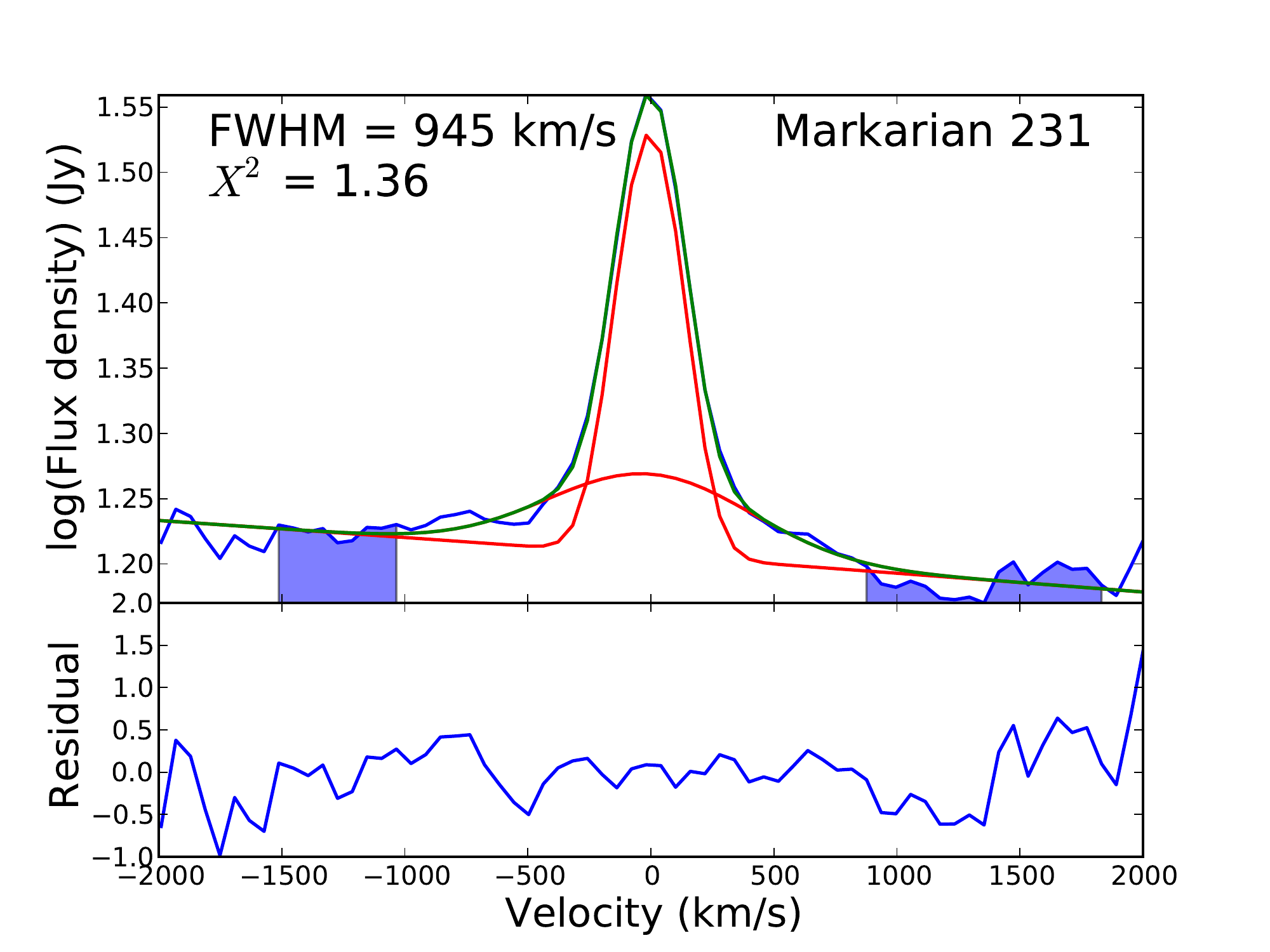}
\includegraphics[scale=0.3]{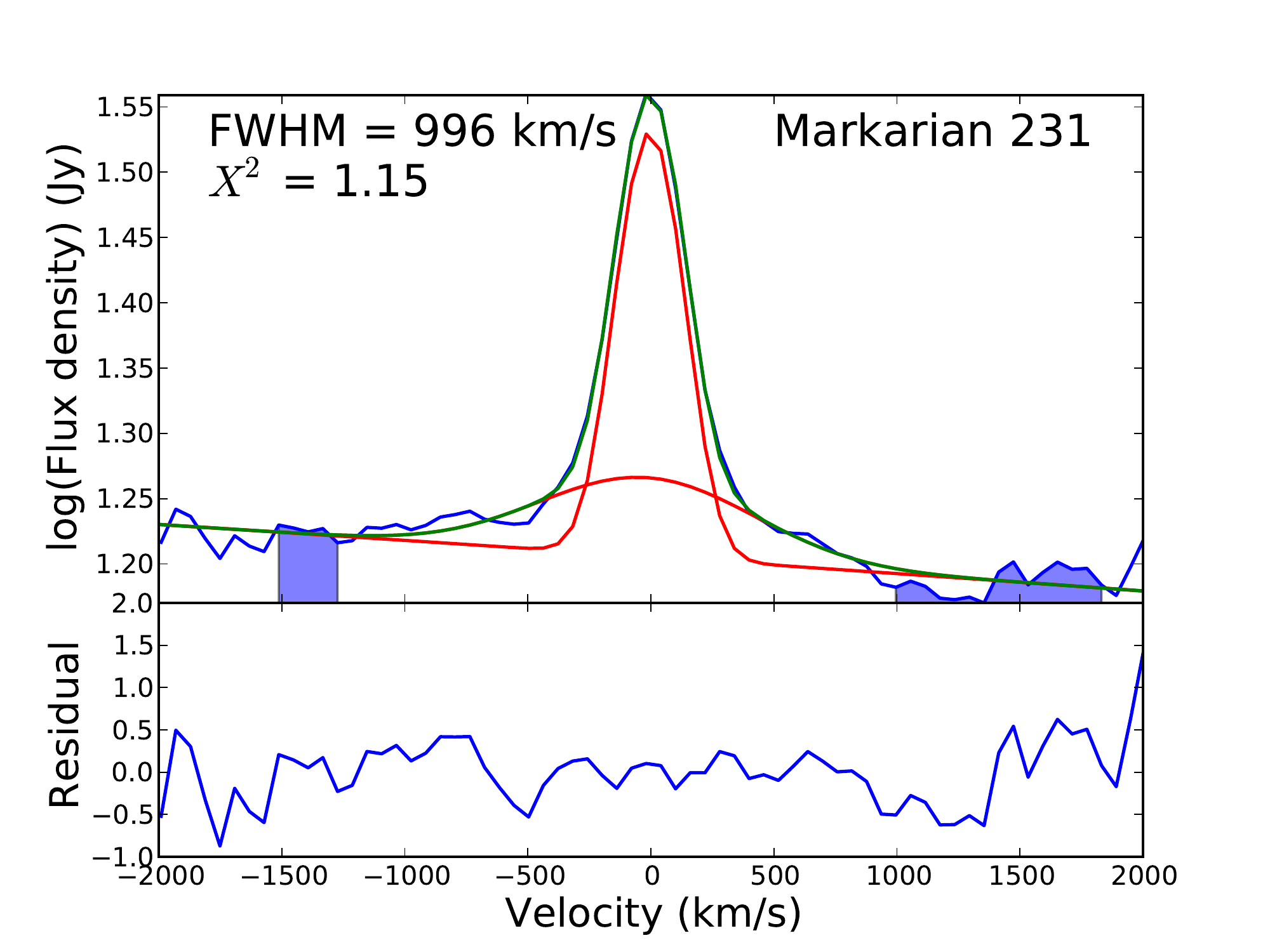}
\includegraphics[scale=0.3]{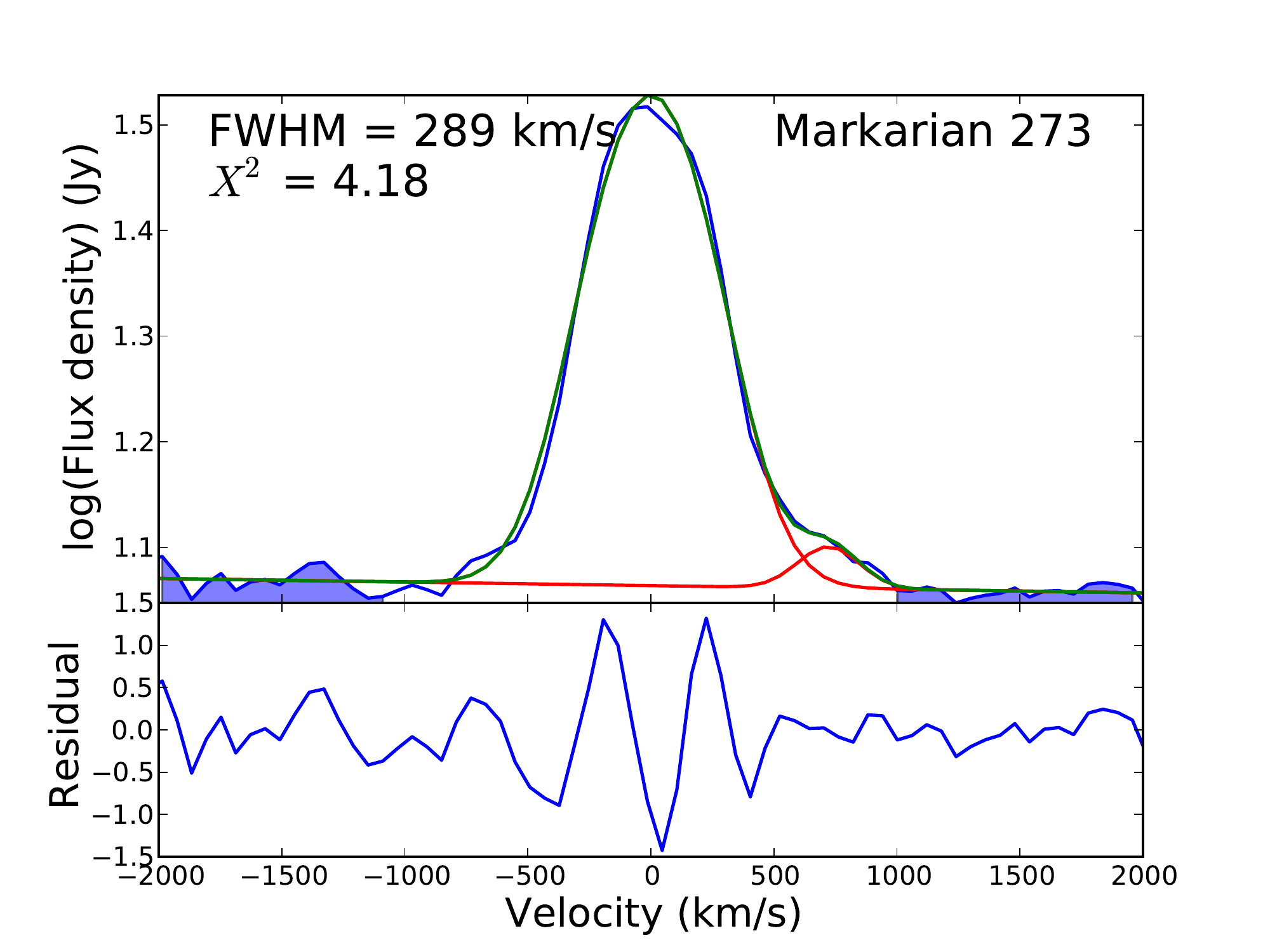}
\includegraphics[scale=0.3]{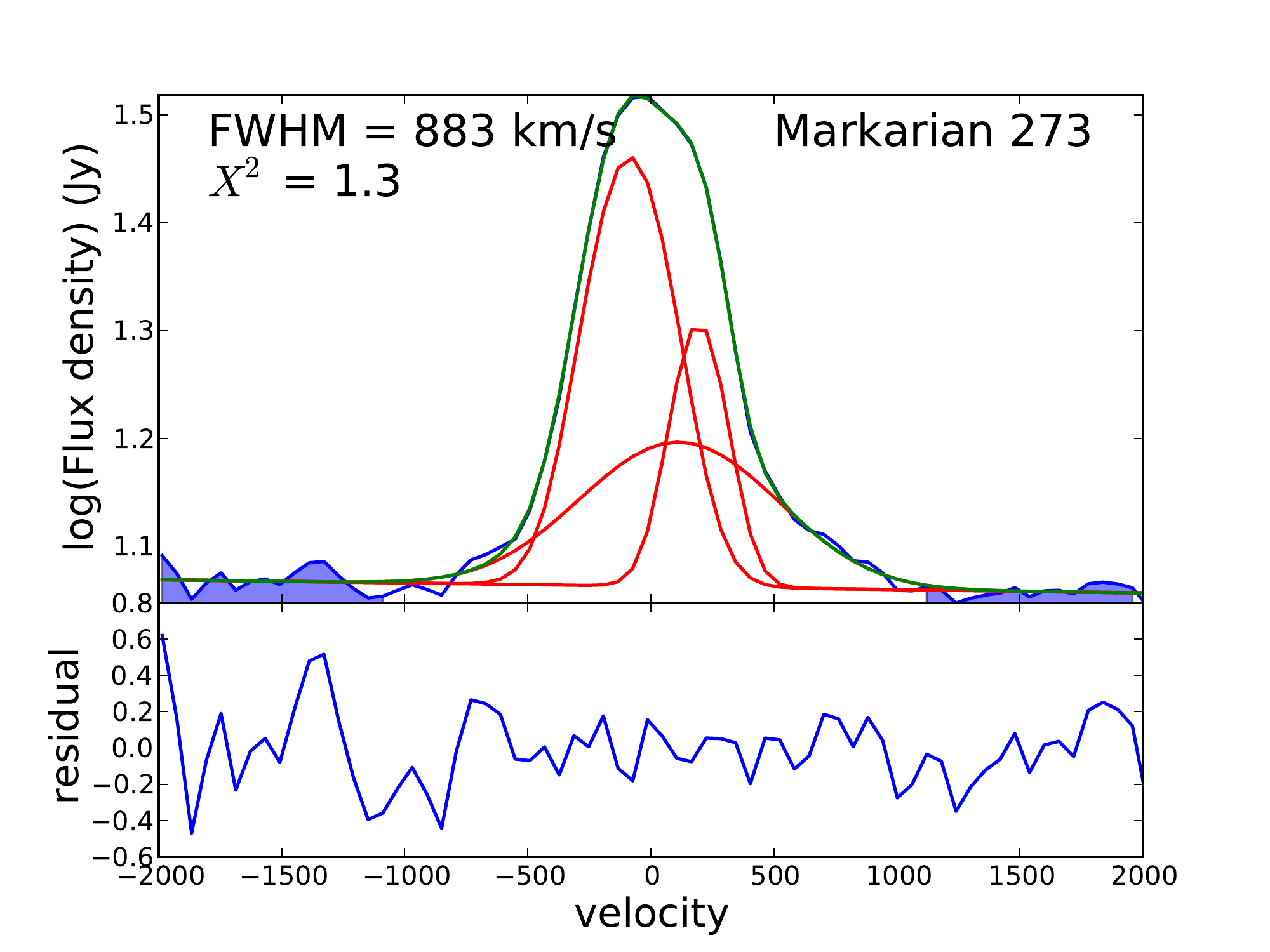}
\includegraphics[scale=0.3]{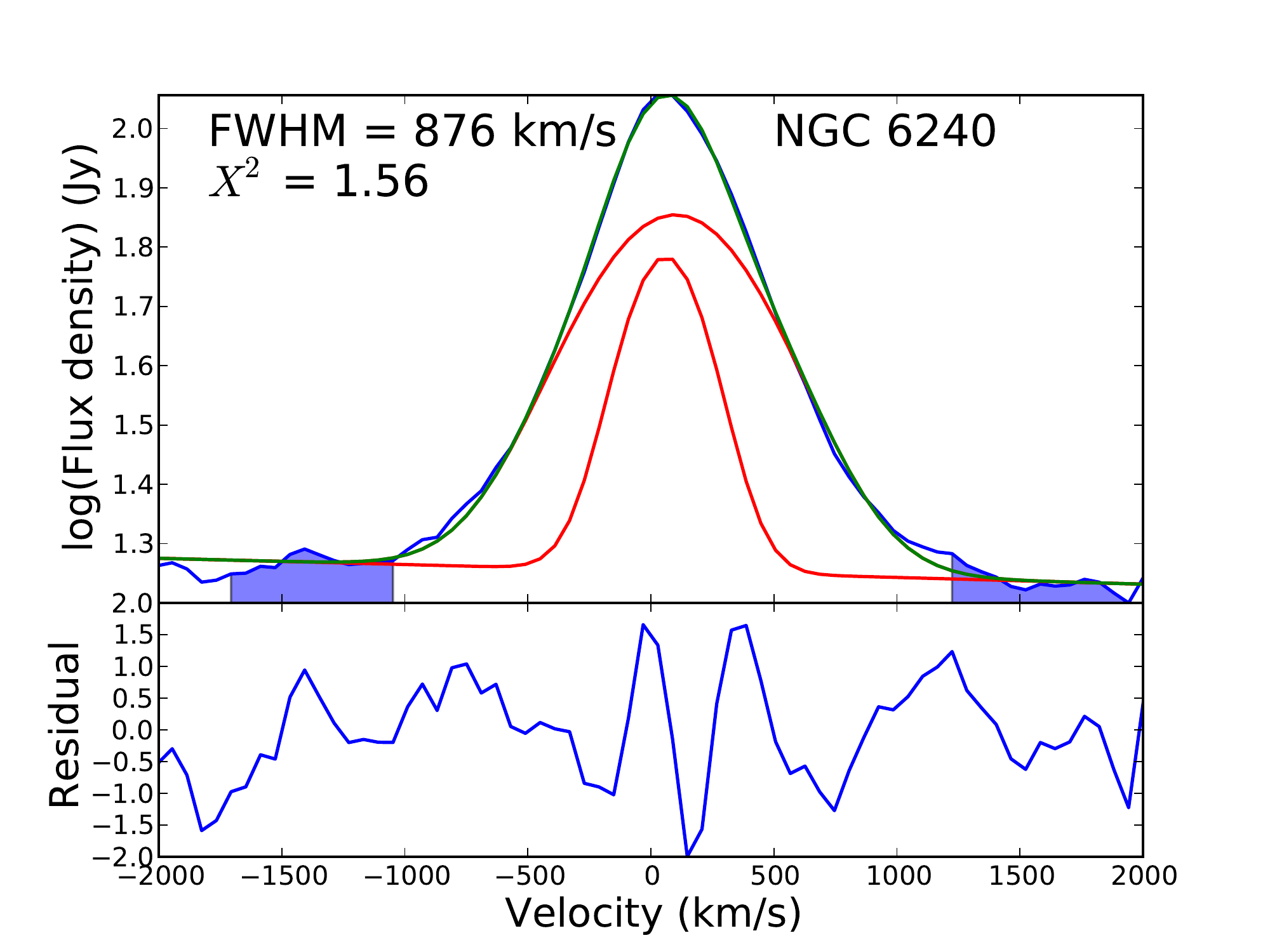}
\includegraphics[scale=0.3]{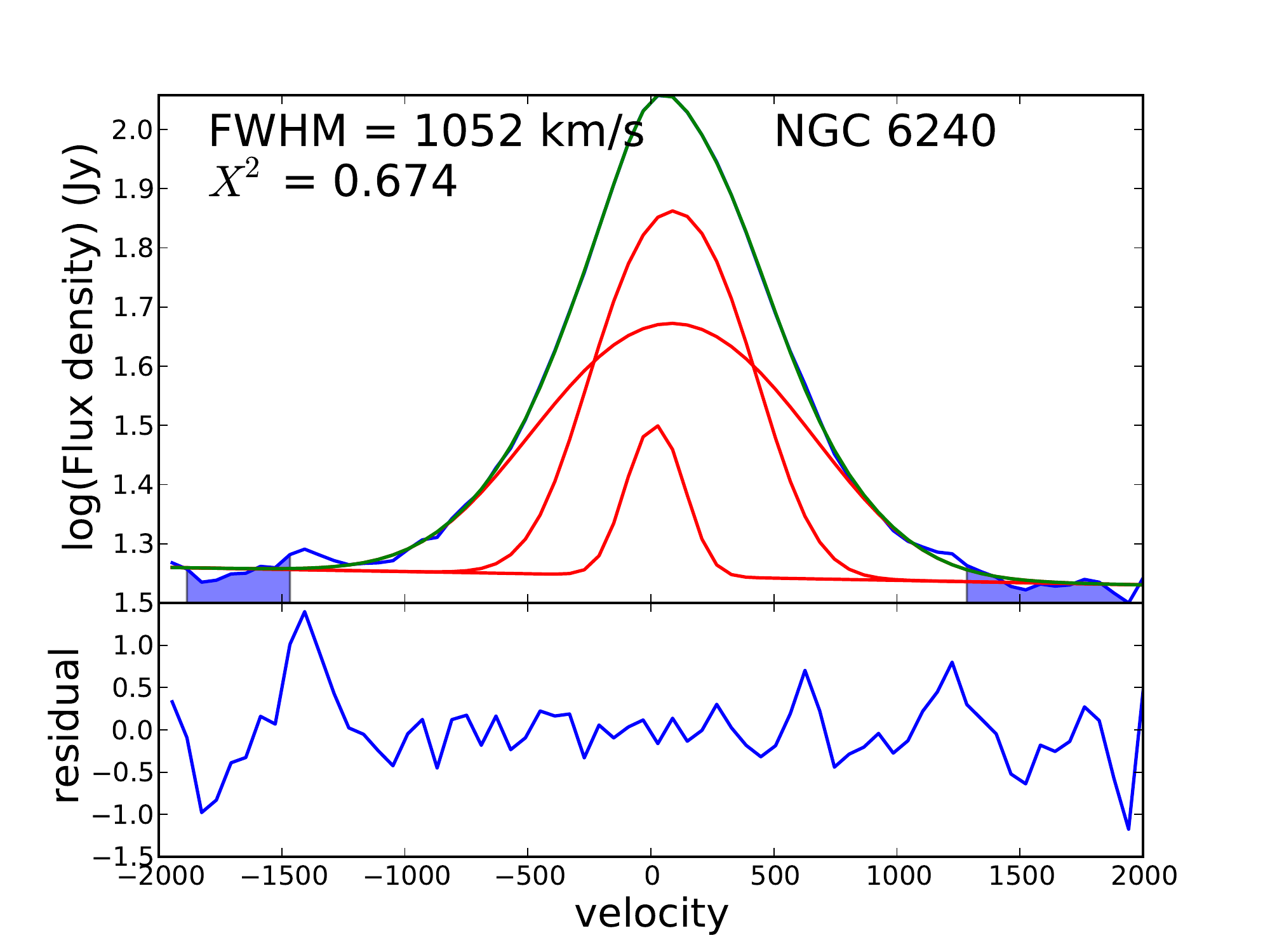}\end{figure} 

\begin{deluxetable}{lcccccccccc}
\tablecaption{\small Gaussian fits}

\tablehead{ \colhead{\small Name} & \colhead{\small h$_{1}$} & \colhead{\small m$_{1}$} & \colhead{\small FWHM$_{1}$} & \colhead{\small h$_{2}$} & \colhead{\small m$_{2}$} & \colhead{\small FWHM$_{2}$} & \colhead{\small h$_{3}$} & \colhead{\small m$_{3}$} & \colhead{\small FHWM$_{3}$} & \colhead{\small $\chi^{2}$}
}
\startdata 
\small
F05189-2524&7.43&6&363&0.86&-609&327&...&...&...&1.43\\ 
F05189-2524&6.82&7&319&0.84&-131&1237&...&...&...&1.45\\ 
07251-0248&4.95&66&449&...&...&...&...&...&...&1.18\\ 
07251-0248&4.76&69&426&0.29&-78&1507&...&...&...&1.29\\ 
F08572+3915&5.6&-31&351&0.29&-52&1372&...&...&...&0.81\\ 
F08572+3915&5.62&-31&352&0.34&-9&1799&...&...&...&1.0\\ 
09022-3615&20.58&51&410&10.35&-106&734&...&...&...&0.76\\ 
09022-3615&20.58&51&410&10.35&-106&734&...&...&...&0.76\\ 
F10565+2448&29.89&0&301&3.37&20&911&...&...&...&0.87\\ 
F10565+2448&29.91&0&301&3.36&16&918&...&...&...&0.88\\ 
13120-5453&51.24&84&414&...&...&...&...&...&...&3.49\\ 
13120-5453&50.72&84&408&1.01&45&1803&...&...&...&2.41\\ 
F14348-1447&12.15&-122&383&2.27&-174&924&...&...&...&1.51\\ 
F14348-1447&12.13&-122&383&2.28&-174&914&...&...&...&1.35\\ 
F14378-3651&5.87&156&308&...&...&...&...&...&...&2.59\\ 
F14378-3651&5.53&157&286&0.4&132&667&...&...&...&3.9\\ 
F15250+3608&5.9&51&362&...&...&...&...&...&...&2.32\\ 
F15250+3608&5.9&51&362&...&...&...&...&...&...&2.32\\ 
F17207-0014&26.32&-1&533&...&...&...&...&...&...&2.71\\ 
F17207-0014&26.32&-1&533&...&...&...&...&...&...&2.71\\ 
F19297-0406&6.76&-28&413&3.33&-40&690&...&...&...&1.04\\ 
F19297-0406&9.46&-35&472&0.67&178&1415&...&...&...&0.9\\ 
19542+1110&3.5&81&366&...&...&...&...&...&...&1.77\\ 
19542+1110&3.5&81&366&...&...&...&...&...&...&1.77\\ 
F20551-4250&21.44&-14&302&2.98&41&737&...&...&...&2.75\\ 
F20551-4250&22.0&-13&307&2.52&59&873&...&...&...&4.12\\ 
F22491-1808&10.33&-30&300&...&...&...&...&...&...&1.22\\ 
F22491-1808&8.34&-28&259&2.19&-40&474&...&...&...&0.89\\ 
F23128-5919&36.73&-1&307&3.08&38&998&...&...&...&0.96\\ 
F23128-5919&36.92&-1&308&3.04&14&1112&...&...&...&1.08\\ 
F23365+3604&9.18&-2&320&...&...&...&...&...&...&1.72\\ 
F23365+3604&9.18&-2&320&...&...&...&...&...&...&1.72\\ 
F12112+0305&12.46&-108&483&...&...&...&...&...&...&1.51\\ 
F12112+0305&12.46&-108&483&...&...&...&...&...&...&1.51\\ 
Mrk 231&17.79&-3&291&2.47&-8&945&...&...&...&1.36\\ 
Mrk 231&17.9&-3&292&2.4&-20&996&...&...&...&1.15\\ 
Mrk 273&22.19&0&561&1.09&722&289&...&...&...&4.18\\ 
Mrk 273&17.33&-85&394&8.74&194&258&4.17&114&885&1.3\\ 
NGC 6240&42.82&59&415&53.63&100&876&...&...&...&1.56\\ 
NGC 6240&14.04&17&248&55.25&90&570&29.45&86&1053&0.67\\ 
UGC 5101&15.91&-186&298&17.65&160&357&...&...&...&0.92\\ 
UGC 5101&15.69&-181&295&16.96&161&340&0.91&118&1182&0.83\\ 
Arp 220 & 28.15&-22&521&...&...&...&...&...&...&14.53\\
Arp 220 & 28.15&-22&521&...&...&...&...&...&...&14.53\\
\enddata
\tablecomments{Parameters of Gaussian fits to the [CII] spectra. The units are Jansky for the height and $\text{km}\,\text{s}^{-1}$ for the mean and FWHM. $\chi^{2} = \frac{1}{N -n -1}  \Sigma_{i=1}^{N} (obs_{i} - fit_{i})^{2} / \sigma^{2}$, with $N$ the number of data points in the spectrum, $n$ the number of free parameters during the fit, and $\sigma$ the RMS in the continuum. Each source has been fitted twice with different placements of the continuum, and appears therefore twice in the table.}
\label{tab:appA}
\end{deluxetable}

\end{document}